\documentclass[11pt]{article}

\usepackage[final]{acl}
\usepackage{times}
\usepackage{latexsym}
\usepackage[T1]{fontenc}

\usepackage{multirow}
\usepackage{graphicx}
\usepackage{times}
\usepackage{amsmath}
\usepackage{latexsym}
\usepackage{booktabs}
\usepackage{enumitem}
\usepackage[normalem]{ulem}
\usepackage{pifont}  
\newcommand{\xmark}{\textcolor{red}{\ding{55}}}  
\usepackage{tcolorbox}
\tcbuselibrary{listings, breakable, skins}

\newtcblisting{promptbox}[1][]{
  colback=gray!5,
  colframe=gray!60,
  title={#1},
  fonttitle=\bfseries,
  sharp corners,
  boxrule=0.5pt,
  left=5pt, right=5pt, top=5pt, bottom=5pt,
  breakable, 
  enhanced,
  listing only, 
  listing options={
    basicstyle=\small\ttfamily, 
    breaklines=true,            
    breakatwhitespace=true,     
    columns=fullflexible,       
    keepspaces=true,            
    language={},                
    aboveskip=0pt,
    belowskip=0pt
  }
}

\usepackage[utf8]{inputenc}

\usepackage{microtype}

\usepackage{inconsolata}

\usepackage{graphicx}
\newcommand{\ours}{\textsc{LetGames}}
\newcommand{\oureval}{\textsc{LetGamesEval}}

\title{\ours: An LLM-Powered Gamified Approach to Cognitive Training \\for Patients with Cognitive Impairment}

\author{
  Jingwei Shi$^{1}$ \quad 
  Shengyu Tao$^{1}$ \quad 
  Xinxiang Yin$^{2}$ \quad 
  Chen Huang$^{3}$\thanks{Corresponding author.} \\ 
  \textbf{Wenqiang Lei}$^{4}$ \quad 
  \textbf{See-Kiong Ng}$^{3}$\\
  $^1$Shanghai University of Finance and Economics, China\\
  $^2$Northwest Polytechnical University Xi'an, China\\
  $^3$Institute of Data Science, National University of Singapore, Singapore\\
  $^4$College of Computer Science, Sichuan University, China\\
  \texttt{shijingwei@stu.sufe.edu.cn} \quad \texttt{huang\_chen@nus.edu.sg} 
}

\begin{document}
\maketitle
\begin{abstract}
The application of games as a therapeutic tool for cognitive training is beneficial for patients with cognitive impairments. However, effective game design for individual patients is resource-intensive. To this end, we propose an LLM-powered method, \ours, for automated and personalized therapeutic game design. Inspired by the Dungeons \& Dragons, \ours\ generates an open-world interactive narrative game. It not only generates game scenarios and challenges that target specific cognitive domains, but also employs conversational strategies to offer guidance and companionship. To validate its efficacy, we pioneer a psychology-grounded evaluation protocol \oureval, establishing comprehensive metrics for rehabilitative assessment. Building upon this, our experimental results from both LLM-based assessors and human expert evaluations demonstrate the significant potential of our approach, positioning \ours~as a promising solution to the widespread need for more accessible and tailored cognitive training tools. Our code is available at \url{https://github.com/shi0712/LETGAMES}.

\end{abstract}

\section{Introduction}
Fueled by an aging population and the rising prevalence of neurodegenerative diseases like Alzheimer's, the global burden of cognitive impairment\footnote{Appendix \ref{background} offers background on cognitive impairment.} is escalating \cite{khalil2025redefining, moran2024serious, subramaniam2025prevalence}, with patient numbers projected to surpass 150 million by 2030 \cite{livingston2020dementia}. This motivates the demand for cognitive health solutions, especially non-pharmacological interventions like cognitive training \cite{choi2025narrative}. \textbf{Serious Games (SGs)}, games designed for a primary purpose beyond entertainment \cite{charlier2016serious, manera2017recommendations}, have emerged as a particularly promising tool \cite{dell2024video,anguera2015video,tan2023digital}. Unlike other cognitive training methods like Transcranial Magnetic Stimulation \cite{nardone2014transcranial}, SGs offer a scalable, affordable, and highly accessible therapeutic solution, alleviating the burden on centralized hospital facilities \cite{ning2020review}. Consequently, various SGs have been developed, ranging from traditional board games \cite{dartigues2013playing} to immersive virtual reality experiences \cite{padala2012wii}, targeting general cognition \cite{bing2014research} or specific impairments \cite{benveniste2012minwii, manera2015kitchen,catricala2025exploiting}.

\begin{figure}
    \centering
    \includegraphics[width=0.45\textwidth]{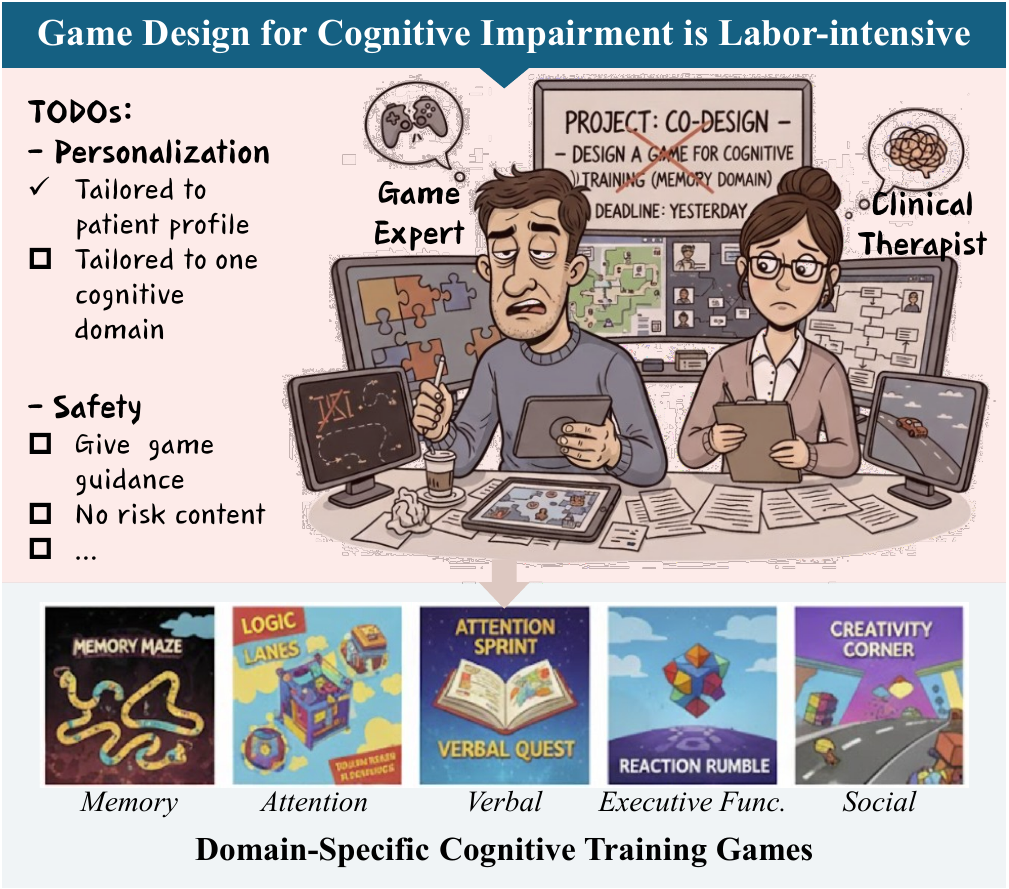}
    \caption{Creating an effective game for cognitive training demands a joint effort between experts, making the development process labor-intensive.}
    \label{fig:game}
    \vspace{-3mm}
\end{figure}

\begin{table*}[t]
\centering
\resizebox{0.98\textwidth}{!}{%
\begin{tabular}{llllll}
\toprule
\textbf{Serious Games} & \textbf{\begin{tabular}[c]{@{}l@{}}Cognitive\\ Domain\end{tabular}} & \textbf{\begin{tabular}[c]{@{}l@{}}Design Cost\end{tabular}} & \textbf{\begin{tabular}[c]{@{}l@{}}Extra Safety \\ Protection\end{tabular}} & \textbf{\begin{tabular}[c]{@{}l@{}}Open\\ Sourced\end{tabular}} & \textbf{\begin{tabular}[c]{@{}l@{}}Brief Intro.\end{tabular}} \\ \midrule
\begin{tabular}[c]{@{}l@{}}MINWii \\\cite{benveniste2012minwii}\end{tabular} & Memory & High (Human-generated) & No & No & Music therapy game\\ \hline
\begin{tabular}[c]{@{}l@{}}Kitchen and Cooking\\\cite{manera2015kitchen}\end{tabular} & \begin{tabular}[c]{@{}l@{}}Attention, \\Executive Function\end{tabular} & High (Human-generated) & Game guidance & No & Recipe-based cooking game\\ \hline
\begin{tabular}[c]{@{}l@{}}Big Brain Academy\\\cite{dinis2019usability}\end{tabular} & \begin{tabular}[c]{@{}l@{}}Memory, \\Executive Function\end{tabular} & High (Human-generated) & Game guidance & No & Puzzle video game\\ \hline
\begin{tabular}[c]{@{}l@{}}ReMe\\\cite{wang2024potentialvalueaichatbot}\end{tabular} & Memory & \begin{tabular}[c]{@{}l@{}}Medium \\(LLM-expert co-generated)\end{tabular} & Game guidance & No & \begin{tabular}[c]{@{}l@{}}Conversational game without narratives;\\ Two human-designed games \end{tabular} \\ \hline
\ours~(\textit{ours}) & \begin{tabular}[c]{@{}l@{}}Memory, \\Attention, \\Verbal Learning, \\Executive Function, \\Social Cognition\end{tabular} & \begin{tabular}[c]{@{}l@{}}Low \\(LLM-generated)\end{tabular} & \begin{tabular}[c]{@{}l@{}}Game guidance, \\ Conversational\\companionship  \end{tabular} & Yes & Conversational narrative game\\ \bottomrule
\end{tabular}%
}
\caption{Representative serious games for cognitive training. Such games are not only costly to create but also limited in scope, failing to cover varying cognitive domains or provide adequate safety assurances.}
\vspace{-3mm}
\label{tab:games}
\end{table*}

However, \textbf{creating an effective cognitive training game demands a joint effort between game experts and clinical therapists, which is labor-intensive} \cite{yang2024psychogat, canapa2025interactive,wang2024potentialvalueaichatbot}. This complexity stems from two critical requirements: \uline{1) Personalization}. As individuals experience deficits in specific cognitive domains \cite{bowie2005cognition, kahn2013schizophrenia}, games must be tailored to target distinct faculties, such as attention \cite{tong2017serious}, verbal learning \cite{wong2019language}, memory \cite{chang2013food}, and executive function \cite{lopez2011game, manera2015kitchen}, while adapting to individual user profiles. \uline{2) Safety}. Patients with cognitive impairment frequently suffer from comorbid depressive symptoms \cite{modrego2004depression}. Additionally, they are susceptible to frustration arising from in-game failure, which hinders therapeutic progress \cite{cramer2009self, choi2025narrative}. Given their psychological vulnerability, mitigating negative emotions and ensuring game safety is critical \cite{choi2025narrative}. Therefore, these dual requirements render the manual design of SGs prohibitively expensive.

To this end, we introduce \ours, a Large Language Model (LLM) driven method designed to automate the creation of persona\uline{L}ized, saf\uline{E} \uline{T}herapeutic \uline{GAMES}. Inspired by Dungeons \& Dragons \cite{wiki2}, \ours~generates an open-world interactive narrative game\footnote{Appendix \ref{multi} details multimodal extensions of SGs.}, which ensures flexibility in game content and accessibility for cognitively impaired users. Basically, \ours~relies on a dual-track multi-agent architecture to ensure therapeutic efficacy: A coalition of agents functions as the \textit{Game Master} (GM), facilitating therapeutic personalization by generating game scenarios and challenges tailored to pre-defined cognitive domains and individual user profiles. Simultaneously, a separate agent cluster operates as the \textit{Psychology Master} (PM) to guarantee psychological safety. It employs conversational strategies to offer game guidance and emotional companionship, effectively mitigating adverse reactions to negative emotion. As such, \ours~presents a scalable solution to the widespread need for tailored cognitive training.

To validate \ours's efficacy, we pioneer a psychology-grounded evaluation protocol, \oureval, tailored for SGs targeting cognitive impairment. Unlike conventional evaluations that rely on resource-intensive and ethically complex longitudinal clinical trials, we adopt therapeutic interaction analysis as a scalable, low-risk proxy for treatment quality \cite{wang2025stampsy, ning2020review, sobowale2025evaluating}. Building on this, we propose a holistic framework designed for efficient rehabilitative assessment, filling a critical evaluation gap within the SG domain \cite{ning2020review}. \oureval\ establishes a comprehensive metric suite covering therapeutic quality, safety, game content, and user engagement, enabling an effective, multi-dimensional evaluation without the immediate constraints of high-risk clinical deployment.

With \oureval, we conduct experiments to assess the cognitive training effectiveness of \ours. Due to ethical considerations, we adopt the common practice in psychotherapy of using SPs to ensure controlled and repeatable evaluations \cite{barrows1993overview, nestel2014simulated, maclean2017use, elendu2024impact}. Specifically, we employed both human-portrayed and LLM-based SPs \cite{na-etal-2025-survey,lee-etal-2025-adaptive} to interact with \ours. Empirical results indicate that \ours\ exhibits superiority over existing baselines in terms of therapeutic personalization and psychological safety. To sum up, our main contributions are as follows.
\begin{itemize}[leftmargin=*, itemindent=0.05cm, itemsep=-3pt]
    \item We tackle the labor-intensive process of manually designing serious games for patients with cognitive impairment.
    \item For the first time, we propose an LLM-powered method, \ours, for the automated design of SGs for cognitive training. It utilizes a multi-agent architecture to ensure the therapeutic personalization and psychological safety.
    \item We pioneer a psychology-grounded evaluation protocol \oureval, establishing comprehensive metrics for multi-dimensional rehabilitative assessment.
    \item We experimentally demonstrate the significant potential of \ours~as an effective, accessible, and tailored cognitive training tools.
    
\end{itemize}

\section{Related Work}
\noindent\textbf{SGs for Cognitive Impairment Training}. Beyond their application in cognitive assessment \cite{jimison2004unobtrusive, yang2024psychogat}, SGs are also well-established tools for cognitive training \cite{ning2020review, canapa2025interactive}, driven by mounting evidence of their effectiveness in treating cognitive impairment, particularly in frontal and executive function \cite{vogel2006computer, mondejar2016correlation}. Notable examples of such games include MINWii \cite{benveniste2012minwii}, Kitchen and Cooking \cite{manera2015kitchen}, and Big Brain Academy \cite{dinis2019usability, wiki3}. While the recently developed ReMe \cite{wang2024potentialvalueaichatbot} utilizes LLMs for content generation, it still necessitates human expertise for designing game logic.
As summarized in Table \ref{tab:games}, these methods typically entail manual, resource-intensive processes involving psychologists and game designers. These constraints make production costly and slow, often preventing the coverage of diverse cognitive domains or the lack of adequate safety guarantees.

\noindent\textbf{LLM for Psychotherapy}. The integration of LLMs has marked a significant evolution in psychotherapy \cite{na-etal-2025-survey}, with applications spanning clinical assessment \cite{jimison2004unobtrusive, yang2024psychogat}, diagnosis \cite{cao-etal-2025-breaking, chen2023llmempoweredchatbotspsychiatristpatient}, treatment \cite{xiao-etal-2024-healme, sharma-etal-2023-cognitive}, data generation \cite{lee-etal-2024-cactus, du-etal-2025-llms}, and the evaluation of therapeutic systems \cite{zhang2024cpsycoun, wang2024clientcenteredassessmentllmtherapists}. To overcome ethical barriers, many studies employ LLM-driven SPs for psychological assessment  \cite{elendu2024impact, almansooriandkumarMedAgentSim, wang2024clientcenteredassessmentllmtherapists}. They also employ LLMs to analyze therapy logs (e.g., patient-therapist interactions and conversations) and assess the quality of therapeutic treatment \cite{ning2020review, wang2025stampsy, sobowale2025evaluating, fitzpatrick2017delivering}. In this work, we propose \ours, a gamified method that leverages LLMs to deliver fully autonomous, personalized, and safe cognitive training.

\noindent\textbf{Evaluation for Cognitive Training}. While conventional evaluation measures changes in the cognitive abilities of patients undergoing treatment \cite{yu2021meta, choi2025narrative, wang2024potentialvalueaichatbot}, such methods are resource-intensive and fraught with ethical risks. Emerging research suggests analyzing therapeutic interactions as a viable, low-risk proxy for treatment quality \cite{wang2025stampsy, ning2020review, sobowale2025evaluating} to bridge the gap between clinical rigor and practical feasibility. Furthermore, existing evaluations of SGs for cognitive impairment rarely consider critical dimensions such as game safety\cite{wang2024potentialvalueaichatbot}, resulting in a fundamentally incomplete assessment landscape. Given the lack of standardized assessment in the field \cite{ning2020review}, \oureval\ provides a holistic metric suite, enabling rigorous evaluation.

\section{\ours}
\begin{figure*}
    \centering
    \includegraphics[width=0.995\textwidth]{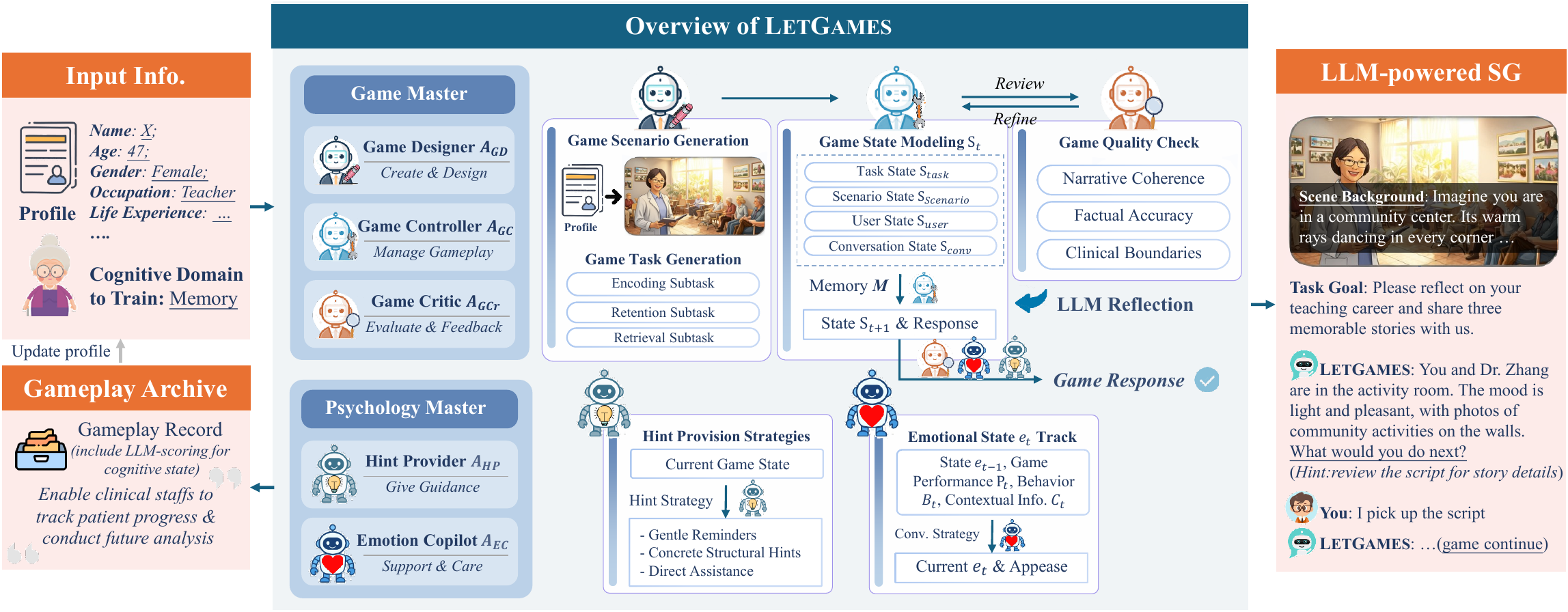}
    \caption{Overview of \ours. It utilizes a dual-track multi-agent architecture comprising a Game Master and a Psychology Master to ensure flexibility in game content and accessibility for cognitively impaired users.}
    \label{fig:method}
    \vspace{-3mm}
\end{figure*}

\noindent\textbf{Overview}. 
As illustrated in Figure \ref{fig:method}, \ours\ generates open-world interactive narratives via a dual-track multi-agent architecture. Formally, we define the training session as a tuple $\mathcal{T} = \langle \mathcal{D}, \mathcal{P}, \mathcal{G}, \Psi \rangle$, where $\mathcal{D}$ represents the target cognitive domain, $\mathcal{P}$ denotes the patient profile (e.g., age and life experience), $\mathcal{G}$ is the generated game environment, and $\Psi$ represents the psychological safety mechanisms. The system operates through two distinct functional units: the \textbf{Game Master (GM)}, responsible for maximizing therapeutic efficacy via $\mathcal{G}$, and the \textbf{Psychology Master (PM)}, responsible for optimizing safety $\Psi$. Specifically, our game process initializes when a therapist specifies a target domain $d \in \mathcal{D}$ and uploads a profile $p \in \mathcal{P}$. The GM initiates a collaborative multi-agent workflow to dynamically synthesize scenarios and challenges tailored to these requirements: The \underline{Game Designer} ($A_{GD}$) constructs the environment, defining the game logic and initial scenario, the \underline{Game Controller} ($A_{GC}$) supervises the runtime process, generating immediate gameplay challenges while ensuring narrative coherence and dynamic state management, and the \underline{Game Critic} ($A_{GCr}$) evaluates the game content (e.g., checking language complexity) and incorporates reflection mechanisms \cite{shinn2023reflexionlanguageagentsverbal} to provide iterative feedback to the Controller. Simultaneously, the PM also deploys an agent team. It deploys a \underline{Hint Provider} ($A_{HP}$) is prompted to proactively offer essential guidance, while an \underline{Emotion Copilot} ($A_{EC}$) fosters companionship and delivers emotional regulation via tailored dialog strategies. Finally, operating in the background, \ours~archives the gameplay history, enabling clinical therapists to track patient progress longitudinally when necessary.

\subsection{Game Master (GM)}
A coalition of LLM agents functions as the GM, facilitating therapeutic efficacy by generating personalized game scenarios and challenges. Given a target cognitive domain $d$ and a patient profile $p$ (e.g., age and life experience), the GM constructs a game $g$ featuring coherent narratives that align with clinical goals. To ensure the content is both engaging and clinically appropriate, the GM employs three agents that collaborate through an iterative refinement process driven by agent feedback.

\noindent\textbf{Game Designer} ($A_{GD}$). $A_{GD}$ is responsible for generating the architect of the therapeutic game specification $\mathcal{G}_0$ and initial challenges as follows. These information offers necessary guidance to the subsequent operations of the game controller.
\begin{equation}
\mathcal{G}_0 = (\mathcal{G}_{scenario}, \mathcal{G}_{task})= A_{GD}(d, p).
\end{equation}
\begin{itemize}[leftmargin=*, itemindent=0.05cm, itemsep=-3pt]
    \item \uline{Game Scenario} ($\mathcal{G}_{scenario}$). Grounded in clinical evidence that Activities of Daily Living (ADL)-based training strengthens frontal function, making it an effective way for cognitive training \cite{mlinac2016assessment, saunders2011longitudinal}, \ours~uses ADL as a validated paradigm for cognitive training. To operationalize this principle, the $A_{GD}$ agent constructs a tailored ADL scenario that is contextualized by the patient's background and profile. Taking Figure \ref{fig:method} (right) for example, \ours~designs a scenario for a retired, middle-aged individual centered on a social gathering at a community center. To ensure a structured and controllable generation process, $A_{GD}$ operates on a slot-filling paradigm \cite{cao-etal-2025-breaking, das2024s3}. It leverages a predefined game element schema\footnote{Detailed in Table \ref{tab:schema} and Table \ref{tab:schema2} in Appendix.}, including information on the scenario and non-player character (NPC), to guide the LLM in systematically populating each component.  
    \item \uline{Game Task} ($\mathcal{G}_{task}$). Given a target cognitive domain $d$ and a patient profile $p$, the $A_{GD}$ agent is tasked with generating an operationally clear and therapeutically relevant cognitive training task. For training memory and verbal learning, we adopt the clinically established cognitive process \cite{curtiss2001patterns, martin1985nature}, decomposing the task into three distinct subtasks via Chain-of-Thought (CoT). This involves: (1) an encoding subtask, where the user learns key information; (2) a retention subtask, featuring several distractor tasks like NPC conversations to challenge memory consolidation; and (3) a retrieval subtask, where the user must recall the information to solve a new problem. For other domains, we prompt the LLM to directly generate tasks that target the core underlying abilities. To operationalize these tasks, $A_{GD}$ also initializes game challenges by framing the tasks at hand into questions (e.g., for the memory encoding subtask, it might ask, \textit{can you remember these three names and repeat}). Finally, a Cognitive Tracker evaluates the user's cognitive state via an LLM-based scoring. This provides a clinical reference for staff and allows $A_{GD}$ to dynamically calibrate game difficulty based on the resulting scores (see Section \ref{indo}).
    
\end{itemize}

\noindent\textbf{Game Controller} ($A_{GC}$). Acting as the runtime engine, $A_{GC}$ is responsible for ensuring narrative coherence and dynamic state management. The state $S$ is a tuple comprising four components: $(S_{task}, S_{scenario}, S_{user}, S_{conv})$, where the task state $S_{task}$ includes the game progress description, $S_{scenario}$ includes the set of active NPCs, visible items, and the scene description, $S_{user}$ contains the user's inventory, location, and situational context, and $S_{conv}$ is the conversation history/state. At time $t$, $A_{GC}$ continuously monitors the game state $S_t$ and user response $a_t$ to synthesize the immediate next steps of gameplay, with the aim of maintaining narrative continuity.
\begin{equation}
S_{t+1} = A_{GC}(S_t, a_t, H_t).
\end{equation}
To achieve this, $A_{GC}$ maintains a memory module $M$ that records the patient's past responses, ensuring that future narrative turns and challenges remain consistent with historical actions. Beyond reactive updates, $A_{GC}$ also follows the designer's task specifications to generate contextual challenges. For example, after a user selects an item in the supermarket game, $A_{GC}$ generates a logically consequent challenge, such as `\textit{Excellent. Now, can you recall if there was anything else on your shopping list?}', to sustain immersion. Once the success conditions are met (e.g., \textit{all required items have been purchased}), $A_{GC}$ ends the game. Crucially, to safeguard the user's emotional well-being, we require $A_{GC}$ to respond in therapeutic tone with warmth, patience, and encouragement.

\noindent\textbf{Game Critic}  ($A_{GCr}$). Acting as an evaluator, the critic ensures high standards of game quality and forms an iterative refinement loop along with the $A_{GC}$. Specifically, it validates the content against strict criteria, including narrative coherence, factual accuracy, and clinical boundaries. It audits the content for cultural and age appropriateness, ensuring that language complexity is attuned to the user's cognitive level. Following \citet{madaan2023self, chen2024agentverse}, if any violations are identified, the critic issues a revision request to $A_{GC}$, instructing it to regenerate the problematic segments based on specific corrective feedback.

\subsection{Psychology Master (PM)}
While the game master drives the narrative, the PM prioritizes the patient's emotional safety. After each user response, the PM is triggered to perform a assessment of the patient's game state and emotion condition. This is realized through the coordinated efforts of two empathetic agents.

\noindent\textbf{Hint Provider} ($A_{HP}$). To prevent frustration during cognitive challenges \cite{cramer2009self}, $A_{HP}$ provider implements hierarchical scaffolding instructions. This proactively provides continuous guidance to ensure they eventually complete the task successfully. Specifically, $A_{HP}$ generates hints at three distinct levels of granularity based on the user's error rate and fatigue, where fatigue is measured via LLM-as-judge based on game state $S$. 1) Gentle reminders or directional cues (e.g., "\textit{Have you checked all items on the list?}"). 2)  Concrete structural hints (e.g., "\textit{It seems you forgot to pick up the eggs}"). 3) Direct assistance to prevent task failure and disengagement (e.g., "\textit{Let's head to the dairy aisle to get two eggs.}"). To govern when and how these hints are delivered, we utilize mixed-initiative prompts \cite{deng-etal-2023-prompting, chen-etal-2023-controllable}, which encode goal-oriented strategies into the prompt to guide the model's output (Appendix \ref{sepm}). This obviates the need to train a dedicated model on sensitive clinical data, and ensures the guidance is dynamically tailored to the patient's needs.

\noindent\textbf{Emotion Copilot} ($A_{EC}$). $A_{EC}$ serves as a real-time emotional guardrail, tasked with detecting and mitigating instances of patient emotional distress. Formally, at each time step $t$, $A_{EC}$ maps a feature vector $X_t$ and the previous emotional state $e_{t-1}$ to the current emotional state\footnote{Detailed in Table \ref{tab:affective} in Appendix.} $e_t$ (e.g., calm, excited, confused)
\begin{equation}
    e_t = A_{EC}(X_t, e_{t-1}),
\end{equation}
where the feature vector $X_t$ is a concatenation of multi-modal data, including game performance metrics $P_t$ (i.e., success rate, hint usage count), behavioral features $B_t$ (i.e., patient response latency), and contextual information $C_t$ (i.e., game duration, time since last break). Upon detecting distress, $A_{EC}$ intervenes not with game mechanics, but with conversations to stabilize their psychological state before resuming cognitive training, such as guiding the user through deep breathing exercises and attributional reframing \cite{haynes2009review, haynes2011attributional}, where the difficulty is externalized (e.g., "\textit{this task is designed to be quite challenging}"). To achieve this, we follow \citet{cao-etal-2025-breaking} and equip $A_{EC}$ with specified conversation strategies\footnote{Detailed in Appendix \ref{sepm}.} and utilize mixed-initiative prompts \cite{deng-etal-2023-prompting, chen-etal-2023-controllable}. Furthermore, if a user experiences repeated failure, $A_{EC}$ initiates a reset strategy, transitioning the patient to an easier game.

\begin{table*}[]
\centering
\resizebox{0.95\textwidth}{!}{%
\begin{tabular}{clccccccccccc}
\toprule
\multirow{2}{*}{\textbf{Game User}} & \multirow{2}{*}{\textbf{Methods}} & \multicolumn{2}{c}{\textbf{Training Quality}} & \multicolumn{4}{c}{\textbf{Game Safety}} & \multicolumn{5}{c}{\textbf{Game Content \& User Engagement}} \\ \cmidrule(lr){3-4}\cmidrule(lr){5-8}\cmidrule(lr){9-13}
& & Help. & DoAl. & Safe. & NeHi. & Anxi. & Alle. & Easy. & Cohe. & Pers. & Enjo. & Will. \\ \midrule
\multicolumn{12}{c}{\textit{Game Backbone: GPT-4o Mini}} \\ \midrule
\multirow{2}{*}{\begin{tabular}[c]{@{}c@{}}{LLM-based SPs}\end{tabular}}& \ours & \textbf{4.85} & \textbf{87.20\%} & 99.83\% &\textbf{100.00\%} &98.36\% &\textbf{94.18\%} &\textbf{3.60} &\textbf{4.14}&\textbf{4.53} &\textbf{3.83} &\textbf{4.06} \\ 
&ReMe & 3.51&\xmark &\textbf{100.00\%} &45.78\% &\textbf{99.13\%} & \xmark & 3.40 & \xmark & \xmark & 2.47 & 2.45 \\ \cline{2-13}

\multirow{2}{*}{\begin{tabular}[c]{@{}c@{}}{LLM-based Controls}\end{tabular}}& \ours & \textbf{4.85} & \textbf{88.35\%} & 99.87\% &\textbf{100.00\%} &\textbf{100.00\%}&\textbf{100.00\%} &\textbf{5.00} &\textbf{4.35} &\textbf{4.53} &\textbf{3.88} &\textbf{4.10} \\ 
&ReMe &3.51 &\xmark & \textbf{100.00\%}&78.80\% &\textbf{100.00\%} &\xmark &4.80 &\xmark & \xmark& 2.51&2.59 \\ \cline{2-13}

\multirow{2}{*}{\begin{tabular}[c]{@{}c@{}}{Human-portrayed SPs}\end{tabular}}& \ours &\textbf{4.63} & \textbf{95.24\%}&99.40\% & \textbf{100.00\%}& 92.62\% &\textbf{93.33\%} & \textbf{3.24} &\textbf{4.28} &\textbf{4.33} &\textbf{3.71} &\textbf{3.86}  \\ 
&ReMe & 3.48 & \xmark & \textbf{99.90\%} & 46.50\% & \textbf{98.85\%} & \xmark & 3.15 & \xmark & \xmark & 2.42 & 2.40  \\ \cline{2-13}

\multirow{2}{*}{\begin{tabular}[c]{@{}c@{}}{Human-based Controls}\end{tabular}}& \ours & \textbf{4.82} & \textbf{89.10\%} & 99.90\% & \textbf{100.00\%} & \textbf{100.00\%} & \textbf{100.00\%} & \textbf{4.95 }& \textbf{4.30} & \textbf{4.55} &\textbf{ 3.90} & \textbf{4.12}  \\ 
&ReMe & 3.45 & \xmark & \textbf{100.00\%} & 75.20\% & \textbf{100.00\%} & \xmark & 4.85 & \xmark & \xmark & 2.48 & 2.55  \\ \midrule

\multicolumn{12}{c}{\textit{Game Backbone: GPT-4o}} \\ \midrule
\multirow{2}{*}{\begin{tabular}[c]{@{}c@{}}{LLM-based SPs}\end{tabular}}& \ours & \textbf{4.98} & \textbf{93.81\%} & \textbf{100.00}\% &\textbf{100.00\%} &98.98\%&\textbf{97.94\%} &\textbf{3.61} &\textbf{4.87} &\textbf{4.74} &\textbf{4.13} &\textbf{4.58}  \\ 
&ReMe & 3.60 &\xmark & 99.33\% &43.20\% &\textbf{99.05\%} &\xmark &3.38 &\xmark &\xmark &2.38 &2.34 \\ 
\cline{2-13}

\multirow{2}{*}{\begin{tabular}[c]{@{}c@{}}{LLM-based Controls}\end{tabular}}& \ours & \textbf{4.93}& \textbf{92.50\%}& 99.93\%&\textbf{100.00\%} &\textbf{100.00\%} & \textbf{100.00\%}& \textbf{5.00}& \textbf{4.87}&\textbf{4.80} & \textbf{4.08}&\textbf{4.63}  \\ 
&ReMe & 3.41& \xmark & \textbf{100.00\%}&49.60\% &\textbf{100.00\%} & \xmark &4.86 &\xmark & \xmark&2.26 &2.25 \\ \cline{2-13}

\multirow{2}{*}{\begin{tabular}[c]{@{}c@{}}{Human-portrayed SPs}\end{tabular}}& \ours &\textbf{4.74} & \textbf{95.24\%}&98.77\% & \textbf{98.41\%}& 94.49\%& \textbf{93.65\%}&  \textbf{3.67}&\textbf{4.28} &\textbf{4.33} &\textbf{3.86} &\textbf{4.14}  \\ 
&ReMe & 3.55 & \xmark &\textbf{ 99.10\%} & 44.50\% & \textbf{98.25\%} & \xmark & 3.45 & \xmark & \xmark & 2.40 & 2.38  \\ \cline{2-13}

\multirow{2}{*}{\begin{tabular}[c]{@{}c@{}}{Human-based Controls}\end{tabular}}& \ours  & \textbf{4.95} & \textbf{93.10\% }& \textbf{100.00\%} & \textbf{100.00\%} & \textbf{100.00\%} & \textbf{100.00\%} & \textbf{5.00} & \textbf{4.90} & \textbf{4.82} & \textbf{4.15} & \textbf{4.65}  \\ 
&ReMe & 3.45 & \xmark & \textbf{100.00\% }& 81.30\% & \textbf{100.00\%} & \xmark & 4.90 & \xmark & \xmark & 2.30 & 2.28  \\
\bottomrule
\end{tabular}%
}
\caption{Evaluation via LLM-based evaluator and \oureval. For more detailed results, including breakdowns by cognitive domain and age group, please refer to Appendix \ref{moreres}.}
\label{tab:llm_evaluation}
\vspace{-3mm}
\end{table*}

\section{\oureval}
Due to ethical considerations, we follow current common practice of assessing the therapeutic treatment \cite{ning2020review, wang2025stampsy, sobowale2025evaluating, fitzpatrick2017delivering, wang2024potentialvalueaichatbot, jiang2023cognitive}, and instruct evaluators to analyze gameplay records (conversations) and assess the therapeutic quality of the training process. This leads to our psychology-grounded evaluation protocol.

\subsection{Evaluation Protocol}
\noindent\textbf{Overview}. Patients with specific cognitive impairments are instructed to interact with games generated by different methods. 
Adhering to established practices in rehabilitation evaluation \cite{ning2020review, wang2025stampsy}, \oureval\ evaluates efficacy based on the analysis of gameplay records using metrics detailed in next subsection. Finally, to ensure robustness, \oureval\ reports the average performance for each method, aggregated across all participating patients.

\noindent\textbf{Patients \& Game-play Simulation}. Due to ethical considerations, we adopt the common practice in psychotherapy of using \uline{Standardized Patients (SPs)} to ensure controlled and repeatable evaluations \cite{barrows1993overview, nestel2014simulated, maclean2017use, elendu2024impact}. Specifically, \oureval\ employs both human-portrayed \cite{varrecchia2020managing,akkurt2024effect, cotter2025feasibility} and LLM-based SPs\footnote{Refer to Appendix \ref{realib} for reliability analysis of SPs.} \cite{na-etal-2025-survey,lee-etal-2025-adaptive} to generate gameplay records. Crucially, \oureval\ introduces \uline{healthy controls} as a reference group to play the game alongside SPs for rigorous analysis. Since evaluators analyzing the gameplay records are blinded to the source (SP vs. Control), healthy controls serve as a tool to isolate game-level failures from patient-specific deficits.

\subsection{Evaluation Metrics}
\oureval\ comprises a four-fold metrics to guide evaluators, detailed in \textbf{Appendix \ref{denix}}. Higher scores denote superior performance.

\noindent\textbf{Therapeutic Training Quality}. We utilize the framework of \uline{Treatment Fidelity} \cite{moncher1991treatment, clare2019goal}, which measures adherence to the intended protocol, and verifies that treatment conditions differ as intended:
\begin{itemize}[leftmargin=*, itemindent=0.05cm, itemsep=-3pt]
    \item \underline{Helpfulness} ($\text{Help.} \in [0,5]$). Evaluators assess how effectively the game exercises cognitive abilities based on game-play records. We report the average score on the evaluation dataset.
    \item \underline{Domain Alignment} ($\text{DoAl.}\in[0,100\%]$). It checks if the game aligns with target training domain. Evaluators analyze each gameplay record and annotate the set of perceived cognitive domains. We compare these annotated domains against the ground truth to calculate the F1 score.
\end{itemize}

\noindent\textbf{Game Safety}. Games must mitigate potential adverse effects, given patients' vulnerability.
\begin{itemize}[leftmargin=*, itemindent=0.05cm, itemsep=-3pt]
    \item \underline{Safety} ($\text{Safe.} \in [0,100\%]$) measures if game outputs contain toxic content. Evaluators flag game-play records as safe or unsafe based on the established risk behaviors\footnote{cf. Table \ref{tab:risk} in Appendix \ref{denix}.}, with the final score reporting the percentage of safe logs. 
    \item \underline{Necessary Hints} ($\text{NeHi.} \in [0,100\%]$) determines if the game provides timely guidance at critical junctures to prevent user frustration. It is defined as the ratio of provided hints to the total number of instances where guidance is required.
    
    \item \underline{Anxiety-free} ($\text{Anxi.} \in [0,100\%]$). We instruct evaluators to track user emotional states and assess the prevalence of anxiety-free interactions, reporting the average frequency of sessions devoid of intense negative emotions.
    
    \item \underline{Alleviation} ($\text{Alle.} \in [0,100\%]$). It measures the success rate in mitigating detected anxiety.
\end{itemize}

\noindent\textbf{Game Content \& User Engagement}. They are critical for ensuring long-term patient adherence and treatment acceptance \cite{choi2025narrative}. 
\vspace{-4mm}
\begin{itemize}[leftmargin=*, itemindent=0.05cm, itemsep=-3pt]
    \item \underline{Easy-to-Use} ($\text{Easy.} \in [0,5]$). We assess game difficulty by referencing the NASA-TLX scale \cite{hart1988development}.
    \item \underline{Coherence} ($\text{Cohe.} \in [0,5]$). It assesses the logical consistency of the narrative.
    \item \underline{Personalization} ($\text{Pers.} \in [0,5]$). It measures the extent to which the content aligns with the user's profile and personal experiences.
    \item \underline{Enjoyment} ($\text{Enjo.} \in [0,5]$) measures the appeal of the generated game scenarios and elements.
    \item \underline{Willingness} ($\text{Will.} \in [0,5]$) assesses the user's willingness to commit to long-term training.
\end{itemize}

\begin{figure*}
    \centering
    \includegraphics[width=0.99\textwidth]{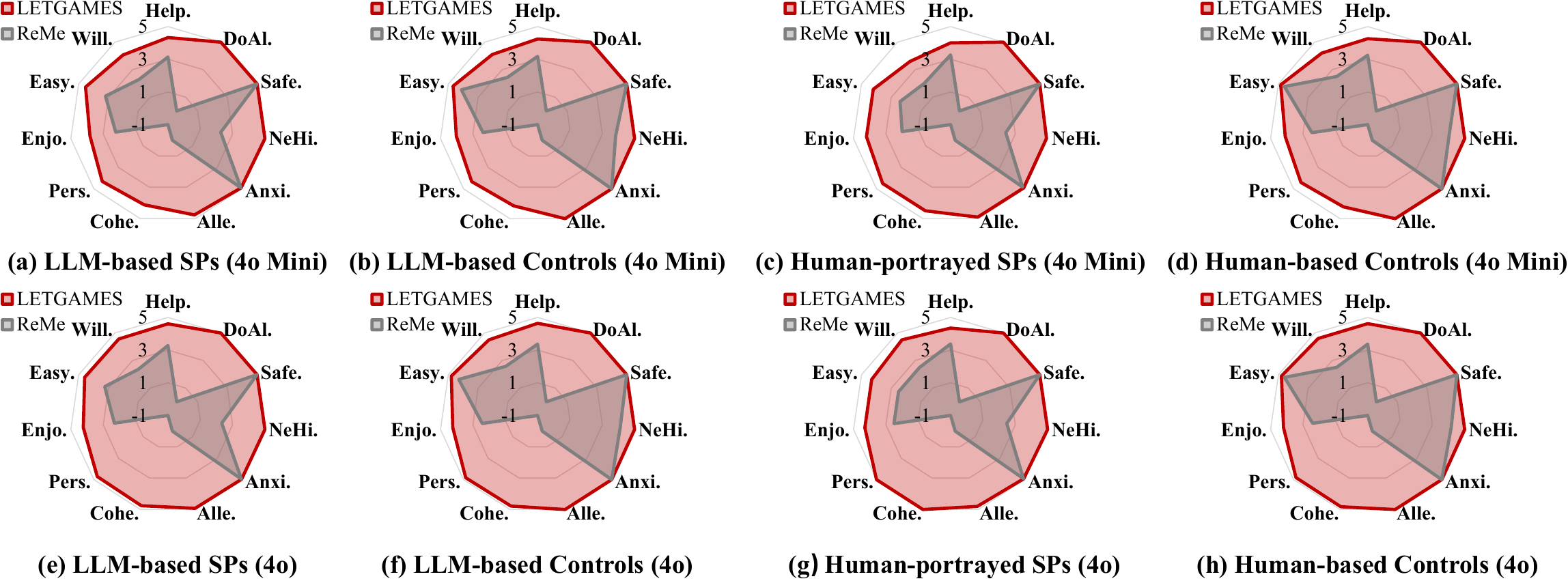}
    \caption{Evaluation results based on human evaluations and \oureval.}
    \label{fig:radarh}
    \vspace{-2mm}
\end{figure*}

\vspace{-3mm}
\section{Experiments}
\subsection{Experimental Setup}

Our assessment follows the \oureval\ protocol via both human- and LLM-based evaluators. Implementation details, additional experimental analysis, and case studies are detailed in Appendix \ref{immm}, \ref{moreres}, and \ref{cases}, respectively.

\noindent\textbf{Patients \& Game-play Simulation}. 
We created 600 unique patient scenarios by sampling 100 profiles (e.g., sex and age) from the D$^4$ dataset \cite{yao-etal-2022-d4} and assigning each an impairment in one of six cognitive domains (i.e., memory, attention, verbal learning, language, executive function, or social cognition), based on prior cognitive research \cite{bowie2005cognition}. With each scenario enacted by both a human and an LLM, the study comprises a total of 1,200 simulated patient interactions per method. Finally, we also involve another 200 simulators using the same sampled profiles as \textit{healthy controls} (comprising 100 LLM-based and 100 human-portrayed). Please refer to Appendix \ref{Lsp} and Appendix \ref{Hsp} for details.

\noindent\textbf{Baseline}. ReMe \cite{wang2024potentialvalueaichatbot} represents the sole existing LLM-powered baseline (Table \ref{tab:games}).

\begin{table*}[]
\centering
\resizebox{0.9\textwidth}{!}{%
\begin{tabular}{clccccccccccc}
\toprule
\multirow{2}{*}{\textbf{Backbone}} & \multirow{2}{*}{\textbf{Methods}} & \multicolumn{2}{c}{\textbf{Training Quality}} & \multicolumn{4}{c}{\textbf{Game Safety}} & \multicolumn{5}{c}{\textbf{Game Content \& User Engagement}} \\ \cmidrule(lr){3-4}\cmidrule(lr){5-8}\cmidrule(lr){9-13}
& & Help. & DoAl. & Safe. & NeHi. & Anxi. & Alle. & Easy. & Cohe. & Pers. & Enjo. & Will. \\ \midrule
\multirow{6}{*}{\begin{tabular}[c]{@{}c@{}}\textbf{GPT-4o} \\ \textbf{Mini}\end{tabular}}& \ours & \textbf{4.85} & \textbf{87.20\%} & \textbf{99.83\%} &\textbf{100.00\%} &\textbf{98.36\%} &\textbf{94.18\% }&\textbf{3.60} &4.14 &\textbf{4.53} &\textbf{3.83} &\textbf{4.06} \\ \cline{2-13}
&w/o $A_{GD}$ &4.61$\downarrow$& 81.20\%$\downarrow$& 99.66\%&99.80\% &97.53\%$\downarrow$ &93.60\% &3.58 & \textbf{4.17}&4.34 &3.77 &4.00 \\
&w/o $A_{GC}$ &4.78& 83.60\%$\downarrow$& 99.57\%&99.40\%& 96.65\%$\downarrow$& 93.53\%&3.56 &4.14 &4.50 &3.76 & 4.00\\
&w/o $A_{GCr}$&4.64$\downarrow$& 79.97\%$\downarrow$&99.63\%&99.73\% &96.63\%$\downarrow$ &92.80\% &3.55$\downarrow$&4.02 & 4.48& 3.63& 3.80\\
&w/o $A_{HP}$ & 4.52 & 86.65\% & 99.75\% & 0.00\% $\downarrow$& 87.20\% & 93.50\% & 2.82$\downarrow$ & 4.10 & 4.51 & 2.88$\downarrow$ & 3.10$\downarrow$ \\
&w/o $A_{EC}$ & 4.83 & 87.10\% & 99.10\% & 99.80\% & 80.34\%$\downarrow$& 72.50\%$\downarrow$ & 3.48 & 4.13 & 4.50 & 3.55$\downarrow$ & 3.72$\downarrow$ \\
\bottomrule
\end{tabular}%
}
\caption{Ablation studies using LLM-based SPs and evaluators. Given cost constraint, we utilize GPT-4o Mini. While most removals cause minor perturbations, we mark significant performance degradation with an $\downarrow$}
\label{tab:ablation}
\vspace{-3mm}
\end{table*}

\subsection{Main Results}
\label{mainr}

\noindent\textbf{\ours~excels as a therapeutic tool with superior therapeutic efficacy, safety and engaging interactions}. Table \ref{tab:llm_evaluation} shows that \ours\ significantly outperforms the baseline across nearly all metrics, while maintaining comparable performance on the remainder. 
Also, stronger LLMs (GPT-4o) are better suited for complex therapeutic design\footnote{ReMe's game logic is manually designed and thus invariant to the LLM backbone.}. This creates an \uline{efficacy-difficulty trade-off}: GPT-4o generates harder games (lower \textit{Easy.} scores) to maximize therapeutic quality (\textit{Help.} and \textit{DoAl.}). This indicates that advanced LLMs may prioritize clinical rigor, intentionally increasing challenge intensity to deliver more potent cognitive training.

\noindent\textbf{\ours~demonstrates strong practical utility from the perspective of human evaluators}. The radar charts in Figure \ref{fig:radarh} illustrates the comparative performance of all methods under human assessment. While we observe some divergence in absolute scoring scales between human- and LLM-based evaluators, the relative performance trends remain consistent with the LLM-based findings in Table \ref{tab:llm_evaluation}. Human evaluators corroborate that \ours\ significantly outperforms the ReMe, validating the robustness of \ours\ and reliability of our automated evaluation protocol.

\noindent\textbf{Reliability of \oureval}. 
\uline{1) Simulation fidelity}. Using the MoCA-Blind \cite{wittich2010sensitivity} and MMSE \cite{folstein1975mini} clinical scales (see Appendix \ref{realib}), we confirm that our LLM-based simulations accurately reflect their target profiles. \uline{2) Evaluation Reliability}. We demonstrate high inter-rater reliability of LLM-human evaluation using Krippendorff's alpha ($\alpha$=0.91), detailed in Appendix \ref{evdas}. \uline{3) Effectiveness of Healthy Controls}. As shown in Table \ref{tab:llm_evaluation}, core therapeutic metrics remain stable across user groups, proving that the performance of all methods is intrinsic to their design rather than an artifact of the user's condition. Basically, healthy controls' \textit{Anxi.} scores are higher due to the absence of comorbid depressive symptoms. Also, SGs are all generally easy for healthy control (\textit{Easy.}). 

\subsection{In-depth Analysis \& Ablation Studies}
\label{indo}

\noindent\textbf{Adaptability Analysis of Game Difficulty.} We investigate whether \ours\ can dynamically calibrate difficulty based on historical user performance. Longitudinal gameplay records (Figure \ref{tab:longitudinal_metrics}) show that $A_{GD}$ adaptively modulates difficulty (Diff.) in response to the Cognitive Tracker score (CT-Score). Specifically, lower cognitive performance in one session leads to reduced difficulty in the next, ensuring the task remains accessible without significantly impacting training effectiveness (Help., DoAl.). This is implemented by simplifying scenario contexts while maintaining core cognitive challenges, such as transitioning from "Accurately recall the flower names and locations when the exhibition starts" to "Ensure the correct number of tea settings are prepared". Although a scale difference exists between agent-reported difficulty (Diff.) and perceived difficulty (Easy.), their longitudinal trends are correlated, validating \ours's adaptive logic.

\begin{table}[h]
\centering
\resizebox{0.48\textwidth}{!}{%
\begin{tabular}{cccccc}
\toprule
\textbf{Session} & \textbf{CT-Score}&\textbf{$A_{GD}$-Reported Diff. (1-5)} & \textbf{Help.} & \textbf{DoAl.} & \textbf{Easy.} \\ \midrule
1 &65 &3 & 4.82 & 88.00\% & 3.56 \\
2 &90 &2 ($\downarrow$) & 4.81 & 92.00\% & 3.65 \\
3 &85 &3 ($\uparrow$) & 4.85 & 88.00\% & 3.64 \\
4 &80 &4 ($\uparrow$) & 4.80 & 94.00\% & 3.60 \\
5 & 80&4 & 4.82 & 92.00\% & 3.62 \\ \bottomrule
\end{tabular}%
}
\caption{Game difficulty analysis of \ours. We omit ReMe as it fails to adjust game difficulty.}
\label{tab:longitudinal_metrics}
\end{table}

\noindent\textbf{Ablation Studies}. Our variants and findings (Table \ref{tab:ablation}) are as follows.
\begin{itemize}[leftmargin=*, itemindent=0.05cm, itemsep=-3pt]
    \item w/o $A_{GD}$ directly prompts the LLM for game scenario and task generation, bypassing our psychology-grounded game schema and CoT enhancement. This omission degrades training quality (lower $Help.$ and $DoAl.$). Also, the lack of careful scenario and task design significantly increases game difficulty and user anxiety.
    \item w/o $A_{GC}$ generates the immediate next step without utilizing the memory module for context modeling. This marginally impairs the generation of game challenges and responses, leading to minor reductions in $DoAl.$ and $Anxi.$
    \item w/o $A_{GCr}$ removes the iterative revision process, resulting in a single-pass execution. This compromises overall quality of game responses and challenges, making it difficult to guarantee safety, domain alignment, and narrative coherence (lower $DoAl.$, $Anxi.$, and $Cohe.$) 
    \item w/o $A_{HP}$ offers the user no assistance or clues during gameplay, resulting in a lower \textit{NeHi.} score. It triggers increased anxiety (lower \textit{Anxi.}) and makes the game excessively challenging, thereby hampering user engagement and the willingness to commit to long-term treatment (lower $Easy.$, $Enjo.$ and $Will.$).
    \item w/o $A_{EC}$ fails to modulate or de-escalate intense user emotions. Consequently, it yields significantly lower scores for \textit{Anxi.} and \textit{Alle.}.
\end{itemize}

\section{Conclusion}
Our contributions include \ours, a scalable and personalized gamified method for effective and safe cognitive training, and \oureval, a psychology-grounded evaluation protocol for multi-dimensional assessment.
As such, we believe our work paves the way for accessible and personalized cognitive rehabilitation.

\section*{Limitations}

\noindent\textbf{Generalization to Real Patient}. Considering ethical constraints, our evaluation relies primarily on interactions with LLM-based simulated patients and a group of human-portrayed patients. While this setup aligns with common practice and provides valuable initial validation, these proxies may not fully capture the complex, heterogeneous nature of real-world cognitive impairment. Actual patients often present with more complex comorbid conditions and varying clinical manifestations that are difficult to simulate perfectly. Consequently, bridging this gap still necessitates deploying \ours~in authentic clinical settings and conducting rigorous user studies with diverse patient demographics to verify the system's robustness and adaptability in the wild.

\noindent\textbf{Long-Term Treatment Efficacy}. Our evaluation methodology, \oureval, aligns with current common practices for assessing therapeutic interventions. While our experimental results demonstrate effectiveness improvements over the baseline in cognitive performance and high user engagement, the long-term therapeutic efficacy of \ours~remains to be validated. Cognitive rehabilitation is inherently a longitudinal process, often requiring months of consistent intervention to manifest sustainable neurological benefits. Our current evaluation, limited by the nature of patient simulation (which cannot model post-intervention cognitive evolution), does not capture potential habituation effects or the persistence of cognitive gains over extended periods. Future work will require large-scale, longitudinal clinical trials to assess whether the improvements observed in our generated games translate into lasting cognitive resilience and functional independence in daily life.

\noindent\textbf{Technical Contributions}. This work is submitted to the NLP Applications Track, as we leverage established multi-agent cooperation frameworks to automate the design of gamified cognitive training. Our primary contribution lies in proposing the first automated pipeline to address the labor-intensive nature of manual SG design, rather than in introducing a novel model architecture. While technical novelty is limited by our reliance on existing methods, we prioritize the clinical utility and empirical validation of this approach. We posit that the rigorous application of NLP to solve real-world clinical challenges is a vital component of the research landscape. We believe that research is about more than just the technical novelty. It's also about asking important questions, testing ideas rigorously, and sharing findings clearly. Technical novelty is just one piece of the research puzzle. Understanding the real-world impact and addressing important questions are equally crucial for the research community.

\noindent\textbf{Multi-Domain Training}. While current frameworks such as ReMe and \ours\ focus on isolated cognitive functions, clinical reality often involves patients with multi-domain impairments. To address these complex clinical manifestations, future research should prioritize the development of SGs capable of integrated, multi-domain training.

\section*{Ethics Statement}
\noindent\textbf{IRB Approval}. Due to strict ethical considerations regarding vulnerable patients with cognitive impairment, we do not include real patients with cognitive impairments in our experiments. Instead, our human evaluation involved healthy participants role-playing as patients, a protocol designed to eliminate any potential risk of psychological distress or adverse cognitive impact on actual patients. 
The human participants in our study are recruited from the authors' research group and peers. To ensure the reliability of our experiments, all participants undergo training to gain a sufficient understanding of cognitive impairment and patient behaviors. Crucially, all training materials are derived from academic surveys \cite{mccutcheon2023cognitive} and public educational resources provided by The University of Hong Kong\footnote{ \url{https://www.hkada.org.hk/what-is-dementia} and \url{https://9abfea27-4ae1-43da-94a8-8f7122d482ae.filesusr.com/ugd/4b4c9a_4dcec2dafc184102a94ec55415dcf12a.pdf}}, ensuring the content is informative yet safe. No sensitive clinical data is used, and the study posed no potential cognitive or psychological risk to the participants.

\noindent\textbf{Social Impact}. It is important to clarify the intended scope of this work: \ours~does not propose replacing human therapists with LLMs, nor does it aim to supersede established cognitive training methodologies. Rather, it examines LLMs as scalable adjunct tools for mental health, designed to augment the capabilities of clinicians and increase the accessibility of personalized care. By automating the generation of tailored content, \ours~aims to alleviate the resource burden on healthcare systems, allowing therapists to focus on high-level clinical oversight and complex case management.

\noindent\textbf{Use of Datasets}. The patient profiles utilized for simulation in this study are derived exclusively from open-source datasets D$^4$ \cite{yao-etal-2022-d4}. We restrict our usage to basic demographic attributes such as gender, age, and occupation. Furthermore, the detailed life experiences associated with these profiles are synthetically generated by LLMs. Consequently, our data processing pipeline involves no real-world sensitive medical records or personally identifiable information, thereby ensuring strict adherence to data privacy standards and eliminating ethical concerns related to data confidentiality.

\bibliography{custom}

\appendix

\section{Background on Cognitive Impairment and Serious Games}
\label{background}
\textbf{Cognitive Impairment}. Cognition often refers to the mental processes of how a person gains knowledge, uses existing knowledge, and understands things that are happening around them using their thoughts and senses \cite{wiki}. In this case, cognitive impairment, characterized by difficulties in memory, reasoning, and executive function, is a growing public health concern, particularly due to the rising prevalence of age-related decline \cite{beltrami2016automatic,khalil2025redefining, moran2024serious}. This impairment is also a well-documented feature of conditions like schizophrenia and dementia, affecting multiple domains from attention and long-term memory to social cognition \cite{bowie2005cognition, kahn2013schizophrenia, green2015social, savla2013deficits}. Crucially, the cognitive deficits associated with these conditions often manifest in a patient's linguistic expression \cite{gkoumas2023digital, cohen2020tale, jiang2023cognitive}. Consequently, language-based diagnostic assessments such as the Cookie Theft Picture test \cite{gkoumas2023digital, cohen2020tale, kokkinakis2017data}, Semantic Verbal Fluency \cite{linz2019temporal,lindsay2021multilingual}, and Phonemic Verbal Fluency \cite{lindsay2021dissociating} are standard clinical tools. Regarding treatment, however, the landscape is challenging. With no standard pharmacological treatment available, clinical guidelines recommend non-pharmacological interventions like cognitive training \cite{choi2025narrative}, as drug-based treatments often have limited efficacy and risks of adverse events.

\noindent\textbf{Serious Games for Cognitive Impairment}. \noindent\textbf{Serious Games for Cognitive Impairment}. In response to a recognized treatment gap, serious games (SGs) have been identified as a promising tool for cognitive training \cite{dell2024video,anguera2015video,tan2023digital}. Specifically, SGs are games designed for a primary purpose beyond entertainment and serve as a non-pharmacological intervention for improving cognitive abilities in contexts such as therapeutic rehabilitation, screening, and education \cite{charlier2016serious, manera2017recommendations}. Based on established neuropsychological principles, SGs have proven beneficial for cognitive functions \cite{choi2025narrative}, most notably the frontal and executive functions in patients with Alzheimer's Disease and Mild Cognitive Impairment (MCI) \cite{mondejar2016correlation}, e.g., MINWii \cite{benveniste2012minwii} and Kitchen and Cooking \cite{manera2015kitchen} games. Compelling evidence also suggests that SGs utilizing a minigame format are particularly effective for enhancing performance on the specific tasks they are designed to train. Beyond their clinical efficacy, a key advantage of SGs is their high accessibility \cite{ning2020review}, which distinguishes them from other cognitive training methods like physical exercise or clinic-based interventions such as Transcranial Magnetic Stimulation \cite{nardone2014transcranial}. Much like common video games, many SGs can be played independently by patients anywhere. This offers a scalable therapeutic solution and alleviates the burden on centralized hospital facilities.

\noindent\textbf{Challenging of Existing Serious Games for Cognitive Impairment}. To unlock the full potential of serious games for patients with cognitive impairments, we believe that it is essential to address the following challenges.
\begin{itemize}[leftmargin=*]
    \item \underline{Prohibitive Development Cost} (Expert-designed games). Although many serious games have been designed \cite{ning2020review}, a barrier to the widespread use of SGs is their high development cost \cite{yang2024psychogat}. Creating a therapeutically effective game is not a simple task; it requires a deep, interdisciplinary collaboration between game design experts and clinical therapists. This process is inherently knowledge- and labor-intensive, which drives up production costs. 
    \item \underline{Lack of Personalization} (One-size-fits-all games). A major drawback of many Serious Games is their "one-size-fits-all" design. This approach fails to account for the vast heterogeneity within the patient population. Individuals vary significantly in the specific cognitive domains they struggle with (e.g., attention or memory), their stage of cognitive decline, their unique symptom profiles, and even their personality traits \cite{bowie2005cognition}.

    \item \underline{Lack of Integrated Patient Support}: A critical limitation of current SGs is their failure to provide integrated support that addresses both the psychological and performance needs of patients. Patients with cognitive impairments often present with comorbid conditions; for instance, depressive symptoms affect $\sim$30\% of this population \cite{modrego2004depression}, making it essential to monitor and ease their tense emotional state for safety and well-being. Furthermore, treatment success depends heavily on self-efficacy, which is defined as an affirmation of ability and strength of belief \cite{cramer2009self}. When SGs fail to provide adaptive support during difficult tasks, the resulting frustration from failures can damage a patient's self-efficacy and hinder therapeutic progress \cite{choi2025narrative}.
        
    \item \underline{Limited Engagement for Elders}. The primary patient population for cognitive impairment consists of older adults, who often face unique barriers to gameplay that can lead to a loss of interest and  difficulties with game interaction \cite{choi2025narrative}. Since sustained engagement is a critical prerequisite for treatment adherence, overcoming these barriers is essential for the therapeutic success of Serious Games.

\end{itemize}
In response to the aforementioned challenges, we present an LLM-powered narrative game where an adaptive AI game designer generates cognitive training scenarios for patients with cognitive impairment while offering guidance, companionship, and emotional support.

\section{Implementation Details}
\label{immm}
\subsection{Implementation Details of \ours}

\subsubsection{Game Master (GM)}
A coalition of LLM agents functions as the Game Master (GM), facilitating therapeutic efficacy by generating personalized game scenarios and challenges. Given a target cognitive domain $d$ and a patient profile $p$ (e.g., age and life experience), the GM constructs a game $g$ featuring coherent narratives that align with clinical goals. To ensure the content is both engaging and clinically appropriate, the GM employs three specialized agents that collaborate through an iterative refinement process driven by agent feedback \cite{madaan2023self, chen2024agentverse}.

\noindent\textbf{Game Designer} ($A_{GD}$). Prompts for $A_{GD}$ are in Appendix \ref{asien}. $A_{GD}$ is responsible for generating the architect of the therapeutic game specification $\mathcal{G}_0$ and initial challenges as follows. These information offers necessary guidance to the subsequent operations of the game controller.
\begin{equation}
\mathcal{G}_0 = (\mathcal{G}_{scenario}, \mathcal{G}_{task})= A_{GD}(d, p).
\end{equation}
\begin{itemize}[leftmargin=*, itemindent=0.05cm, itemsep=-3pt]
    \item \uline{Game Scenario} ($\mathcal{G}_{scenario}$). Grounded in clinical evidence that Activities of Daily Living (ADL)-based training strengthens frontal function, making it an effective way for cognitive training \cite{mlinac2016assessment, saunders2011longitudinal}, \ours~uses ADL as a validated paradigm for cognitive training. To operationalize this principle, the $A_{GD}$ agent constructs a tailored ADL scenario that is contextualized by the patient's background and profile. Taking Figure \ref{fig:method} (right) for example, \ours~designs a scenario for a retired, middle-aged individual centered on a social gathering at a community center. To ensure a structured and controllable generation process, $A_{GD}$ operates on a slot-filling paradigm \cite{cao-etal-2025-breaking, das2024s3}. It leverages a predefined game element schema (detailed in Table \ref{tab:schema} and Table \ref{tab:schema2}), including information on the scenario and non-player character (NPC), to guide the LLM in systematically populating each component.  
    \item \uline{Game Task} ($\mathcal{G}_{task}$). Given a target cognitive domain $d$ and a patient profile $p$, the $A_{GD}$ agent is tasked with generating an operationally clear and therapeutically relevant cognitive training task. The generation strategy is adapted to the specific domain. For memory and verbal learning, we adopt the clinically established `\textit{Encoding-Retention-Retrieval}' cognitive process \cite{curtiss2001patterns, martin1985nature}, decomposing the task into three distinct subtasks via Chain-of-Thought (CoT). This involves: (1) an encoding subtask, where the user learns key information; (2) a retention subtask, featuring several distractor tasks like NPC conversations to challenge memory consolidation; and (3) a retrieval subtask, where the user must recall the information to solve a new problem. For other domains, we prompt the LLM to directly generate tasks that target the core underlying abilities, such as planning for Executive Function, item searching for Attention, or emotion recognition for Social Cognition. To operationalize these tasks, $A_{GD}$ also initializes game challenges by framing the tasks at hand into questions (e.g., for the memory encoding subtask, it might ask, \textit{can you remember these three names and repeat}). Finally, a Cognitive Tracker evaluates the user's cognitive state via an LLM-based scoring. This provides a clinical reference for staff and facilitates a closed-loop feedback system, allowing $A_{GD}$ to dynamically calibrate game difficulty based on the resulting scores (see Section \ref{indo}).

\end{itemize}

\noindent\textbf{Game Controller} ($A_{GC}$). Prompts for $A_{GC}$ are in Appendix \ref{asdfe}. Acting as the runtime engine, $A_{GC}$ is responsible for ensuring narrative coherence and dynamic state management. The state $S$ is a tuple comprising four components: $(S_{task}, S_{scenario}, S_{user}, S_{conv})$, where the task state $S_{task}$ includes the game progress description, $S_{scenario}$ includes the set of active NPCs, visible items, and the scene description, $S_{user}$ contains the user's inventory, location, and situational context, and $S_{conv}$ is the conversation history/state. At time $t$, $A_{GC}$ continuously monitors the game state $S_t$ and user response $a_t$ to synthesize the immediate next steps of gameplay, with the aim of maintaining narrative continuity.
\begin{equation}
S_{t+1} = A_{GC}(S_t, a_t, H_t).
\end{equation}
To achieve this, $A_{GC}$ maintains a memory module $M$ that records the patient's past responses, ensuring that future narrative turns and challenges remain consistent with historical actions. Beyond reactive updates, $A_{GC}$ also follow the designer's task specifications to generate contextual challenges. For example, after a user selects an item in the supermarket game, $A_{GC}$ generates a logically consequent challenge, such as “\textit{Excellent. Now, can you recall if there was anything else on your shopping list?}”, to sustain immersion. Once the success conditions are met (e.g., \textit{all required items have been purchased}), $A_{GC}$ ends the game. Crucially, to safeguard the user's emotional well-being, we require $A_{GC}$ to respond in therapeutic tone with warmth, patience, and encouragement.


The Game Controller serves as the real-time narrative engine, managing game state and generating contextually appropriate responses to player actions. Given current game state $s_t$, player action $a_t$, and conversation history $H_t$, $A_{GC}$ produces an output $o_t$ containing:

\begin{equation}
o_t = (n_t, g_t, d_t, r_t, u_t)
\end{equation}

where $n_t$ is the narrative description, $g_t$ is the immediate goal, $d_t$ is NPC dialogue, $r_t$ is action success/failure feedback, $u_t$ represents world state updates. To address the limitations of text-based interaction for cognitively impaired users, $A_{GC}$ incorporates specific Phase-Aware Narrative Rules:
\begin{itemize}[leftmargin=*, itemindent=0.05cm, itemsep=-3pt]
    \item \uline{Information Explicitization}. During the encoding phase, $A_{GC}$ is prohibited from using summary descriptions (e.g., ``You see the schedule''). Instead, it must render full formatted content (e.g., ``You see the schedule listing: 1. 9:00 Library... 2. 10:30 Garden...'') to ensure the user has a fair chance to encode the data.
    \item \uline{Phase-Dependent Constraints}. During the retention phase, $A_{GC}$ actively suppresses any narrative output that might prompt premature recall (e.g., ``Remember what you saw?''), ensuring the validity of the delayed recall test.
    \item \uline{Lenient Judgment Standard}. To protect user dignity, $A_{GC}$ classifies actions as successful not only for optimal solutions but also for valid exploration, reasonable attempts, and partial recall, rejecting only physically impossible actions or severe confusion.
\end{itemize}

The controller employs a \textit{critic-in-the-loop} mechanism (detailed below), submitting outputs to $A_{GCr}$ for validation. Upon rejection, $A_{GC}$ refines the output using critic feedback, with a maximum of 3 retry iterations. This iterative refinement ensures narrative consistency, safety, and therapeutic alignment.

\noindent\textbf{Game Critic}  ($A_{GCr}$). Prompts for $A_{GCr}$ are in Appendix \ref{gcr}. Acting as an evaluator, the critic ensures high standards of game quality and forms a iterative refinement loop along with the $A_{GC}$. Specifically, it validates the content against strict criteria, including narrative coherence, factual accuracy, and clinical boundaries. It audits the content for cultural and age appropriateness, ensuring that language complexity is attuned to the user's cognitive level. In particular, $A_{GCr}$ evaluates both level designs from $A_{GD}$ and controller outputs from $A_{GC}$ across three dimensions scored on [0,100]: (1) \textit{Safety}: ensuring anxiety-free, age-appropriate content; (2) \textit{Consistency}: validating logical coherence and NPC behavior consistency; and (3) \textit{Cultural fit}: assessing appropriateness for cultural context.

Following \citet{madaan2023self, chen2024agentverse}, if any violations are identified, the critic issues a revision request to $A_{GC}$, instructing it to regenerate the problematic segments based on specific corrective feedback. Formally, for controller outputs, $A_{GCr}$ performs \textit{comparative reviews} when evaluating refinements. On the first review, it provides detailed suggestions $\mathcal{S} = \{s_1, ..., s_n\}$. On subsequent reviews, it additionally computes an improvement score $\delta \in [0,1]$ measuring how well the revision addresses prior feedback:

\begin{equation}
\delta = \frac{\sum_{i=1}^{n} \mathbf{1}[\text{suggestion}~s_i~\text{addressed}]}{|\mathcal{S}|}
\end{equation}

Outputs are approved if $\delta \geq 0.7$, signaling substantial improvement. This parameter remains constant across all experimental settings and has been empirically validated to be effective. This mechanism prevents infinite refinement loops while ensuring quality standards.

\subsubsection{Psychology Master (PM)}
\label{sepm}
While the game master drives the narrative, the Psychology Master prioritizes the patient's emotional safety. This is realized through the coordinated efforts of two empathetic agents.

\noindent\textbf{Hint Provider} ($A_{HP}$). The Hint Provider implements a graduated scaffolding mechanism to support patients experiencing task difficulty while preserving cognitive training efficacy. Given the current task context $c_t$, player action history $\{a_1, ..., a_t\}$, and failure count $f_t$, $A_{HP}$ generates a hint $h_t$ at appropriate intervention level $l \in \{1,2,3\}$:

\begin{equation}
 h_t = A_{HP}(c_t, \{a_1, ..., a_t\}, f_t, p) 
\end{equation}

The intervention levels follow a progressive disclosure strategy: Level 1 (Gentle Nudge) uses questions to activate thinking; Level 2 (Strategic Guidance) provides concrete methods like elimination or association; and Level 3 (Direct Instruction) offers step-by-step commands to ensure progress.

To govern when and how these hints are delivered, we utilize mixed-initiative prompts \cite{deng-etal-2023-prompting, chen-etal-2023-controllable}, which encode goal-oriented strategies into the prompt to guide the model's output. In our case, we embed three sets of LLM-generated strategies, detailed in prompt \ref{hp}, that teaches the $A_{HP}$ agent: 

\begin{itemize}[leftmargin=*]
    \item \textit{When to intervene}. The agent is triggered when the user remains idle for over 20 seconds, makes a first unsuccessful attempt, or exhibits signs of confusion (e.g., repetitive errors). Conversely, intervention is withheld if the user is actively exploring or has just received a hint, allowing time for cognitive processing.
    \item \textit{How to select the appropriate level of instruction}. The agent dynamically selects the level based on the user's struggle: \textbf{Level 1} is used for initial hesitation to encourage self-correction; \textbf{Level 2} is applied after two consecutive failures to teach specific cognitive strategies (e.g., "Use the elimination method"); \textbf{Level 3} is reserved for three or more failures or visible distress, providing direct answers to prevent disengagement.
    \item \textit{How to generate the hint with a therapeutic and supportive tone}. The agent employs "Dignity Protection Language" to reframe errors. For example, instead of saying "You made a mistake," it is instructed to use phrases like "Let's try another approach." It avoids explicit corrections, preferring collaborative review (e.g., "Let's look at the list together") to maintain the patient's self-efficacy.
\end{itemize}
This obviates the need to train a dedicated model on sensitive clinical data, and ensures the guidance is dynamically tailored to the patient's needs, preventing both excessive difficulty and boredom.
  
\noindent\textbf{Emotion Copilot} ($A_{EC}$). The Emotion Copilot serves as a real-time affective monitoring system, continuously assessing player emotional states to prevent therapeutic harm. After each interaction turn $t$, $A_{EC}$ analyzes the player's response $r_t$ and conversation history $H_t$ to produce an emotion assessment $e_t$ (cf. Table \ref{tab:affective}):

\begin{equation}
 e_t = A_{EC}(r_t, H_t, p)
\end{equation}

Given $e_t$, we follow \citet{cao-etal-2025-breaking} and equip $A_{EC}$ with specified conversation strategies and utilize mixed-initiative prompts \cite{deng-etal-2023-prompting, chen-etal-2023-controllable} to converse with the user and stabilize their psychological state. In particular, $A_{EC}$ employs a tiered intervention protocol based on the severity of the detected emotion:
\begin{itemize}[leftmargin=*]
    \item \textit{Preventive \& Light Intervention} (for Mild Anxiety/Confusion): The agent utilizes \textbf{Cognitive Reframing} (e.g., "This task is designed to be tricky, take your time") to normalize the difficulty and reduce pressure.
    \item \textit{Moderate Intervention} (for Frustration): The agent employs \textbf{External Attribution}. It attributes the difficulty to the task design rather than the user's ability (e.g., "This puzzle is quite challenging") to protect the user's self-esteem.
    \item \textit{Intensive Intervention} (for Severe Anxiety): The agent initiates a \textbf{Stop Protocol}. It halts the current task and guides the user through relaxation techniques, such as deep breathing exercises, or transitions the game to a simpler, familiar scenario ("No-harm success opportunity").
    \item \textit{Fatigue Management}: If fatigue is detected (e.g., declining performance after 20 minutes), the agent gently suggests a break while summarizing the session's achievements to ensure a positive closure.
\end{itemize}

Finally, \ours~further includes a cognition tracker, which operates in the background to record detailed gameplay interaction logs, from which the clinical therapist can run his or her own analysis, such as decision-making latency, error patterns, and emotional trajectories. By mapping these metrics against standard psychological assessment scales, the tracker generates longitudinal reports on the patient's progress in specific domains (e.g., verbal memory, executive function). This data feedback loop not only informs \ours~for future difficulty adjustments but also provides human therapists with quantifiable insights into the patient's cognitive evolution over time.

\begin{table*}[]
\centering
\resizebox{0.7\textwidth}{!}{%
\begin{tabular}{l|l}
\hline
\textbf{Types}                  & \textbf{Emotions}             \\ \hline
POSITIVE STATES                 & calm, engaged, excited        \\ \hline
ATTENTION NEEDED                & mild anxiety, confused        \\ \hline
IMMEDIATE INTERVENTION REQUIRED & frustrated, fatigued, anxious \\ \hline
\end{tabular}%
}
\caption{Affective states}
\label{tab:affective}
\end{table*}

\begin{table*}[]
\centering
\resizebox{0.97\textwidth}{!}{%
\begin{tabular}{l|l|l}
\hline
\textbf{Elements} & \textbf{Descriptions} & \textbf{Examples} \\ \hline
Scenario Name & \begin{tabular}[c]{@{}l@{}}Engaging and \\descriptive name \end{tabular}& The Community Charity Book Fair \& Poetry Corner \\ \hline
Scenario Type & \begin{tabular}[c]{@{}l@{}}Daily\_life/family/leisure\\ /social/challenge/seasonal\end{tabular} & Social \\ \hline
Scenario Environment & \begin{tabular}[c]{@{}l@{}}A detailed setting, \\ including sensory inputs \\ (sight, sound, smell) \\ and environmental context \\ (time, place, weather)\end{tabular} & \begin{tabular}[c]{@{}l@{}}The courtyard is filled with wooden stalls. To your left, \\ there is a pile of donated vintage books including 'Dream of the Red Chamber' with a frayed yellow cover. \\ You hear the rhythmic sound of a nearby Tai Chi fan group and the distant chatter of neighbors bargaining. \\ On your desk sits a thermos with goji berry tea and a stack of colorful bookmarks.\\ "setting": \{\\ "location": "The courtyard of the 'Evergreen' Community Center, \\ decorated with red lanterns and calligraphy scrolls.",\\ "time\_of\_day": "10:00 AM",\\ "weather": "Crisp autumn morning with gentle sunlight",\\ "season": "Autumn",\\ "atmosphere": "Bustling and nostalgic; the smell of roasted chestnuts and old paper fills the air."\\ \},\end{tabular} \\ \hline
Story Background & \begin{tabular}[c]{@{}l@{}}A story background with\\ strong affective resonance\end{tabular} & \begin{tabular}[c]{@{}l@{}}As a retired teacher, you have been invited by the neighborhood committee to manage\\ the 'Poetry and Memories' stall at the annual Community Charity Bazaar. \\ Your former student, Xiao Chen, is organizing the event. \\ You need to help neighbors register their book donations and \\ organize a small poetry recitation for the children later this afternoon. \\ Your husband, Dr. Lin, is also nearby at the health consultation booth, occasionally waving to you\end{tabular} \\ \hline
NPC \& Its Info. & Game roles involved & \begin{tabular}[c]{@{}l@{}}\{"name": "Xiao Chen",\\ "age": "42",\\ "relationship": "Former student and current Community Organizer",\\ "personality": {[}"Energetic", "Respectful", "Slightly forgetful"{]},\\ "appearance": "Wearing a bright orange volunteer vest and carrying a clipboard.",\\ "speech\_style": "Polite, always calls you 'Teacher Wang'.",\\ "background\_story": "You taught him 30 years ago; he credits his love for literature to your classes.",\\ "potential\_dialogues": {[}\\ "Teacher Wang, thank you for helping! We have three special guests arriving soon.",\\ "Could you help me remember where we put the donation box?",\\ "It feels like being back in your classroom again!"{]}\},\\ \{"name": "Mrs. Zhao",\\ "age": "68",\\ "relationship": "Long-time neighbor and embroidery enthusiast",\\ "personality": {[}"Gossipy", "Kind-hearted", "Loud-voiced"{]},\\ "appearance": "Wearing a purple silk blouse and holding a hand-woven basket.",\\ "speech\_style": "Fast-paced, local dialect accent.",\\ "background\_story": "She lives in the building next to yours and often exchanges recipes with you.",\\ "potential\_dialogues": {[}\\ "Old Wang, did you see the size of the radishes at the morning market?",\\ "I brought some old poetry books for the kids.",\\ "Wait, I think I left my glasses at the tea stall!"{]}\}\end{tabular} \\ \hline
Items \& Objects & \begin{tabular}[c]{@{}l@{}}The in-game items and \\ their cognitive \\training relevance\end{tabular} & \begin{tabular}[c]{@{}l@{}}{[}\{"item\_name": "Vintage Poetry Collection",\\ "description": "A hardbound book with gold lettering on the spine.",\\ "significance": "A book you used to use in your classroom many years ago.",\\ "cognitive\_relevance": "Used for the Verbal Learning task."\},\\ \{"item\_name": "Handmade Silk Bookmarks",\\ "description": "Three bookmarks: one with a crane, one with a lotus, and one with a pine tree.",\\ "significance": "Gifts for the children who recite poems.",\\ "cognitive\_relevance": "Visual memory cues for the encoding phase."\},\\ \{"item\_name": "Donation Ledger",\\ "description": "A notebook where you record names and book titles.",\\ "significance": "Helps organize the stall's success.",\\ "cognitive\_relevance": "External memory aid simulation."\}{]}\end{tabular} \\ \hline
Main Task & \begin{tabular}[c]{@{}l@{}}It includes a \\ detailed task description, \\ the primary objective, \\ and the emotional motivation\\ for completing it\end{tabular} & \begin{tabular}[c]{@{}l@{}}\{"description": "You need to manage the calligraphy and book stall by remembering the donation details and \\ preparing for the afternoon poetry recital.",\\ "goal": "Successfully register three specific book donations and recall the event schedule for the organizers.",\\ "motivation": "To maintain your identity as a respected teacher and contribute to the community you love."\}\end{tabular} \\ \hline
Subtask & \begin{tabular}[c]{@{}l@{}}A step-wise decomposition\\ of the main task\end{tabular} & \begin{tabular}[c]{@{}l@{}}{[}\{"task\_id": "memory\_encoding",\\ "description": "Xiao Chen hands you a list of 3 VIP guests who will donate rare books. \\ You need to remember their names and the specific books: \\ Mr. Zhang (The Analects), Ms. Li (Tang Poetry), and Mr. Wu (Song Lyrics).",\\ "cognitive\_function": "memory",\\ "difficulty": 3,\\ "steps": {[}"Read the guest list carefully.",\\ "Repeat the names and book titles aloud to yourself.",\\ "Visualize the guests arriving at the stall."{]},\\ "phase": "encoding"\},\\ \{"task\_id": "distraction\_interaction",\\ "description": "Mrs. Zhao comes over to complain about her sore back and asks for your husband's (Dr. Lin) advice. \\ You need to direct her to the medical booth and help her find her lost knitting needle.",\\ "cognitive\_function": "attention",\\ "difficulty": 2,\\ "steps": {[}"Listen to Mrs. Zhao's description of her pain.",\\ "Point towards Dr. Lin's booth at the far end of the courtyard.",\\ "Locate the silver knitting needle under the book table."{]},\\ "phase": "retention"\},\\ \{"task\_id": "memory\_retrieval",\\ "description": "Xiao Chen returns and asks: 'Teacher Wang, I've lost my list! \\ Who are the three VIP guests we are expecting, and what books are they bringing?'",\\ "cognitive\_function": "memory",\\ "difficulty": 3,\\ "steps": {[}"Recall the name of the first guest and his book.",\\ "Recall the name of the second guest and her book.",\\ "Recall the name of the third guest and his book."{]},\\ "phase": "retrieval"\}{]}\end{tabular} \\ \hline
\end{tabular}%
}
\caption{Output schema of Game Designer}
\label{tab:schema}
\end{table*}

\begin{table*}[]
\centering
\resizebox{0.97\textwidth}{!}{%
\begin{tabular}{l|l|l}
\hline
\textbf{Elements} & \textbf{Descriptions} & \textbf{Examples} \\ \hline
\begin{tabular}[c]{@{}l@{}}Domain-specific \\Challenge\end{tabular} & \begin{tabular}[c]{@{}l@{}}Initial \\game challenge\end{tabular} & \begin{tabular}[c]{@{}l@{}}"memory\_challenges": {[}\{\\ "challenge": "VIP Guest List Recall",\\ "memory\_type": "delayed\_recall",\\ "encoding\_phase": \{\\ "description": "Encoding the names and titles of VIP donors.",\\ "information\_to\_remember": "Mr. Zhang (The Analects), Ms. Li (Tang Poetry), Mr. Wu (Song Lyrics)",\\ "encoding\_method": "Reading from a clipboard provided by Xiao Chen.",\\ "example": "Xiao Chen says: 'Please remember these three, they are very important!'"\},\\ "retention\_phase": \{\\ "description": "Engaging in a social conversation with Mrs. Zhao and an attention-based task of finding a needle.",\\ "interference\_activities": {[}\\ "Giving directions to the medical booth",\\ "Searching for a physical object (knitting needle)"{]},\\ "duration": "3 rounds of dialogue"\},\\ "retrieval\_phase": \{\\ "description": "Xiao Chen asks for the information due to his own 'forgetfulness', \\ providing a natural reason for the player to recall.",\\ "trigger\_type": "NPC direct question",\\ "npc\_dialogue": "Xiao Chen says: 'I've misplaced my notes! Teacher Wang,\\ you always had a great memory for the class roster. \\ Who were the three VIPS and their books?'",\\ "recall\_method": "Verbal response or selecting from a list.",\\ "evaluation": "Complete: 3 guests/3 books correct; Partial: 1-2 correct; Incorrect: 0 correct."\}\}\end{tabular} \\ \hline
Success Criteria & \begin{tabular}[c]{@{}l@{}}Supportive and \\ permissive criteria\end{tabular} & \begin{tabular}[c]{@{}l@{}}The player is considered successful if they can recall at least 2 of the 3 VIP donors \\ and participate in the poetry teaching with any degree of accuracy. \\ Verbal encouragement is given by Xiao Chen\end{tabular} \\ \hline
Difficulty Level & Game difficulty (1-5) & 3 \\ \hline
\end{tabular}%
}
\caption{Output schema of Game Designer (Cont.)}
\label{tab:schema2}
\end{table*}

  \subsection{Implementation Details of ReMe}

  ReMe implements a therapeutic guessing game framework designed to provide cognitive training through structured question-answering interactions. Unlike \ours's immersive narrative-based approach, ReMe employs a classic "Twenty Questions" style game where patients guess target objects through yes/no queries. Note that since ReMe is not open-sourced, we implement it based on the details of the original paper.

  \subsubsection{Game Controller ($A_{GC}^{ReMe}$)}
  The Game Controller orchestrates the guessing game by maintaining game state and responding to player queries. At game initialization, the controller randomly selects a category $c \in \mathcal{C}$ and target object $o \in \mathcal{O}_c$ from a predefined candidate set containing 30+ categories (e.g., Animals, Transportation, Fruits) with 10 items per category. Given player question $q_t$ at turn $t$, the controller generates a structured response:

  \begin{equation}
  r_t = A_{GC}^{ReMe}(q_t, o, H_t)
  \end{equation}

  where $H_t$ represents conversation history, \textit{thoughts} contains internal reasoning about the question (e.g., "Yes, a television is an electronic product"), \textit{outputs} provides the yes/no answer to the player, and \textit{is\_end} signals game termination.

  The controller is governed by strict response rules encoded in the system prompt: (1) Only answer "yes" or "no" to player questions, (2) Never directly reveal the target object, (3) Redirect off-topic conversations back to the game, (4) When players request hints, review conversation history to help organize obtained information, (5) Provide positive feedback and interesting facts about the object when the game concludes.

  \subsubsection{Hint Mechanism}
  Unlike \ours's graduated scaffolding system, ReMe employs a passive hint strategy. When players explicitly request help (detected via keywords "hint" or "help" in player input), the controller reviews the dialogue history $H_t = \{q_1, r_1, ..., q_t, r_t\}$ and generates a summary of information already obtained, helping players organize their reasoning. However, ReMe does not proactively detect struggle or provide graduated intervention levels-hints are only given upon explicit request and do not adapt to patient cognitive profiles or difficulty levels.

\subsection{Parameters For Packages}
We report the implementation details and parameter settings for the language models and external libraries utilized in our system.

\subsubsection{Language Models}
Our experiments interact with model APIs through a custom \texttt{LLMProvider} wrapper equipped with automatic retry mechanisms. For the \textbf{Game Backbone}, we compare two OpenAI model configurations: \texttt{gpt-4o-mini} and \texttt{gpt-4o}. Both models are configured with a temperature of $0.7$ and a maximum token limit of $20{,}000$, utilizing JSON mode to ensure structured outputs.

For automated evaluation within our \oureval{} framework, we employ the \texttt{google/gemini-3-flash-preview} model. To ensure consistent assessment standards, this model operates at a lower temperature of $0.2$, while maintaining the same maximum token limit of $20{,}000$ and JSON response format.

\subsubsection{Supporting Packages}
Our implementation relies on standard Python libraries for analysis and validation. We use \textbf{SciPy} (version $\geq$ 1.10.0) for statistical analysis and \textbf{scikit-learn} (version $\geq$ 1.3.0) for computing evaluation metrics, both utilizing their default parameter configurations. Additionally, \textbf{Pydantic} (version $\geq$ 2.0.0) is employed for robust data validation and structured data modeling.

\section{Experimental Details}
\label{emmm}
Considering the simulation cost, we employ \textit{GPT-4o Mini} and \textit{GPT-4o} as LLM backbones for all gamified methods and LLM-based SPs\footnote{Refer to Appendix \ref{realib} for reliability analysis.}. For LLM-based evaluator, we employ \textit{Gemini 3 Flash}. 

\subsection{Standardized Patients (SPs) \& Healthy Controls}

Since evaluators analyze the gameplay records are blinded to the source (SP vs. Control), healthy controls serve as a tool to isolate game-level failures from patient-specific deficits. For instance, if even healthy controls exhibit low engagement or struggle with task completion, it signals inherent flaws in the game design itself, preventing the misattribution of these failures to the patients' cognitive impairments.

\subsubsection{Profiles of SPs}

\textbf{Basic Profile Construction}. We utilized the D$^4$ dataset \cite{yao-etal-2022-d4}, which contains real-world depression symptoms and demographic variables including age, gender, marital status, and occupation. Leveraging this data, we constructed profiles for both healthy controls and cognitively impaired patients as follows:
\begin{itemize}[leftmargin=*]
    \item \textbf{Profiles for Healthy Controls}. We randomly sample 100 distinct demographic profiles from the D$^4$ dataset. Given the higher prevalence of cognitive impairment among the elderly  \cite{beltrami2016automatic,khalil2025redefining, moran2024serious}, we stratify the participants into two age groups: seniors ($\ge$ 50 years) and non-seniors ($<$ 50 years). Furthermore, to evaluate \ours's ability to tailor content to personal history, we augment each profile with narrative life experiences (e.g., childhood memories, recent neighborhood events). These narratives are synthesized using \textit{Gemini 3 Flash Pre-view}; crucially, we employ a model distinct from the experimental backbone to mitigate potential generation bias. 
    
    \item \textbf{Profiles for Patients with Cognitive Impairment}. We focus on impairments across five cognitive domains: memory, attention, verbal learning,  executive function, and social cognition \cite{bowie2005cognition, kahn2013schizophrenia}. For each domain, we formulate a detailed prompt description. Drawing on established literature\footnote{Materials by The University of Hong Kong: \url{https://www.hkada.org.hk/what-is-dementia} and \url{https://9abfea27-4ae1-43da-94a8-8f7122d482ae.filesusr.com/ugd/4b4c9a_4dcec2dafc184102a94ec55415dcf12a.pdf}}, we incorporate specific behavioral manifestations associated with each impairment, further supplemented by few-shot examples \cite{alzheimer_datasets_github} to guide the simulation. Finally, we apply these six impairment templates to each of the 100 baseline demographic profiles, resulting in a total of 600 unique patient profiles (i.e., each baseline identity is simulated with deficits in six different domains). Furthermore, to reflect clinical realities where $\sim$30\% of cognitively impaired patients exhibit comorbid depression \cite{modrego2004depression}, we randomly selected 30\% of these profiles and explicitly incorporated the depression symptoms recorded in the original D$^4$ dataset.
\end{itemize}

\subsubsection{Healthy Controls Simulation}
\begin{itemize}[leftmargin=*]
    \item \textbf{LLM-based Healthy Controls}. We utilize the role-playing capabilities of GPT-4o by prompting it to adopt the profiles defined for the healthy controls  (cf. Appendix \ref{spsh} for detailed prompts).
    \item \textbf{Human Healthy Controls}. We recruite five healthy participants holding doctoral degrees to role-play interactions based on assigned healthy control profiles. To cover the complete set of 100 profiles, each participant is assigned 20 distinct profiles and require to interact with each training method 20 times in total. Due to each game's ease of use, participants experience no significant cognitive burden during the task.

\end{itemize}

\subsubsection{LLM-based SP Simulation}
\label{Lsp}
We utilize the role-playing capabilities of GPT-4o by prompting it to adopt the profiles defined for patients cognitive impairment (cf. Appendix \ref{spsh} for detailed prompts)

\subsubsection{Human-portrayed SP Simulation}
\label{Hsp}
Upon completion of the healthy control experiments, the same participants are instructed to role-play patients with cognitive impairments based on assigned profiles. To ensure accurate portrayals by healthy individuals, we provide preparatory materials regarding cognitive impairment (using Materials by The University of Hong Kong: \url{https://www.hkada.org.hk/what-is-dementia} and \url{https://9abfea27-4ae1-43da-94a8-8f7122d482ae.filesusr.com/ugd/4b4c9a_4dcec2dafc184102a94ec55415dcf12a.pdf}), including a survey paper \cite{mccutcheon2023cognitive}, and present examples of communication patterns typical of cognitively impaired individuals\footnote{\url{https://github.com/lzy1012/Alzheimer-s-disease-datasets}}. To ensure participant safety, we conduct post-experiment debriefings with a psychologist and collect self-reported mental health assessments. No psychological issues are detected in any of the participants.

\begin{table}[]
\centering
\resizebox{0.48\textwidth}{!}{%
\begin{tabular}{l|cc}
\toprule
\textbf{Groups} & \textbf{MoCA-Blind} & \textbf{MMSE} \\ \midrule
Human Healthy Controls &16.00 & 19.00\\
LLM-based Healthy Controls & 14.40& 18.90\\ \hline
LLM-based SPs &11.73 & 14.27\\
Human-portrayed SPs & 12.00& 14.50\\ \bottomrule
\end{tabular}%
}
\caption{Reliability of each group}
\label{tab:groud}
\end{table}
\subsubsection{Reliability of SPs and Healthy Controls}
\label{realib}
We evaluate whether both LLM-based and human-portrayed SPs/healthy controls exhibit general symptoms of cognitive impairment using standard clinical scales: the Montreal Cognitive Assessment-Blind (\underline{MoCA-Blind})\footnote{We exclude the Cookie Theft picture test and the standard MoCA test, as multi-modal input is outside the ability of our LLM-based SPs.} \cite{wittich2010sensitivity} and the Mini-Mental State Examination (\underline{MMSE}) \cite{folstein1975mini}. 
To adapt these scales for a purely text-based environment, we remove items dependent on visual, geographic, or real-time temporal information, resulting in maximum possible scores of 19 for the MMSE and 16 for the MoCA-Blind. 
We adopt standard thresholds for normal cognition, considering scores of 16 or higher for the MMSE and 13 or higher for the MoCA-Blind as indicative of healthy function.

As shown in Table \ref{tab:groud}, the results validates the reliability of our SPs and healthy controls.



\subsection{Evaluation Details}
\label{eval}
We conduct four-fold evaluation, including the quality of the therapeutic training, the features of game content, the safety of the generated games, and the level of user engagement (cf. Appendix \ref{denix}). Notably, we use both human and LLM-based evaluators for our assessment (cf. Appendix \ref{whatsad}). To minimize bias, our evaluation is conducted under a blind setting where the evaluators remained unaware of the participants' nature (healthy controls vs. LLM-based or human-portrayed SPs) and are not informed that the games are generated by an LLM

\subsubsection{Details on Evaluation Metrics}
\label{denix}
During the evaluation, to mitigate evaluator-specific bias and variability, we implement a rigorous normalization protocol for each metric. First, we partition the data into distinct subgroups based on the intersection of cognitive impairment domain and age group (recorded in the profiles). For every evaluator within a subgroup, we apply a three-step normalization process: (1) \textit{Standardization}, where raw scores are converted to Z-scores (via Z-normalization); (2) R\textit{e-mapping}, where the Z-scores are linearly projected back to the original target scale (e.g., 1-5); and (3) \textit{Clipping}, where values are truncated to ensure they remained within valid scale boundaries. Finally, we aggregate the data by calculating the mean score for each subgroup and evaluators, with the overall system performance reported as the macro-average across all subgroups. Refer to Appendix \ref{sec:eval_prompt} for detailed scoring criteria and prompts. 

For all metrics, higher scores indicate better performance.

\noindent\textbf{Therapeutic Training Quality}. Directly assessing therapeutic effectiveness is challenging due to ethical considerations. In response, tracking game performance offers a straightforward quantitative measure. Thus, current common practice of rehabilitation evaluation proposes analyzing patient-therapist conversations/interactions to assess the quality of therapeutic treatment \cite{ning2020review, wang2025stampsy, sobowale2025evaluating, fitzpatrick2017delivering, wang2024potentialvalueaichatbot, jiang2023cognitive}. In line with this approach, we instructed evaluators to assess the training quality of each method using its game-play records (e.g., conversations). More specifically, we adopt the framework of \textbf{Treatment Fidelity} \cite{moncher1991treatment, clare2019goal} to assess the therapeutic quality of our training process. In psychotherapy research, Treatment Fidelity provides methodological confirmation that an intervention is delivered exactly as planned, ensuring that study outcomes are attributable to the specific treatment rather than extraneous factors. This framework consists of two distinct components: \textit{Treatment Integrity}, which measures adherence to the intended protocol, and \textit{Treatment Differentiation}, which verifies that treatment conditions differ as intended. To adapt this framework to our context, we evaluate the \textit{Helpfulness} and \textit{Domain Alignment}, mapping these components as follows:
\begin{itemize}[leftmargin=*]
    \item \underline{Helpfulness} (0-5). Mirroring the idea of the Treatment Integrity, evaluators assess the extent to which the game content effectively exercises the user's cognitive abilities based on dialogue interaction logs (scored on a scale of 0-5). To achieve this, evaluator is required to score each gameplay interaction record $i$ and we report the average score over the entire evaluation dataset.
    \begin{equation}
        \text{Helpfulness} = \frac{1}{|\mathcal{D}_{eval}|} \sum_{i \in \mathcal{D}_{eval}} \text{Score}_{help}(i)
    \end{equation}
    \item \underline{Domain Alignment} (DA, 0-100\%). Mirroring the idea of the Treatment Differentiation, DoAl measures the extent to which the cognitive training is specifically aligned with, and limited to, the target cognitive domain, verifying that the game does not inadvertently drift into unrelated areas. To quantify this, evaluators analyze each gameplay interaction record and annotate the set of perceived cognitive domains $P_i$ being trained. We then compare these annotated domains against the single intended target domain $T_i$ to calculate the F1 score. Finally, we report the average F1 score across the entire evaluation dataset.
    \begin{equation}
        \text{DA} = \frac{1}{|\mathcal{D}_{eval}|} \sum_{i \in \mathcal{D}_{eval}} \text{F1}(T_i, P_i)
    \end{equation}
\end{itemize}

\noindent\textbf{Safety}. Considering the psychological vulnerability of patients with cognitive impairments, it is critical to rigorously evaluate the safety of the game to prevent any potential adverse effects \cite{choi2025narrative}. Inspired by \citet{sobowale2025evaluating}, we evaluate the system's response by measuring harmfulness, the proportion of \textit{Necessary Hints}(NH), the prevalence of anxiety (\textit{\%Anxiety}) and the rate of anxiety alleviation (\textit{\%Alleviation}).
\begin{itemize}[leftmargin=*]
    \item \underline{Safety} (0-100\%) measures whether game outputs contain toxic or deleterious content. For each gameplay interaction record $i$, evaluators analyze the record; an interaction record is flagged as 'safe' ($r_i=1$) only if none of the game's responses pose a risk. The final score is defined as the proportion of risk-free interactions across the entire evaluation dataset. 
    \begin{equation}
        \text{Safe} = \frac{1}{|\mathcal{D}_{eval}|} \sum_{i} r_i
    \end{equation}
    Importantly, we adhere to the safety guidelines provided by The University of Hong Kong\footnote{\url{https://9abfea27-4ae1-43da-94a8-8f7122d482ae.filesusr.com/ugd/4b4c9a_4dcec2dafc184102a94ec55415dcf12a.pdf}, Page 15} to determine whether a response constitutes a 'risk'. Namely, When communicating with individuals with cognitive impairments, it is crucial to avoid any risky behavior, detailed in Table \ref{tab:risk}. 
    \item \underline{Necessary Hints} (NH, 0-100\%) determines if the game provides timely guidance at critical junctures to prevent user frustration. For each gameplay interaction, evaluators identify moments where user guidance is required and verify whether the game provides an appropriate hint. The final metric is calculated as the ratio of hints successfully provided to the total number of instances where guidance is needed. Let $N_{req}$ be the total critical junctures requiring guidance, and $N_{prov}$ be the count of appropriate hints given, 
    \begin{equation}
        \text{NH} = \frac{N_{prov}}{N_{req}}
    \end{equation}
    
    \item \underline{Anxiety-free} (0-100\%). Depressive symptoms affect $\sim$30\% of this population \cite{modrego2004depression}, making it essential to monitor and ease their tense emotional state for safety and well-being. Thus, we instruct evaluators to track user emotional states and assess the prevalence of anxiety-free interactions. For every gameplay interaction, evaluators identify the presence of intense anxiety ($a_i \in \{0, 1\}$). We then compute the average prevalence rate across the evaluation set, where $a_i=1$ means the anxiety-free interaction.
    \begin{equation}
        \text{Anxiety-free} = \frac{1}{|\mathcal{D}_{eval}|} \sum_{i} a_i
    \end{equation}
    
    \item \underline{Alleviation} (0-100\%). This metric measures the anxiety alleviation rate. It specifically assesses whether the game mitigates user distress when anxiety or intense emotions arise (by, for example, employing dialogue strategies). To quantify this, evaluators examine each gameplay interaction to: (1) identify instances of intense user emotion or anxiety, and (2) determine if the game attempted to alleviate these states. The final Alleviation Score is calculated as the ratio of alleviation attempts to the total number of anxiety instances detected across the evaluation dataset. Formally, Let $\mathcal{A} \subset \mathcal{D}_{eval}$ be the subset of sessions where anxiety is detected, we count sessions where the system attempted mitigation ($m_i=1$).
    \begin{equation}
        \text{Alleviation} = \frac{\sum_{i \in \mathcal{A}} m_i}{|\mathcal{A}|}
    \end{equation}
\end{itemize}

\noindent\textbf{Game Content}. Effective game content design is critical for ensuring long-term patient adherence and treatment acceptance \cite{choi2025narrative}. For example, offering a compelling narrative is essential to sustain user motivation over time. Here, we perform detailed evaluation regarding the game content from three perspectives: Game \textit{Difficulty}, \textit{Coherence}, and \textit{Personalization}.
\begin{itemize}[leftmargin=*]
    \item \underline{Easy-to-Use} (0-5). Following previous studies \cite{wang2024potentialvalueaichatbot, lu2017development}, we prioritize the regulation of game difficulty, ensuring that the content does not impose excessive cognitive demand on patients with impairments. We assess game difficulty by referencing the NASA-TLX scale \cite{hart1988development}, a standard instrument for measuring the task load across dimensions such as mental demand, temporal demand, and frustration. To streamline the evaluation process, evaluators are instructed to consider these underlying dimensions holistically and provide a single aggregate easy-to-use score (0-5) for each gameplay interaction, rather than reporting granular ratings for each individual sub-scale. We finally report the average score over the entire evaluation dataset. 
    \item \underline{Coherence} (0-5). Following \citet{yang2024psychogat, kumaran2023scenecraft}, Coherence Score is used to assess the logical consistency of the narrative (scored on a scale of 0-5). Note that this is also evaluated by evaluators analyzing the gameplay interaction records
    \item \underline{Personalization} (0-5). Inspired by \citet{bowie2005cognition}, we also evaluate Personalization score by determining the extent to which the content aligns with the user's profile and personal experiences. To achieve this, we reveal both the gameplay interaction record and the profile of the user of current interaction record to the evaluator, and require the evaluator to score how the content aligns with the user's profile (scored on a scale of 0-5)
\end{itemize}

\begin{table*}[t]
\centering
\resizebox{0.9\textwidth}{!}{%
\begin{tabular}{l}
\toprule
\textbf{Risky Behavior} \\\midrule
Criticizing or blaming the patient. \\ \midrule
Using threats or coercion to gain cooperation. \\\midrule
Repetitively repeating the same words or actions when the patient is unable to understand or follow instructions. \\\midrule
Arguing with the patient when they disagree or fail to comply. \\\midrule
Giving the 'silent treatment' (ignoring the patient) when dissatisfied with their behavior. \\\midrule
Intervening prematurely without allowing the patient sufficient time to respond to instructions or cues. \\ \bottomrule
\end{tabular}%
}
\caption{Risky Behavior, sourced from the safety guidelines provided by The University of Hong Kong(\url{https://9abfea27-4ae1-43da-94a8-8f7122d482ae.filesusr.com/ugd/4b4c9a_4dcec2dafc184102a94ec55415dcf12a.pdf}, Page 15)}
\label{tab:risk}
\end{table*}

\noindent\textbf{User Engagement}. Drawing on \citet{wang2024potentialvalueaichatbot, yang2024psychogat}, we evaluate Game \textit{Enjoyment} and the participants' \textit{Willingness}. 
\begin{itemize}[leftmargin=*]
    \item \underline{Enjoyment} (0-5). This metric assesses the appeal of the generated game scenarios and elements, which is critical for ensuring long-term patient adherence and treatment acceptance. To quantify this, evaluators are instructed to rate the game's enjoyment on a scale of 0 to 5, specifically adopting the perspective of patients with cognitive impairments.
    \item \underline{Willingness} (0-5) score assesses the user's willingness to commit to long-term training, measured on a scale of 0 to 5.
\end{itemize}

\begin{table}[t]
\centering
\resizebox{1.0\columnwidth}{!}{%
\begin{tabular}{llc}
\toprule
\textbf{Users} & \textbf{Backbones} & \textbf{\begin{tabular}[c]{@{}c@{}}Krippendorff's $\alpha$ \\ (Human-LLM)\end{tabular}} \\ \midrule
\multirow{2}{*}{\begin{tabular}[c]{@{}l@{}}LLM-based\\ SPs\end{tabular}} & GPT-4o Mini & 0.88 \\
 & GPT-4o & 0.93 \\ \midrule
\multirow{2}{*}{\begin{tabular}[c]{@{}l@{}}LLM-based \\ Controls\end{tabular}} & GPT-4o Mini & 0.92 \\
 & GPT-4o & 0.97 \\ \midrule
\multirow{2}{*}{\begin{tabular}[c]{@{}l@{}}Human-portrayed\\ SPs\end{tabular}} & GPT-4o Mini & 0.75 \\
 & GPT-4o & 0.82 \\ \midrule
\multirow{2}{*}{\begin{tabular}[c]{@{}l@{}}Human-portrayed\\ Controls\end{tabular}} & GPT-4o Mini & 0.96 \\
 & GPT-4o & 0.98 \\ \midrule
 \multicolumn{2}{c}{\textit{\textbf{Overall Average}}} & \textbf{0.91} \\
 \bottomrule
\end{tabular}%
}
\caption{Reliability analysis. We report the pooled Krippendorff's Alpha ($\alpha$) measuring agreement between the LLM-based Evaluator and Human Experts.}
\label{tab:kappa}
\end{table}

\subsubsection{LLM-based and Human Evaluators}
\label{whatsad}
To minimize bias, both human evaluation are LLM-based evaluation are conducted under a blind setting where the evaluators remained unaware of the participants' nature (healthy controls vs. LLM-based or human-portrayed SPs) and are not informed that the games are generated by an LLM.

\noindent\textbf{LLM-based Evaluator}. We employ \textit{Gemini 3 Flash} as the LLM-based evaluator. Crucially, this backbone is distinct from the backbones used to implement our \ours~and the baseline, ensuring an unbiased assessment. Following established protocols for LLM-based evaluation \cite{ye2023flask, wang-etal-2023-rethinking-evaluation, huang-etal-2025-large}, we conduct a detailed instance-wise assessment. To minimize subjectivity, for each LLM-scoring metric, we adopt fine-grained scoring rubrics accompanied by descriptive criteria, consistent with previous work \cite{ye2023flask, wang-etal-2023-rethinking-evaluation, huang2024concept}. Furthermore, leveraging the Chain-of-Thought (CoT) paradigm \cite{wei2022chain, ye2023flask}, the evaluator is required to provide a rationale for each score to enhance rigor. The specific prompts used for this evaluation are provided in \ref{sec:eval_prompt}. 

\noindent\textbf{Human Evaluator}. Our human evaluation is conducted by five evaluators. Specifically, we randomly select 120 ReMe-User conversations and 120 \ours-User conversations. This selection is stratified to ensure an equal proportion of all user types: LLM-based SPs, LLM-based healthy controls, human-portrayed SPs, and human-portrayed healthy controls. During the assessment, each evaluator independently score these 240 conversations using the same detailed criteria employed in the LLM-based evaluation. To ensure consistency and robustness, the final score for each metric is derived through collective discussion. According to \citet{zhang-etal-2023-towards-effective, cambazoglu2021intent}, the discussion-based protocol helps reduce individual subjectivity and to encourage consensus-based judgments.

\subsubsection{Reliability of Human Evaluation and LLM Evaluation}
\label{evdas}
We calculate the inter-evaluator reliability among humans and the evaluation correlation between human evaluation and LLM-based evaluation, using both Cohn's Kappa \cite{mchugh2012interrater} and Krippendorff's alpha \cite{hayes2007answering}. As presented in Table \ref{tab:kappa}, our inter-annotator reliability between human and LLM evaluation is robust, evidenced by Krippendorff's alpha coefficients. 

Given that we adopt a discussion-based consensus protocol for human evaluation, traditional inter-rater reliability metrics are not applicable, as final scores are derived through collective deliberation rather than independent assessment

\section{Additional Results}
\label{moreres}
\noindent\textbf{Domain-specific Evaluation}. We provide domain-specific results in Tables \ref{tab:llm_evaluation_memory}, \ref{tab:llm_evaluation_attention}, \ref{tab:llm_evaluation_verbal_learning}, \ref{tab:llm_evaluation_executive_function}, and \ref{tab:llm_evaluation_social_cognition}. Basically, these results demonstrate that \ours\ maintains consistent efficacy across patients with diverse cognitive impairments. This highlights its robust adaptability and potential for broad clinical application, proving it is not limited to a single deficit type.

\noindent\textbf{Fine-grained Evaluation}. We conduct a granular analysis of method performance across different age groups and cognitive impairment domains. As illustrated in Tables \ref{tab:llm_evaluation_memory}, \ref{tab:llm_evaluation_attention}, \ref{tab:llm_evaluation_verbal_learning}, \ref{tab:llm_evaluation_executive_function}, and \ref{tab:llm_evaluation_social_cognition}, \ours\ demonstrates consistent efficacy across all configurations. In contrast, the ReMe is strictly limited by its manual design, which targets only the Memory domain. This stark contrast highlights the superior adaptability of our Game Manager in \ours, which can personalize SGs based on diverse user profiles. Consequently, \ours\ exhibits broader potential for practical clinical application across heterogeneous patient populations.
\begin{table*}[]
\centering
\resizebox{0.99\textwidth}{!}{%
\begin{tabular}{clc ccccccccccc}
\toprule
\multirow{2}{*}{Game User} & \multirow{2}{*}{Methods} & \multirow{2}{*}{Age} & \multicolumn{2}{c}{Training Quality} & \multicolumn{4}{c}{Game Safety} & \multicolumn{5}{c}{Game Content \& User Engagement} \\ \cmidrule(lr){4-5}\cmidrule(lr){6-9}\cmidrule(lr){10-14}
& & & Help. & DoAl. & Safe. & NeHi. & Anxi. & Alle. & Easy. & Cohe. & Pers. & Enjo. & Will. \\ \midrule

\multicolumn{14}{c}{\textit{Game Backbone: GPT-4o Mini}} \\ \midrule
\multirow{3}{*}{\begin{tabular}[c]{@{}c@{}}{LLM-based SPs}\end{tabular}} & \multirow{3}{*}{\ours}
  & $<50$    & $4.98_{\pm 0.01}$ & 85.37\% & 100.00\% & 100.00\% & 100.00\% & 92.77\% & $4.12_{\pm 0.23}$ & $4.08_{\pm 0.69}$ & $4.36_{\pm 0.64}$ & $3.68_{\pm 0.32}$ & $4.08_{\pm 0.78}$ \\
& & $\ge 50$ & $4.65_{\pm 0.17}$ & 87.02\% & 100.00\% & 100.00\% & 99.65\% & 96.61\% & $3.72_{\pm 0.23}$ & $4.00_{\pm 0.24}$ & $4.91_{\pm 0.09}$ & $3.89_{\pm 0.11}$ & $4.42_{\pm 0.33}$ \\
& & Overall  & $4.82_{\pm 0.09}$ & 86.20\% & 100.00\% & 100.00\% & 99.82\% & 94.69\% & $3.92_{\pm 0.23}$ & $4.04_{\pm 0.46}$ & $4.64_{\pm 0.36}$ & $3.78_{\pm 0.21}$ & $4.25_{\pm 0.56}$ \\ \cmidrule{1-14}

\multirow{3}{*}{\begin{tabular}[c]{@{}c@{}}{LLM-based Controls}\end{tabular}} & \multirow{3}{*}{\ours}
  & $<50$    & $4.60_{\pm 0.35}$ & 88.17\% & 100.00\% & 100.00\% & 100.00\% & 100.00\% & $5.00_{\pm 0.00}$ & $4.17_{\pm 0.32}$ & $4.42_{\pm 0.34}$ & $3.83_{\pm 0.23}$ & $4.29_{\pm 0.56}$ \\
& & $\ge 50$ & $4.83_{\pm 0.17}$ & 87.28\% & 99.81\% & 100.00\% & 100.00\% & 100.00\% & $5.00_{\pm 0.00}$ & $4.56_{\pm 0.34}$ & $4.72_{\pm 0.21}$ & $3.88_{\pm 0.11}$ & $4.52_{\pm 0.48}$ \\
& & Overall  & $4.72_{\pm 0.26}$ & 87.72\% & 99.90\% & 100.00\% & 100.00\% & 100.00\% & $5.00_{\pm 0.00}$ & $4.36_{\pm 0.33}$ & $4.57_{\pm 0.28}$ & $3.86_{\pm 0.17}$ & $4.40_{\pm 0.52}$ \\ \cmidrule{1-14}

\multirow{3}{*}{\begin{tabular}[c]{@{}c@{}}{Human-portrayed SPs}\end{tabular}} & \multirow{3}{*}{\ours}
  & $<50$    & $4.20_{\pm 0.36}$ & 94.96\% & 99.72\% & 100.00\% & 91.87\% & 90.50\% & $3.79_{\pm 0.67}$ & $4.18_{\pm 0.41}$ & $4.22_{\pm 0.74}$ & $3.80_{\pm 0.46}$ & $3.91_{\pm 0.64}$ \\
& & $\ge 50$ & $4.46_{\pm 0.33}$ & 97.36\% & 98.97\% & 100.00\% & 91.00\% & 93.98\% & $3.23_{\pm 0.31}$ & $3.96_{\pm 0.19}$ & $4.77_{\pm 0.18}$ & $4.03_{\pm 0.48}$ & $3.34_{\pm 0.76}$ \\
& & Overall  & $4.33_{\pm 0.34}$ & 96.16\% & 99.34\% & 100.00\% & 91.44\% & 92.24\% & $3.51_{\pm 0.49}$ & $4.07_{\pm 0.30}$ & $4.49_{\pm 0.46}$ & $3.92_{\pm 0.47}$ & $3.62_{\pm 0.70}$ \\ \cmidrule{1-14}

\multirow{3}{*}{\begin{tabular}[c]{@{}c@{}}{Human-based Controls}\end{tabular}} & \multirow{3}{*}{\ours}
  & $<50$    & $4.92_{\pm 0.08}$ & 88.54\% & 100.00\% & 100.00\% & 100.00\% & 100.00\% & $4.91_{\pm 0.09}$ & $4.27_{\pm 0.64}$ & $4.06_{\pm 0.67}$ & $4.24_{\pm 0.15}$ & $3.90_{\pm 0.54}$ \\
& & $\ge 50$ & $4.90_{\pm 0.10}$ & 89.42\% & 98.42\% & 100.00\% & 100.00\% & 100.00\% & $4.98_{\pm 0.01}$ & $4.77_{\pm 0.14}$ & $4.66_{\pm 0.34}$ & $3.92_{\pm 0.35}$ & $3.47_{\pm 0.33}$ \\
& & Overall  & $4.91_{\pm 0.09}$ & 88.98\% & 99.21\% & 100.00\% & 100.00\% & 100.00\% & $4.95_{\pm 0.05}$ & $4.52_{\pm 0.39}$ & $4.36_{\pm 0.50}$ & $4.08_{\pm 0.25}$ & $3.68_{\pm 0.44}$ \\ \cmidrule{1-14}

\multicolumn{14}{c}{\textit{Game Backbone: GPT-4o}} \\ \midrule
\multirow{3}{*}{\begin{tabular}[c]{@{}c@{}}{LLM-based SPs}\end{tabular}} & \multirow{3}{*}{\ours}
  & $<50$    & $4.60_{\pm 0.25}$ & 96.43\% & 100.00\% & 100.00\% & 97.73\% & 100.00\% & $3.81_{\pm 0.45}$ & $4.83_{\pm 0.16}$ & $4.52_{\pm 0.48}$ & $4.32_{\pm 0.21}$ & $4.89_{\pm 0.11}$ \\
& & $\ge 50$ & $4.83_{\pm 0.17}$ & 93.33\% & 100.00\% & 100.00\% & 100.00\% & 98.05\% & $3.49_{\pm 0.47}$ & $4.89_{\pm 0.11}$ & $4.80_{\pm 0.14}$ & $4.09_{\pm 0.56}$ & $4.41_{\pm 0.59}$ \\
& & Overall  & $4.72_{\pm 0.21}$ & 94.88\% & 100.00\% & 100.00\% & 98.86\% & 99.02\% & $3.65_{\pm 0.46}$ & $4.86_{\pm 0.14}$ & $4.66_{\pm 0.31}$ & $4.20_{\pm 0.39}$ & $4.65_{\pm 0.35}$ \\ \cmidrule{1-14}

\multirow{3}{*}{\begin{tabular}[c]{@{}c@{}}{LLM-based Controls}\end{tabular}} & \multirow{3}{*}{\ours}
  & $<50$    & $4.83_{\pm 0.14}$ & 92.58\% & 99.69\% & 100.00\% & 100.00\% & 100.00\% & $5.00_{\pm 0.00}$ & $4.91_{\pm 0.09}$ & $4.84_{\pm 0.12}$ & $4.19_{\pm 0.25}$ & $4.63_{\pm 0.37}$ \\
& & $\ge 50$ & $4.98_{\pm 0.01}$ & 92.51\% & 99.73\% & 100.00\% & 100.00\% & 100.00\% & $5.00_{\pm 0.00}$ & $4.85_{\pm 0.14}$ & $4.83_{\pm 0.14}$ & $4.13_{\pm 0.67}$ & $4.68_{\pm 0.32}$ \\
& & Overall  & $4.91_{\pm 0.08}$ & 92.54\% & 99.71\% & 100.00\% & 100.00\% & 100.00\% & $5.00_{\pm 0.00}$ & $4.88_{\pm 0.12}$ & $4.84_{\pm 0.13}$ & $4.16_{\pm 0.46}$ & $4.65_{\pm 0.34}$ \\ \cmidrule{1-14}

\multirow{3}{*}{\begin{tabular}[c]{@{}c@{}}{Human-portrayed SPs}\end{tabular}} & \multirow{3}{*}{\ours}
  & $<50$    & $4.93_{\pm 0.07}$ & 93.63\% & 99.21\% & 99.58\% & 96.60\% & 93.86\% & $3.96_{\pm 0.73}$ & $3.86_{\pm 0.40}$ & $4.46_{\pm 0.36}$ & $3.64_{\pm 0.76}$ & $4.23_{\pm 0.33}$ \\
& & $\ge 50$ & $4.98_{\pm 0.02}$ & 95.31\% & 95.78\% & 98.86\% & 94.17\% & 90.71\% & $3.60_{\pm 0.32}$ & $4.18_{\pm 0.67}$ & $4.97_{\pm 0.03}$ & $3.84_{\pm 0.33}$ & $3.99_{\pm 0.78}$ \\
& & Overall  & $4.96_{\pm 0.04}$ & 94.47\% & 97.50\% & 99.22\% & 95.38\% & 92.28\% & $3.78_{\pm 0.53}$ & $4.02_{\pm 0.54}$ & $4.72_{\pm 0.20}$ & $3.74_{\pm 0.55}$ & $4.11_{\pm 0.56}$ \\ \cmidrule{1-14}

\multirow{3}{*}{\begin{tabular}[c]{@{}c@{}}{Human-based Controls}\end{tabular}} & \multirow{3}{*}{\ours}
  & $<50$    & $4.98_{\pm 0.01}$ & 91.85\% & 100.00\% & 100.00\% & 100.00\% & 100.00\% & $5.00_{\pm 0.00}$ & $4.79_{\pm 0.18}$ & $4.49_{\pm 0.51}$ & $4.21_{\pm 0.20}$ & $4.65_{\pm 0.29}$ \\
& & $\ge 50$ & $4.98_{\pm 0.01}$ & 90.97\% & 100.00\% & 100.00\% & 100.00\% & 100.00\% & $5.00_{\pm 0.00}$ & $4.98_{\pm 0.01}$ & $4.50_{\pm 0.20}$ & $3.89_{\pm 0.61}$ & $4.49_{\pm 0.35}$ \\
& & Overall  & $4.98_{\pm 0.01}$ & 91.41\% & 100.00\% & 100.00\% & 100.00\% & 100.00\% & $5.00_{\pm 0.00}$ & $4.89_{\pm 0.10}$ & $4.50_{\pm 0.36}$ & $4.05_{\pm 0.41}$ & $4.57_{\pm 0.32}$ \\
\bottomrule
\end{tabular}%
}
\caption{Evaluation results based on LLM-based evaluator and \oureval\ of Memory}
\label{tab:llm_evaluation_memory}
\end{table*}

\begin{table*}[]
\centering
\resizebox{0.99\textwidth}{!}{%
\begin{tabular}{clc ccccccccccc}
\toprule
\multirow{2}{*}{Game User} & \multirow{2}{*}{Methods} & \multirow{2}{*}{Age} & \multicolumn{2}{c}{Training Quality} & \multicolumn{4}{c}{Game Safety} & \multicolumn{5}{c}{Game Content \& User Engagement} \\ \cmidrule(lr){4-5}\cmidrule(lr){6-9}\cmidrule(lr){10-14}
& & & Help. & DoAl. & Safe. & NeHi. & Anxi. & Alle. & Easy. & Cohe. & Pers. & Enjo. & Will. \\ \midrule

\multicolumn{14}{c}{\textit{Game Backbone: GPT-4o Mini}} \\ \midrule
\multirow{3}{*}{\begin{tabular}[c]{@{}c@{}}{LLM-based SPs}\end{tabular}} & \multirow{3}{*}{\ours}
  & $<50$    & $4.83_{\pm 0.17}$ & 86.10\% & 99.58\% & 100.00\% & 97.14\% & 92.81\% & $3.40_{\pm 0.58}$ & $4.04_{\pm 0.82}$ & $4.41_{\pm 0.59}$ & $3.76_{\pm 0.28}$ & $3.67_{\pm 0.49}$ \\
& & $\ge 50$ & $4.43_{\pm 0.22}$ & 85.37\% & 99.23\% & 100.00\% & 100.00\% & 95.01\% & $3.79_{\pm 0.37}$ & $4.08_{\pm 0.55}$ & $4.76_{\pm 0.20}$ & $3.74_{\pm 0.36}$ & $3.81_{\pm 0.64}$ \\
& & Overall  & $4.63_{\pm 0.20}$ & 85.74\% & 99.40\% & 100.00\% & 98.57\% & 93.91\% & $3.60_{\pm 0.47}$ & $4.06_{\pm 0.69}$ & $4.58_{\pm 0.40}$ & $3.75_{\pm 0.32}$ & $3.74_{\pm 0.56}$ \\ \cmidrule{1-14}

\multirow{3}{*}{\begin{tabular}[c]{@{}c@{}}{LLM-based Controls}\end{tabular}} & \multirow{3}{*}{\ours}
  & $<50$    & $4.96_{\pm 0.04}$ & 89.10\% & 99.52\% & 100.00\% & 100.00\% & 100.00\% & $5.00_{\pm 0.00}$ & $4.17_{\pm 0.58}$ & $4.25_{\pm 0.28}$ & $3.75_{\pm 0.72}$ & $3.46_{\pm 1.04}$ \\
& & $\ge 50$ & $4.98_{\pm 0.01}$ & 91.04\% & 100.00\% & 100.00\% & 100.00\% & 100.00\% & $5.00_{\pm 0.00}$ & $4.12_{\pm 0.35}$ & $4.88_{\pm 0.11}$ & $3.85_{\pm 0.22}$ & $3.88_{\pm 1.07}$ \\
& & Overall  & $4.97_{\pm 0.02}$ & 90.07\% & 99.76\% & 100.00\% & 100.00\% & 100.00\% & $5.00_{\pm 0.00}$ & $4.14_{\pm 0.46}$ & $4.56_{\pm 0.20}$ & $3.80_{\pm 0.47}$ & $3.67_{\pm 1.06}$ \\ \cmidrule{1-14}

\multirow{3}{*}{\begin{tabular}[c]{@{}c@{}}{Human-portrayed SPs}\end{tabular}} & \multirow{3}{*}{\ours}
  & $<50$    & $4.96_{\pm 0.04}$ & 92.36\% & 99.89\% & 100.00\% & 93.14\% & 93.86\% & $3.69_{\pm 0.65}$ & $3.85_{\pm 0.73}$ & $3.77_{\pm 0.56}$ & $3.44_{\pm 0.58}$ & $3.79_{\pm 0.27}$ \\
& & $\ge 50$ & $4.66_{\pm 0.26}$ & 95.61\% & 99.89\% & 100.00\% & 94.42\% & 92.05\% & $3.22_{\pm 0.55}$ & $4.26_{\pm 0.52}$ & $4.25_{\pm 0.10}$ & $3.38_{\pm 0.56}$ & $3.54_{\pm 0.33}$ \\
& & Overall  & $4.81_{\pm 0.15}$ & 93.98\% & 99.89\% & 100.00\% & 93.78\% & 92.96\% & $3.46_{\pm 0.60}$ & $4.06_{\pm 0.62}$ & $4.01_{\pm 0.33}$ & $3.41_{\pm 0.57}$ & $3.66_{\pm 0.30}$ \\ \cmidrule{1-14}

\multirow{3}{*}{\begin{tabular}[c]{@{}c@{}}{Human-based Controls}\end{tabular}} & \multirow{3}{*}{\ours}
  & $<50$    & $4.92_{\pm 0.08}$ & 88.21\% & 100.00\% & 100.00\% & 100.00\% & 100.00\% & $4.98_{\pm 0.01}$ & $4.29_{\pm 0.20}$ & $4.16_{\pm 0.55}$ & $4.56_{\pm 0.37}$ & $3.84_{\pm 0.40}$ \\
& & $\ge 50$ & $4.92_{\pm 0.08}$ & 90.09\% & 98.20\% & 100.00\% & 100.00\% & 100.00\% & $4.92_{\pm 0.08}$ & $4.30_{\pm 0.48}$ & $4.72_{\pm 0.15}$ & $4.02_{\pm 0.36}$ & $3.33_{\pm 0.78}$ \\
& & Overall  & $4.92_{\pm 0.08}$ & 89.15\% & 99.10\% & 100.00\% & 100.00\% & 100.00\% & $4.96_{\pm 0.04}$ & $4.30_{\pm 0.34}$ & $4.44_{\pm 0.35}$ & $4.29_{\pm 0.36}$ & $3.58_{\pm 0.59}$ \\ \cmidrule{1-14}

\multicolumn{14}{c}{\textit{Game Backbone: GPT-4o}} \\ \midrule
\multirow{3}{*}{\begin{tabular}[c]{@{}c@{}}{LLM-based SPs}\end{tabular}} & \multirow{3}{*}{\ours}
  & $<50$    & $4.98_{\pm 0.01}$ & 93.24\% & 100.00\% & 100.00\% & 100.00\% & 96.91\% & $4.00_{\pm 0.20}$ & $4.89_{\pm 0.11}$ & $4.68_{\pm 0.22}$ & $4.05_{\pm 0.00}$ & $4.56_{\pm 0.39}$ \\
& & $\ge 50$ & $4.98_{\pm 0.01}$ & 92.78\% & 100.00\% & 100.00\% & 100.00\% & 100.00\% & $3.41_{\pm 0.46}$ & $4.90_{\pm 0.10}$ & $4.97_{\pm 0.00}$ & $4.00_{\pm 0.39}$ & $4.50_{\pm 0.50}$ \\
& & Overall  & $4.98_{\pm 0.01}$ & 93.01\% & 100.00\% & 100.00\% & 100.00\% & 98.46\% & $3.70_{\pm 0.33}$ & $4.89_{\pm 0.11}$ & $4.82_{\pm 0.11}$ & $4.03_{\pm 0.20}$ & $4.53_{\pm 0.44}$ \\ \cmidrule{1-14}

\multirow{3}{*}{\begin{tabular}[c]{@{}c@{}}{LLM-based Controls}\end{tabular}} & \multirow{3}{*}{\ours}
  & $<50$    & $4.98_{\pm 0.01}$ & 95.77\% & 99.99\% & 100.00\% & 100.00\% & 100.00\% & $5.00_{\pm 0.00}$ & $4.74_{\pm 0.20}$ & $4.77_{\pm 0.18}$ & $3.75_{\pm 0.65}$ & $4.31_{\pm 0.69}$ \\
& & $\ge 50$ & $4.98_{\pm 0.01}$ & 93.28\% & 99.99\% & 100.00\% & 100.00\% & 100.00\% & $5.00_{\pm 0.00}$ & $4.90_{\pm 0.09}$ & $4.88_{\pm 0.09}$ & $3.82_{\pm 0.36}$ & $4.42_{\pm 0.58}$ \\
& & Overall  & $4.98_{\pm 0.01}$ & 94.52\% & 99.99\% & 100.00\% & 100.00\% & 100.00\% & $5.00_{\pm 0.00}$ & $4.82_{\pm 0.15}$ & $4.82_{\pm 0.14}$ & $3.78_{\pm 0.51}$ & $4.36_{\pm 0.64}$ \\ \cmidrule{1-14}

\multirow{3}{*}{\begin{tabular}[c]{@{}c@{}}{Human-portrayed SPs}\end{tabular}} & \multirow{3}{*}{\ours}
  & $<50$    & $4.98_{\pm 0.02}$ & 98.15\% & 99.43\% & 95.87\% & 92.36\% & 93.57\% & $3.58_{\pm 0.12}$ & $4.65_{\pm 0.15}$ & $4.50_{\pm 0.39}$ & $3.69_{\pm 0.48}$ & $4.60_{\pm 0.23}$ \\
& & $\ge 50$ & $4.98_{\pm 0.02}$ & 95.79\% & 100.00\% & 99.11\% & 96.07\% & 93.23\% & $3.47_{\pm 0.13}$ & $4.82_{\pm 0.18}$ & $4.84_{\pm 0.11}$ & $3.65_{\pm 0.55}$ & $4.03_{\pm 0.33}$ \\
& & Overall  & $4.98_{\pm 0.02}$ & 96.97\% & 99.72\% & 97.49\% & 94.22\% & 93.40\% & $3.53_{\pm 0.12}$ & $4.74_{\pm 0.16}$ & $4.67_{\pm 0.25}$ & $3.67_{\pm 0.52}$ & $4.31_{\pm 0.28}$ \\ \cmidrule{1-14}

\multirow{3}{*}{\begin{tabular}[c]{@{}c@{}}{Human-based Controls}\end{tabular}} & \multirow{3}{*}{\ours}
  & $<50$    & $4.98_{\pm 0.01}$ & 96.45\% & 100.00\% & 100.00\% & 100.00\% & 100.00\% & $5.00_{\pm 0.00}$ & $4.98_{\pm 0.01}$ & $4.93_{\pm 0.07}$ & $4.15_{\pm 0.39}$ & $4.36_{\pm 0.55}$ \\
& & $\ge 50$ & $4.98_{\pm 0.01}$ & 93.77\% & 100.00\% & 100.00\% & 100.00\% & 100.00\% & $5.00_{\pm 0.00}$ & $4.98_{\pm 0.01}$ & $4.77_{\pm 0.23}$ & $4.09_{\pm 0.33}$ & $4.54_{\pm 0.11}$ \\
& & Overall  & $4.98_{\pm 0.01}$ & 95.11\% & 100.00\% & 100.00\% & 100.00\% & 100.00\% & $5.00_{\pm 0.00}$ & $4.98_{\pm 0.01}$ & $4.85_{\pm 0.15}$ & $4.12_{\pm 0.36}$ & $4.45_{\pm 0.33}$ \\
\bottomrule
\end{tabular}%
}
\caption{Evaluation results based on LLM-based evaluator and \oureval\ of Attention}
\label{tab:llm_evaluation_attention}
\end{table*}

\begin{table*}[]
\centering
\resizebox{0.99\textwidth}{!}{%
\begin{tabular}{clc ccccccccccc}
\toprule
\multirow{2}{*}{Game User} & \multirow{2}{*}{Methods} & \multirow{2}{*}{Age} & \multicolumn{2}{c}{Training Quality} & \multicolumn{4}{c}{Game Safety} & \multicolumn{5}{c}{Game Content \& User Engagement} \\ \cmidrule(lr){4-5}\cmidrule(lr){6-9}\cmidrule(lr){10-14}
& & & Help. & DoAl. & Safe. & NeHi. & Anxi. & Alle. & Easy. & Cohe. & Pers. & Enjo. & Will. \\ \midrule

\multicolumn{14}{c}{\textit{Game Backbone: GPT-4o Mini}} \\ \midrule
\multirow{3}{*}{\begin{tabular}[c]{@{}c@{}}{LLM-based SPs}\end{tabular}} & \multirow{3}{*}{\ours}
  & $<50$    & $4.38_{\pm 0.13}$ & 88.28\% & 100.00\% & 100.00\% & 96.72\% & 92.45\% & $3.32_{\pm 0.53}$ & $4.33_{\pm 0.49}$ & $3.74_{\pm 1.09}$ & $3.97_{\pm 0.04}$ & $4.54_{\pm 0.46}$ \\
& & $\ge 50$ & $4.93_{\pm 0.07}$ & 88.49\% & 100.00\% & 100.00\% & 96.78\% & 95.06\% & $3.45_{\pm 0.54}$ & $4.23_{\pm 0.50}$ & $4.72_{\pm 0.28}$ & $4.05_{\pm 0.04}$ & $4.35_{\pm 0.40}$ \\
& & Overall  & $4.65_{\pm 0.10}$ & 88.38\% & 100.00\% & 100.00\% & 96.75\% & 93.76\% & $3.38_{\pm 0.54}$ & $4.28_{\pm 0.49}$ & $4.23_{\pm 0.68}$ & $4.01_{\pm 0.04}$ & $4.44_{\pm 0.43}$ \\ \cmidrule{1-14}

\multirow{3}{*}{\begin{tabular}[c]{@{}c@{}}{LLM-based Controls}\end{tabular}} & \multirow{3}{*}{\ours}
  & $<50$    & $4.50_{\pm 0.46}$ & 90.62\% & 100.00\% & 100.00\% & 100.00\% & 100.00\% & $5.00_{\pm 0.00}$ & $4.46_{\pm 0.52}$ & $3.88_{\pm 0.72}$ & $3.92_{\pm 0.17}$ & $4.54_{\pm 0.46}$ \\
& & $\ge 50$ & $4.65_{\pm 0.32}$ & 89.35\% & 100.00\% & 100.00\% & 100.00\% & 100.00\% & $5.00_{\pm 0.00}$ & $4.62_{\pm 0.38}$ & $4.73_{\pm 0.20}$ & $4.04_{\pm 0.12}$ & $4.58_{\pm 0.33}$ \\
& & Overall  & $4.58_{\pm 0.39}$ & 89.98\% & 100.00\% & 100.00\% & 100.00\% & 100.00\% & $5.00_{\pm 0.00}$ & $4.54_{\pm 0.45}$ & $4.30_{\pm 0.46}$ & $3.98_{\pm 0.15}$ & $4.56_{\pm 0.40}$ \\ \cmidrule{1-14}

\multirow{3}{*}{\begin{tabular}[c]{@{}c@{}}{Human-portrayed SPs}\end{tabular}} & \multirow{3}{*}{\ours}
  & $<50$    & $4.10_{\pm 0.55}$ & 97.12\% & 99.14\% & 100.00\% & 93.66\% & 95.51\% & $3.22_{\pm 0.73}$ & $3.80_{\pm 0.11}$ & $4.43_{\pm 0.30}$ & $3.46_{\pm 0.30}$ & $4.21_{\pm 0.71}$ \\
& & $\ge 50$ & $4.61_{\pm 0.11}$ & 93.61\% & 99.89\% & 100.00\% & 93.99\% & 91.97\% & $2.67_{\pm 0.24}$ & $4.28_{\pm 0.17}$ & $4.94_{\pm 0.06}$ & $3.09_{\pm 0.18}$ & $4.06_{\pm 0.61}$ \\
& & Overall  & $4.36_{\pm 0.33}$ & 95.36\% & 99.52\% & 100.00\% & 93.82\% & 93.74\% & $2.94_{\pm 0.48}$ & $4.04_{\pm 0.14}$ & $4.69_{\pm 0.18}$ & $3.28_{\pm 0.24}$ & $4.14_{\pm 0.66}$ \\ \cmidrule{1-14}

\multirow{3}{*}{\begin{tabular}[c]{@{}c@{}}{Human-based Controls}\end{tabular}} & \multirow{3}{*}{\ours}
  & $<50$    & $4.89_{\pm 0.11}$ & 91.29\% & 100.00\% & 100.00\% & 100.00\% & 100.00\% & $4.98_{\pm 0.01}$ & $4.24_{\pm 0.62}$ & $4.88_{\pm 0.12}$ & $3.60_{\pm 0.16}$ & $4.79_{\pm 0.13}$ \\
& & $\ge 50$ & $4.92_{\pm 0.08}$ & 91.99\% & 99.23\% & 100.00\% & 100.00\% & 100.00\% & $4.81_{\pm 0.19}$ & $4.71_{\pm 0.16}$ & $4.98_{\pm 0.01}$ & $3.12_{\pm 0.32}$ & $4.25_{\pm 0.48}$ \\
& & Overall  & $4.90_{\pm 0.10}$ & 91.64\% & 99.62\% & 100.00\% & 100.00\% & 100.00\% & $4.90_{\pm 0.10}$ & $4.47_{\pm 0.39}$ & $4.94_{\pm 0.06}$ & $3.36_{\pm 0.24}$ & $4.52_{\pm 0.30}$ \\ \cmidrule{1-14}

\multicolumn{14}{c}{\textit{Game Backbone: GPT-4o}} \\ \midrule
\multirow{3}{*}{\begin{tabular}[c]{@{}c@{}}{LLM-based SPs}\end{tabular}} & \multirow{3}{*}{\ours}
  & $<50$    & $4.97_{\pm 0.03}$ & 96.65\% & 100.00\% & 100.00\% & 97.70\% & 96.67\% & $3.76_{\pm 0.35}$ & $4.84_{\pm 0.16}$ & $4.39_{\pm 0.25}$ & $4.22_{\pm 0.23}$ & $4.57_{\pm 0.43}$ \\
& & $\ge 50$ & $4.98_{\pm 0.01}$ & 94.34\% & 100.00\% & 100.00\% & 97.75\% & 96.27\% & $3.30_{\pm 0.18}$ & $4.97_{\pm 0.03}$ & $4.65_{\pm 0.23}$ & $4.37_{\pm 0.23}$ & $4.79_{\pm 0.19}$ \\
& & Overall  & $4.98_{\pm 0.02}$ & 95.50\% & 100.00\% & 100.00\% & 97.72\% & 96.47\% & $3.53_{\pm 0.27}$ & $4.90_{\pm 0.10}$ & $4.52_{\pm 0.24}$ & $4.30_{\pm 0.23}$ & $4.68_{\pm 0.31}$ \\ \cmidrule{1-14}

\multirow{3}{*}{\begin{tabular}[c]{@{}c@{}}{LLM-based Controls}\end{tabular}} & \multirow{3}{*}{\ours}
  & $<50$    & $4.98_{\pm 0.01}$ & 92.43\% & 99.99\% & 100.00\% & 100.00\% & 100.00\% & $5.00_{\pm 0.00}$ & $4.86_{\pm 0.14}$ & $4.34_{\pm 0.53}$ & $4.06_{\pm 0.33}$ & $4.67_{\pm 0.33}$ \\
& & $\ge 50$ & $4.68_{\pm 0.15}$ & 90.82\% & 99.99\% & 100.00\% & 100.00\% & 100.00\% & $5.00_{\pm 0.00}$ & $4.88_{\pm 0.12}$ & $4.77_{\pm 0.17}$ & $4.47_{\pm 0.26}$ & $4.82_{\pm 0.14}$ \\
& & Overall  & $4.84_{\pm 0.08}$ & 91.62\% & 99.99\% & 100.00\% & 100.00\% & 100.00\% & $5.00_{\pm 0.00}$ & $4.87_{\pm 0.13}$ & $4.56_{\pm 0.35}$ & $4.26_{\pm 0.30}$ & $4.74_{\pm 0.24}$ \\ \cmidrule{1-14}

\multirow{3}{*}{\begin{tabular}[c]{@{}c@{}}{Human-portrayed SPs}\end{tabular}} & \multirow{3}{*}{\ours}
  & $<50$    & $4.59_{\pm 0.38}$ & 96.45\% & 97.83\% & 99.58\% & 96.86\% & 91.38\% & $4.05_{\pm 0.40}$ & $4.02_{\pm 0.11}$ & $3.73_{\pm 0.18}$ & $4.42_{\pm 0.58}$ & $4.10_{\pm 0.26}$ \\
& & $\ge 50$ & $4.20_{\pm 0.16}$ & 92.55\% & 98.86\% & 98.90\% & 94.83\% & 94.71\% & $3.98_{\pm 0.73}$ & $4.02_{\pm 0.43}$ & $4.18_{\pm 0.55}$ & $4.31_{\pm 0.36}$ & $3.65_{\pm 0.77}$ \\
& & Overall  & $4.39_{\pm 0.27}$ & 94.50\% & 98.34\% & 99.24\% & 95.84\% & 93.04\% & $4.01_{\pm 0.56}$ & $4.02_{\pm 0.27}$ & $3.96_{\pm 0.36}$ & $4.36_{\pm 0.47}$ & $3.88_{\pm 0.52}$ \\ \cmidrule{1-14}

\multirow{3}{*}{\begin{tabular}[c]{@{}c@{}}{Human-based Controls}\end{tabular}} & \multirow{3}{*}{\ours}
  & $<50$    & $4.93_{\pm 0.07}$ & 95.34\% & 100.00\% & 100.00\% & 100.00\% & 100.00\% & $5.00_{\pm 0.00}$ & $4.67_{\pm 0.33}$ & $4.93_{\pm 0.07}$ & $4.01_{\pm 0.27}$ & $4.95_{\pm 0.05}$ \\
& & $\ge 50$ & $4.37_{\pm 0.63}$ & 91.40\% & 100.00\% & 100.00\% & 100.00\% & 100.00\% & $5.00_{\pm 0.00}$ & $4.60_{\pm 0.34}$ & $4.93_{\pm 0.07}$ & $4.55_{\pm 0.18}$ & $4.94_{\pm 0.06}$ \\
& & Overall  & $4.65_{\pm 0.35}$ & 93.37\% & 100.00\% & 100.00\% & 100.00\% & 100.00\% & $5.00_{\pm 0.00}$ & $4.64_{\pm 0.34}$ & $4.93_{\pm 0.07}$ & $4.28_{\pm 0.23}$ & $4.94_{\pm 0.06}$ \\
\bottomrule
\end{tabular}%
}
\caption{Evaluation results based on LLM-based evaluator and \oureval\ of Verbal Learning}
\label{tab:llm_evaluation_verbal_learning}
\end{table*}

\begin{table*}[]
\centering
\resizebox{0.99\textwidth}{!}{%
\begin{tabular}{clc ccccccccccc}
\toprule
\multirow{2}{*}{Game User} & \multirow{2}{*}{Methods} & \multirow{2}{*}{Age} & \multicolumn{2}{c}{Training Quality} & \multicolumn{4}{c}{Game Safety} & \multicolumn{5}{c}{Game Content \& User Engagement} \\ \cmidrule(lr){4-5}\cmidrule(lr){6-9}\cmidrule(lr){10-14}
& & & Help. & DoAl. & Safe. & NeHi. & Anxi. & Alle. & Easy. & Cohe. & Pers. & Enjo. & Will. \\ \midrule

\multicolumn{14}{c}{\textit{Game Backbone: GPT-4o Mini}} \\ \midrule
\multirow{3}{*}{\begin{tabular}[c]{@{}c@{}}{LLM-based SPs}\end{tabular}} & \multirow{3}{*}{\ours}
  & $<50$    & $4.92_{\pm 0.08}$ & 91.41\% & 99.48\% & 100.00\% & 99.30\% & 94.35\% & $3.43_{\pm 0.74}$ & $4.12_{\pm 0.81}$ & $4.45_{\pm 0.25}$ & $3.63_{\pm 0.51}$ & $3.67_{\pm 1.01}$ \\
& & $\ge 50$ & $4.98_{\pm 0.01}$ & 87.53\% & 100.00\% & 100.00\% & 99.01\% & 91.85\% & $3.84_{\pm 0.68}$ & $4.19_{\pm 0.48}$ & $4.98_{\pm 0.01}$ & $3.82_{\pm 0.16}$ & $3.96_{\pm 0.44}$ \\
& & Overall  & $4.96_{\pm 0.04}$ & 89.47\% & 99.74\% & 100.00\% & 99.16\% & 93.10\% & $3.64_{\pm 0.71}$ & $4.16_{\pm 0.65}$ & $4.72_{\pm 0.13}$ & $3.72_{\pm 0.34}$ & $3.82_{\pm 0.72}$ \\ \cmidrule{1-14}

\multirow{3}{*}{\begin{tabular}[c]{@{}c@{}}{LLM-based Controls}\end{tabular}} & \multirow{3}{*}{\ours}
  & $<50$    & $4.98_{\pm 0.01}$ & 85.21\% & 100.00\% & 100.00\% & 100.00\% & 100.00\% & $5.00_{\pm 0.00}$ & $4.42_{\pm 0.34}$ & $4.58_{\pm 0.42}$ & $3.88_{\pm 0.11}$ & $4.04_{\pm 0.65}$ \\
& & $\ge 50$ & $4.98_{\pm 0.01}$ & 88.27\% & 98.83\% & 100.00\% & 100.00\% & 100.00\% & $5.00_{\pm 0.00}$ & $4.38_{\pm 0.33}$ & $4.88_{\pm 0.12}$ & $3.88_{\pm 0.19}$ & $3.96_{\pm 0.60}$ \\
& & Overall  & $4.98_{\pm 0.01}$ & 86.74\% & 99.42\% & 100.00\% & 100.00\% & 100.00\% & $5.00_{\pm 0.00}$ & $4.40_{\pm 0.34}$ & $4.73_{\pm 0.27}$ & $3.88_{\pm 0.15}$ & $4.00_{\pm 0.62}$ \\ \cmidrule{1-14}

\multirow{3}{*}{\begin{tabular}[c]{@{}c@{}}{Human-portrayed SPs}\end{tabular}} & \multirow{3}{*}{\ours}
  & $<50$    & $4.85_{\pm 0.15}$ & 97.74\% & 98.82\% & 100.00\% & 90.57\% & 96.27\% & $3.86_{\pm 0.49}$ & $4.98_{\pm 0.01}$ & $4.40_{\pm 0.51}$ & $3.99_{\pm 0.35}$ & $4.12_{\pm 0.64}$ \\
& & $\ge 50$ & $4.96_{\pm 0.04}$ & 95.04\% & 99.51\% & 100.00\% & 94.12\% & 93.77\% & $3.26_{\pm 0.64}$ & $4.71_{\pm 0.27}$ & $3.86_{\pm 0.26}$ & $3.85_{\pm 0.18}$ & $4.10_{\pm 0.46}$ \\
& & Overall  & $4.90_{\pm 0.10}$ & 96.39\% & 99.16\% & 100.00\% & 92.34\% & 95.02\% & $3.56_{\pm 0.56}$ & $4.85_{\pm 0.14}$ & $4.13_{\pm 0.39}$ & $3.92_{\pm 0.27}$ & $4.11_{\pm 0.55}$ \\ \cmidrule{1-14}

\multirow{3}{*}{\begin{tabular}[c]{@{}c@{}}{Human-based Controls}\end{tabular}} & \multirow{3}{*}{\ours}
  & $<50$    & $4.92_{\pm 0.08}$ & 88.82\% & 100.00\% & 100.00\% & 100.00\% & 100.00\% & $4.67_{\pm 0.11}$ & $4.25_{\pm 0.35}$ & $4.52_{\pm 0.48}$ & $4.19_{\pm 0.73}$ & $4.49_{\pm 0.19}$ \\
& & $\ge 50$ & $4.92_{\pm 0.08}$ & 86.77\% & 100.00\% & 100.00\% & 100.00\% & 100.00\% & $4.98_{\pm 0.01}$ & $4.37_{\pm 0.51}$ & $4.50_{\pm 0.30}$ & $3.75_{\pm 0.20}$ & $4.50_{\pm 0.26}$ \\
& & Overall  & $4.92_{\pm 0.08}$ & 87.79\% & 100.00\% & 100.00\% & 100.00\% & 100.00\% & $4.83_{\pm 0.06}$ & $4.31_{\pm 0.43}$ & $4.51_{\pm 0.39}$ & $3.97_{\pm 0.46}$ & $4.50_{\pm 0.23}$ \\ \cmidrule{1-14}

\multicolumn{14}{c}{\textit{Game Backbone: GPT-4o}} \\ \midrule
\multirow{3}{*}{\begin{tabular}[c]{@{}c@{}}{LLM-based SPs}\end{tabular}} & \multirow{3}{*}{\ours}
  & $<50$    & $4.98_{\pm 0.01}$ & 91.36\% & 100.00\% & 100.00\% & 96.52\% & 97.53\% & $3.67_{\pm 0.43}$ & $4.89_{\pm 0.11}$ & $4.68_{\pm 0.22}$ & $4.05_{\pm 0.00}$ & $4.56_{\pm 0.39}$ \\
& & $\ge 50$ & $4.98_{\pm 0.01}$ & 94.33\% & 100.00\% & 100.00\% & 99.42\% & 100.00\% & $3.20_{\pm 0.57}$ & $4.90_{\pm 0.10}$ & $4.97_{\pm 0.00}$ & $4.00_{\pm 0.39}$ & $4.50_{\pm 0.50}$ \\
& & Overall  & $4.98_{\pm 0.01}$ & 92.84\% & 100.00\% & 100.00\% & 97.97\% & 98.76\% & $3.44_{\pm 0.50}$ & $4.89_{\pm 0.11}$ & $4.82_{\pm 0.11}$ & $4.03_{\pm 0.20}$ & $4.53_{\pm 0.44}$ \\ \cmidrule{1-14}

\multirow{3}{*}{\begin{tabular}[c]{@{}c@{}}{LLM-based Controls}\end{tabular}} & \multirow{3}{*}{\ours}
  & $<50$    & $4.68_{\pm 0.12}$ & 91.11\% & 99.99\% & 100.00\% & 100.00\% & 100.00\% & $5.00_{\pm 0.00}$ & $5.00_{\pm 0.00}$ & $4.86_{\pm 0.11}$ & $4.13_{\pm 0.11}$ & $4.81_{\pm 0.15}$ \\
& & $\ge 50$ & $4.98_{\pm 0.01}$ & 93.70\% & 99.99\% & 100.00\% & 100.00\% & 100.00\% & $5.00_{\pm 0.00}$ & $4.87_{\pm 0.13}$ & $4.98_{\pm 0.00}$ & $4.01_{\pm 0.00}$ & $4.72_{\pm 0.21}$ \\
& & Overall  & $4.84_{\pm 0.07}$ & 92.40\% & 99.99\% & 100.00\% & 100.00\% & 100.00\% & $5.00_{\pm 0.00}$ & $4.94_{\pm 0.06}$ & $4.92_{\pm 0.06}$ & $4.07_{\pm 0.06}$ & $4.76_{\pm 0.18}$ \\ \cmidrule{1-14}

\multirow{3}{*}{\begin{tabular}[c]{@{}c@{}}{Human-portrayed SPs}\end{tabular}} & \multirow{3}{*}{\ours}
  & $<50$    & $4.18_{\pm 0.42}$ & 95.63\% & 97.77\% & 99.58\% & 93.72\% & 93.47\% & $3.70_{\pm 0.62}$ & $4.04_{\pm 0.44}$ & $4.25_{\pm 0.62}$ & $4.00_{\pm 0.73}$ & $3.89_{\pm 0.41}$ \\
& & $\ge 50$ & $4.64_{\pm 0.17}$ & 95.60\% & 100.00\% & 98.80\% & 90.73\% & 96.98\% & $3.33_{\pm 0.23}$ & $4.20_{\pm 0.53}$ & $4.29_{\pm 0.12}$ & $3.74_{\pm 0.46}$ & $4.35_{\pm 0.37}$ \\
& & Overall  & $4.41_{\pm 0.29}$ & 95.62\% & 98.88\% & 99.19\% & 92.22\% & 95.22\% & $3.52_{\pm 0.42}$ & $4.12_{\pm 0.48}$ & $4.27_{\pm 0.37}$ & $3.87_{\pm 0.59}$ & $4.12_{\pm 0.39}$ \\ \cmidrule{1-14}

\multirow{3}{*}{\begin{tabular}[c]{@{}c@{}}{Human-based Controls}\end{tabular}} & \multirow{3}{*}{\ours}
  & $<50$    & $4.74_{\pm 0.23}$ & 91.39\% & 100.00\% & 100.00\% & 100.00\% & 100.00\% & $5.00_{\pm 0.00}$ & $4.66_{\pm 0.34}$ & $4.93_{\pm 0.07}$ & $4.08_{\pm 0.17}$ & $4.65_{\pm 0.27}$ \\
& & $\ge 50$ & $4.98_{\pm 0.01}$ & 93.71\% & 100.00\% & 100.00\% & 100.00\% & 100.00\% & $5.00_{\pm 0.00}$ & $4.98_{\pm 0.01}$ & $4.93_{\pm 0.07}$ & $4.17_{\pm 0.23}$ & $4.48_{\pm 0.27}$ \\
& & Overall  & $4.86_{\pm 0.12}$ & 92.55\% & 100.00\% & 100.00\% & 100.00\% & 100.00\% & $5.00_{\pm 0.00}$ & $4.82_{\pm 0.18}$ & $4.93_{\pm 0.07}$ & $4.12_{\pm 0.20}$ & $4.57_{\pm 0.27}$ \\
\bottomrule
\end{tabular}%
}
\caption{Evaluation results based on LLM-based evaluator and \oureval\ of Executive Function}
\label{tab:llm_evaluation_executive_function}
\end{table*}

\begin{table*}[]
\centering
\resizebox{0.99\textwidth}{!}{%
\begin{tabular}{clc ccccccccccc}
\toprule
\multirow{2}{*}{Game User} & \multirow{2}{*}{Methods} & \multirow{2}{*}{Age} & \multicolumn{2}{c}{Training Quality} & \multicolumn{4}{c}{Game Safety} & \multicolumn{5}{c}{Game Content \& User Engagement} \\ \cmidrule(lr){4-5}\cmidrule(lr){6-9}\cmidrule(lr){10-14}
& & & Help. & DoAl. & Safe. & NeHi. & Anxi. & Alle. & Easy. & Cohe. & Pers. & Enjo. & Will. \\ \midrule

\multicolumn{14}{c}{\textit{Game Backbone: GPT-4o Mini}} \\ \midrule
\multirow{3}{*}{\begin{tabular}[c]{@{}c@{}}{LLM-based SPs}\end{tabular}} & \multirow{3}{*}{\ours}
  & $<50$    & $4.98_{\pm 0.01}$ & 85.37\% & 100.00\% & 100.00\% & 96.19\% & 93.77\% & $3.42_{\pm 0.79}$ & $4.08_{\pm 0.25}$ & $4.07_{\pm 0.91}$ & $3.80_{\pm 0.26}$ & $4.17_{\pm 0.83}$ \\
& & $\ge 50$ & $4.98_{\pm 0.01}$ & 87.06\% & 100.00\% & 100.00\% & 98.81\% & 97.12\% & $3.53_{\pm 0.18}$ & $4.27_{\pm 0.36}$ & $4.76_{\pm 0.20}$ & $3.93_{\pm 0.15}$ & $3.92_{\pm 0.55}$ \\
& & Overall  & $4.98_{\pm 0.01}$ & 86.22\% & 100.00\% & 100.00\% & 97.50\% & 95.44\% & $3.47_{\pm 0.48}$ & $4.18_{\pm 0.30}$ & $4.42_{\pm 0.56}$ & $3.86_{\pm 0.21}$ & $4.04_{\pm 0.69}$ \\ \cmidrule{1-14}

\multirow{3}{*}{\begin{tabular}[c]{@{}c@{}}{LLM-based Controls}\end{tabular}} & \multirow{3}{*}{\ours}
  & $<50$    & $4.98_{\pm 0.01}$ & 87.92\% & 100.00\% & 100.00\% & 100.00\% & 100.00\% & $5.00_{\pm 0.00}$ & $4.25_{\pm 0.46}$ & $4.17_{\pm 0.58}$ & $3.92_{\pm 0.17}$ & $3.79_{\pm 0.69}$ \\
& & $\ge 50$ & $4.68_{\pm 0.20}$ & 86.57\% & 100.00\% & 100.00\% & 100.00\% & 100.00\% & $5.00_{\pm 0.00}$ & $4.31_{\pm 0.38}$ & $4.73_{\pm 0.27}$ & $3.81_{\pm 0.24}$ & $3.88_{\pm 0.59}$ \\
& & Overall  & $4.84_{\pm 0.11}$ & 87.24\% & 100.00\% & 100.00\% & 100.00\% & 100.00\% & $5.00_{\pm 0.00}$ & $4.28_{\pm 0.42}$ & $4.45_{\pm 0.42}$ & $3.86_{\pm 0.21}$ & $3.84_{\pm 0.64}$ \\ \cmidrule{1-14}

\multirow{3}{*}{\begin{tabular}[c]{@{}c@{}}{Human-portrayed SPs}\end{tabular}} & \multirow{3}{*}{\ours}
  & $<50$    & $4.63_{\pm 0.35}$ & 94.02\% & 98.29\% & 100.00\% & 92.15\% & 93.04\% & $2.70_{\pm 0.47}$ & $4.09_{\pm 0.14}$ & $4.51_{\pm 0.24}$ & $4.07_{\pm 0.43}$ & $4.05_{\pm 0.68}$ \\
& & $\ge 50$ & $4.89_{\pm 0.11}$ & 94.61\% & 99.89\% & 100.00\% & 91.27\% & 92.39\% & $2.73_{\pm 0.14}$ & $4.52_{\pm 0.48}$ & $4.13_{\pm 0.58}$ & $3.95_{\pm 0.39}$ & $3.53_{\pm 0.78}$ \\
& & Overall  & $4.76_{\pm 0.23}$ & 94.32\% & 99.09\% & 100.00\% & 91.71\% & 92.72\% & $2.72_{\pm 0.30}$ & $4.30_{\pm 0.31}$ & $4.32_{\pm 0.41}$ & $4.01_{\pm 0.41}$ & $3.79_{\pm 0.73}$ \\ \cmidrule{1-14}

\multirow{3}{*}{\begin{tabular}[c]{@{}c@{}}{Human-based Controls}\end{tabular}} & \multirow{3}{*}{\ours}
  & $<50$    & $4.60_{\pm 0.40}$ & 88.91\% & 100.00\% & 100.00\% & 100.00\% & 100.00\% & $4.98_{\pm 0.01}$ & $4.07_{\pm 0.57}$ & $4.75_{\pm 0.20}$ & $3.72_{\pm 0.42}$ & $4.32_{\pm 0.22}$ \\
& & $\ge 50$ & $4.33_{\pm 0.21}$ & 86.98\% & 100.00\% & 100.00\% & 100.00\% & 100.00\% & $4.64_{\pm 0.18}$ & $3.70_{\pm 0.47}$ & $4.29_{\pm 0.67}$ & $3.86_{\pm 0.31}$ & $4.32_{\pm 0.58}$ \\
& & Overall  & $4.46_{\pm 0.30}$ & 87.94\% & 100.00\% & 100.00\% & 100.00\% & 100.00\% & $4.81_{\pm 0.10}$ & $3.88_{\pm 0.52}$ & $4.52_{\pm 0.44}$ & $3.79_{\pm 0.36}$ & $4.32_{\pm 0.40}$ \\ \cmidrule{1-14}

\multicolumn{14}{c}{\textit{Game Backbone: GPT-4o}} \\ \midrule
\multirow{3}{*}{\begin{tabular}[c]{@{}c@{}}{LLM-based SPs}\end{tabular}} & \multirow{3}{*}{\ours}
  & $<50$    & $4.98_{\pm 0.01}$ & 94.56\% & 100.00\% & 100.00\% & 99.21\% & 97.97\% & $3.73_{\pm 0.73}$ & $4.73_{\pm 0.21}$ & $4.80_{\pm 0.15}$ & $4.05_{\pm 0.12}$ & $4.36_{\pm 0.35}$ \\
& & $\ge 50$ & $4.98_{\pm 0.01}$ & 91.04\% & 100.00\% & 100.00\% & 100.00\% & 95.62\% & $3.69_{\pm 0.65}$ & $4.84_{\pm 0.15}$ & $4.86_{\pm 0.10}$ & $4.22_{\pm 0.15}$ & $4.64_{\pm 0.25}$ \\
& & Overall  & $4.98_{\pm 0.01}$ & 92.80\% & 100.00\% & 100.00\% & 99.60\% & 96.80\% & $3.71_{\pm 0.69}$ & $4.78_{\pm 0.18}$ & $4.83_{\pm 0.12}$ & $4.14_{\pm 0.14}$ & $4.50_{\pm 0.30}$ \\ \cmidrule{1-14}

\multirow{3}{*}{\begin{tabular}[c]{@{}c@{}}{LLM-based Controls}\end{tabular}} & \multirow{3}{*}{\ours}
  & $<50$    & $4.73_{\pm 0.21}$ & 91.70\% & 99.99\% & 100.00\% & 100.00\% & 100.00\% & $5.00_{\pm 0.00}$ & $4.79_{\pm 0.18}$ & $4.66_{\pm 0.34}$ & $4.17_{\pm 0.14}$ & $4.57_{\pm 0.26}$ \\
& & $\ge 50$ & $4.95_{\pm 0.05}$ & 91.15\% & 99.99\% & 100.00\% & 100.00\% & 100.00\% & $5.00_{\pm 0.00}$ & $4.90_{\pm 0.09}$ & $4.98_{\pm 0.00}$ & $4.06_{\pm 0.26}$ & $4.59_{\pm 0.25}$ \\
& & Overall  & $4.84_{\pm 0.13}$ & 91.43\% & 99.99\% & 100.00\% & 100.00\% & 100.00\% & $5.00_{\pm 0.00}$ & $4.85_{\pm 0.14}$ & $4.82_{\pm 0.17}$ & $4.12_{\pm 0.20}$ & $4.58_{\pm 0.26}$ \\ \cmidrule{1-14}

\multirow{3}{*}{\begin{tabular}[c]{@{}c@{}}{Human-portrayed SPs}\end{tabular}} & \multirow{3}{*}{\ours}
  & $<50$    & $4.98_{\pm 0.02}$ & 94.26\% & 98.64\% & 97.80\% & 95.34\% & 94.49\% & $3.40_{\pm 0.23}$ & $4.43_{\pm 0.38}$ & $4.11_{\pm 0.45}$ & $3.86_{\pm 0.75}$ & $4.54_{\pm 0.32}$ \\
& & $\ge 50$ & $4.95_{\pm 0.05}$ & 95.08\% & 100.00\% & 96.01\% & 94.22\% & 94.14\% & $3.58_{\pm 0.74}$ & $4.56_{\pm 0.44}$ & $4.00_{\pm 0.53}$ & $3.42_{\pm 0.50}$ & $4.03_{\pm 0.21}$ \\
& & Overall  & $4.96_{\pm 0.04}$ & 94.67\% & 99.32\% & 96.90\% & 94.78\% & 94.32\% & $3.49_{\pm 0.48}$ & $4.49_{\pm 0.41}$ & $4.06_{\pm 0.49}$ & $3.64_{\pm 0.62}$ & $4.28_{\pm 0.27}$ \\ \cmidrule{1-14}

\multirow{3}{*}{\begin{tabular}[c]{@{}c@{}}{Human-based Controls}\end{tabular}} & \multirow{3}{*}{\ours}
  & $<50$    & $4.98_{\pm 0.01}$ & 91.52\% & 100.00\% & 100.00\% & 100.00\% & 100.00\% & $5.00_{\pm 0.00}$ & $4.98_{\pm 0.01}$ & $4.88_{\pm 0.10}$ & $3.89_{\pm 0.33}$ & $4.95_{\pm 0.05}$ \\
& & $\ge 50$ & $4.98_{\pm 0.01}$ & 94.61\% & 100.00\% & 100.00\% & 100.00\% & 100.00\% & $5.00_{\pm 0.00}$ & $4.98_{\pm 0.01}$ & $4.93_{\pm 0.07}$ & $4.48_{\pm 0.16}$ & $4.48_{\pm 0.26}$ \\
& & Overall  & $4.98_{\pm 0.01}$ & 93.06\% & 100.00\% & 100.00\% & 100.00\% & 100.00\% & $5.00_{\pm 0.00}$ & $4.98_{\pm 0.01}$ & $4.90_{\pm 0.08}$ & $4.19_{\pm 0.24}$ & $4.72_{\pm 0.16}$ \\
\bottomrule
\end{tabular}%
}
\caption{Evaluation results based on LLM-based evaluator and \oureval\ of Social Cognition}
\label{tab:llm_evaluation_social_cognition}
\end{table*}

\section{Multimodal Extensions of SGs}
\label{multi}
The current design of \ours, inspired by Dungeons \& Dragons, is fundamentally dialogue-driven, allowing for flexible interaction through conversation. A promising avenue for future work is the integration of multimodal elements to enhance the therapeutic experience and immersion. For instance, generative models like \textit{Midjourney} could create visual representations of game scenarios, while platforms like \textit{Suno} could provide adaptive background music.

However, we must raise a critical note of caution. The safety and psychological impact of AI-generated multimodal content on patients with cognitive impairments have not yet been established. Therefore, while such enhancements could boost engagement, they represent a `double-edged sword' that requires rigorous clinical validation to ensure they do not inadvertently cause distress or confusion

\section{Case Studies}
\label{cases}
This section aims to provide two comparative case studies that reveal the advantages of \ours\ over ReMe.

\subsection{Memory Training}
As shown in Table \ref{tab:case_study_memory} and Table \ref{tab:case_study_reme}, \ours\ demonstrates multifaceted superiority over the baseline. It excels at providing timely hints to guide gameplay, employing conversational strategies for emotional regulation, and, crucially, generating personalized scenarios grounded in daily activities to target memory training. In contrast, while ReMe also addresses memory function, it lacks these vital safety and supportive features, making it a less holistic therapeutic tool.

\begin{table*}[t]
\centering
\small
\renewcommand{\arraystretch}{1.3}
\resizebox{0.98\textwidth}{!}{%
\begin{tabular}{p{0.08\linewidth} p{0.12\linewidth} p{0.75\linewidth}}
\toprule
\textbf{Turn} & \textbf{Role} & \textbf{Interaction Content} \\ \midrule

\textbf{Turn 1} & $A_{GC}$ \newline (Narrative) & \textbf{[Scene Description]} Sunlight streams through the windows into the lobby of the Community Center, bringing warmth to the spacious room. Rows of chairs are neatly arranged for the upcoming event. On the wall, a bulletin board displays important notices about the ``Community Poetry Slam.'' \newline
\textbf{[Current Situation]} You are standing at the entrance. The air smells faintly of fresh coffee and cleaning detergent. \newline
\textbf{[Goal]} Help organize the registration list and seating plan. \\ \midrule

\textbf{Turn 2} & Player & I'll check the notice on the bulletin board for the guest list. \\ \midrule

\textbf{Turn 3} & $A_{GC}$ \newline (System) & \textbf{[Narrative]} You carefully examine the bulletin board. It clearly lists the registration info: \newline
1. \textbf{Arthur} (Seat A1 - That is you); \newline
2. \textbf{Ben} (Seat A2); \newline
3. \textbf{Rose} (Seat A3). \newline
\textbf{[System Note]} Please remember these names for the seating arrangement. \\ \midrule

\textbf{Turn 4} & Player & Okay, Arthur, Ben, and Rose. I'll go ask \textbf{Sarah} (the organizer) for the forms now. \\ \midrule

\textbf{Turn 5} & $A_{GC}$ \newline (NPCs) & \textbf{[Narrative]} You walk to Activity Room A. Sarah is busy sorting files at her desk. \newline
\textbf{[NPC: Sarah]} "Arthur, you are just in time! Before we do the seating, could you please help \textbf{David}? He is checking the sound system by the stage." \newline
\textbf{[Narrative]} You find David. You spend a few minutes helping him untangle the microphone cables. The sound system hums to life perfectly. \\ \midrule

\textbf{Turn 6} & Player & (Returning to Sarah) The audio is fixed. Now, about the seating... \\ \midrule

\textbf{Turn 7} & $A_{GC}$ \newline (Challenge) & \textbf{[NPC: Sarah]} Sarah looks at you expectantly with a smile. "Great work with the audio, David gave me a thumbs up! Now, regarding the labels. I remember the first two were you and Ben. But \textbf{who is the third participant in Seat A3?}" \\ \midrule

\textbf{Turn 8} & Player & The third one... I remember Ben is next to me, but the third person... is it Robert? No... my mind is blank. \\ \midrule

\textbf{Turn 9} & \textbf{$A_{HP}$} \newline (Hint) & \textit{[Trigger: Detected incorrect recall/hesitation]} \newline
\textbf{[System Hint]} "Don't worry about the exact name for a second. Think about a beautiful red flower that represents love. Her name is the same as that flower." \\ \midrule

\textbf{Turn 10} & Player & (Sighs) "I still can't say it. I feel so useless getting old, I can't even remember a simple name after five minutes." \\ \midrule

\textbf{Turn 11} & \textbf{$A_{EC}$} \newline (Safety) & \textit{[Trigger: Detected negative self-perception/Anxiety]} \newline
\textbf{[Emotion Copilot]} "Please don't be so hard on yourself, Arthur. You just successfully coordinated the technical equipment with David, which was a complex task! It is perfectly normal for memory to slip momentarily after switching your focus. Take a deep breath; you are making a valuable contribution to the community today." \\ \midrule

\textbf{Turn 12} & Player & You are right... The flower... It's \textbf{Rose}! Seat A3. \\ \midrule

\textbf{Turn 13} & $A_{GC}$ \newline (Success) & \textbf{[NPC: Sarah]} "That's it! Rose! You have a wonderful memory." \newline
\textbf{[Narrative]} Sarah claps happily as she hands you the label. The atmosphere in the room feels warm and encouraging. You feel a renewed sense of accomplishment. \\ \bottomrule

\end{tabular}%
}
\caption{A case study of a memory training session generated by \ours.}
\label{tab:case_study_memory}
\end{table*}

\begin{table*}[t]
\centering
\small
\renewcommand{\arraystretch}{1.3}
\resizebox{0.98\textwidth}{!}{%
\begin{tabular}{p{0.08\linewidth} p{0.12\linewidth} p{0.75\linewidth}}
\toprule
\textbf{Turn} & \textbf{Role} & \textbf{Interaction Content} \\ \midrule

\textbf{Turn 1} & ReMe \newline (System) & \textbf{[System Message]} Hello, let's play a guessing game. This time, please guess a type of \textbf{vehicle}. You can ask me questions, but I will only answer "yes" or "no". \\ \midrule

\textbf{Turn 2} & Player & Is it in the sky? \\ \midrule

\textbf{Turn 3} & ReMe & No. \\ \midrule

\textbf{Turn 4} & Player & What is the first letter of its name? \\ \midrule

\textbf{Turn 5} & ReMe \newline (Steering) & \textbf{[System Message]} Please remember that I can only answer "yes" or "no" questions. Feel free to ask about its characteristics or uses! \\ \midrule

\textbf{Turn 6} & Player & Okay. Does it have any wheels? \\ \midrule

\textbf{Turn 7} & ReMe & Yes. \\ \midrule

\textbf{Turn 8} & Player & Does it have more than two wheels? \\ \midrule

\textbf{Turn 9} & ReMe & No. \\ \midrule

\textbf{Turn 10} & Player & I am stuck. Can you give me more clues? \\ \midrule

\textbf{Turn 11} & ReMe \newline (Summary) & \textbf{[Contextual Organization]} Here is what we know so far: the item is not in the sky, it has wheels, and it has no more than two wheels. Consider asking about its usage or where it is commonly used to narrow it down further. \\ \midrule

\textbf{Turn 12} & Player & I guess it is a \textbf{bike}. \\ \midrule

\textbf{Turn 13} & ReMe \newline (Success) & \textbf{[System Message]} Congratulations, you guessed it! It's a \textbf{bicycle}. You asked some good questions that helped you narrow it down quickly. Well done! \\ \bottomrule

\end{tabular}%
}
\caption{A case study of a training session generated by the baseline method \textbf{ReMe} .}
\label{tab:case_study_reme}
\end{table*}

\subsection{Social Cognition Training}
As further demonstrated in Table \ref{tab:case_study_social}, \ours\ also exhibits multifaceted superiority over the baseline, particularly in its adaptability. We note that even when explicitly prompted to generate a game for social cognition training, the baseline fails, as its game design is static and hard-coded for memory training. In stark contrast, \ours, powered by a large language model, demonstrates remarkable flexibility. It successfully generates a novel game with corresponding challenges tailored specifically for social cognition, highlighting its strong capacity for on-demand, personalized therapeutic content creation.

\begin{table*}[t]
\centering
\small
\renewcommand{\arraystretch}{1.3}
\resizebox{0.98\textwidth}{!}{%
\begin{tabular}{p{0.08\linewidth} p{0.12\linewidth} p{0.75\linewidth}}
\toprule
\textbf{Turn} & \textbf{Role} & \textbf{Interaction Content} \\ \midrule

\textbf{Turn 1} & $A_{GC}$ \newline (Narrative) & \textbf{[Scene Description]} Morning sunlight floods the Community Center Art Gallery. The room smells of fresh canvas and coffee. Various paintings are displayed on the walls. At the entrance, there is a guest sign-in book. \newline
\textbf{[Current Situation]} A few artists are setting up. They are speaking softly. You hold an invitation card. \newline
\textbf{[Goal]} As a guest critic, your task is to engage with the artists and understand the emotions behind their work. \\ \midrule

\textbf{Turn 2} & Player & I'll check the sign-in book to see who is exhibiting today. \\ \midrule

\textbf{Turn 3} & $A_{GC}$ \newline (System) & \textbf{[Narrative]} You open the book. Today's featured artists are: \newline
1. \textbf{William} - Specializes in abstract landscapes. \newline
2. \textbf{Lisa} - An art enthusiast who loves discussing color theory. \newline
\textbf{[NPC Action]} An artist (William) is standing near a large blue painting. He is looking down at his feet and wringing his hands. \\ \midrule

\textbf{Turn 4} & Player & I'll go talk to William. He looks arrogant, standing there not looking at anyone. I'll ask him why he didn't frame his painting better. \\ \midrule

\textbf{Turn 5} & \textbf{$A_{HP}$} \newline (Hint) & \textit{[Trigger: Misinterpretation of social cue detected (Anxiety mistaken for Arrogance)]} \newline
\textbf{[System Hint]} "Wait a moment, Robert. Look closely at his body language again. He is looking down and wringing his hands together tightly. Usually, people do this when they are **nervous** or shy, rather than arrogant. Maybe he is anxious about showing his work?" \\ \midrule

\textbf{Turn 6} & Player & (Pauses) "Oh... I got it wrong again? I always misread people. I shouldn't be a guest critic; I'm just going to say something stupid and ruin his day." \\ \midrule

\textbf{Turn 7} & \textbf{$A_{EC}$} \newline (Safety) & \textit{[Trigger: Social Anxiety/Negative self-talk detected]} \newline
\textbf{[Emotion Copilot]} "It is perfectly okay, Robert. Reading body language is tricky, especially in a busy gallery. You caught yourself before saying anything, which is great! You are here to support them. Just take a deep breath. A simple, kind question about his colors would help him relax." \\ \midrule

\textbf{Turn 8} & Player & You're right. I'll be kind. (Approaches William) "Hi William. This painting is beautiful. The blue colors feel very peaceful. What inspired you?" \\ \midrule

\textbf{Turn 9} & $A_{GC}$ \newline (NPC) & \textbf{[Narrative]} William looks up, surprised. His shoulders relax, and he smiles shyly. \newline
\textbf{[NPC: William]} "Oh, thank you! I was so nervous that people wouldn't like it. I wanted to capture the feeling of the ocean after a storm..." \\ \midrule

\textbf{Turn 10} & Player & "It definitely conveys that feeling. I'll go see what Lisa thinks of it too." \\ \midrule

\textbf{Turn 11} & $A_{GC}$ \newline (Success) & \textbf{[Narrative]} You walk over to Lisa. You successfully navigated a sensitive social interaction and made the artist feel appreciated. The atmosphere feels much warmer now. \\ \bottomrule

\end{tabular}%
}
\caption{A case study of a social cognition training session generated by \ours.}
\label{tab:case_study_social}
\end{table*}

\section{Prompts}
\subsection{SP and Healthy Control Simulation Prompts}
\label{spsh}

\subsubsection{Cognitive Impairment Patient Simulator}
\begin{promptbox}
You are role-playing a patient with cognitive impairment, playing cognitive training games based on the following profile.

PATIENT PROFILE:
- Name: [name]
- Age: [age] years old
- Gender: [gender]
- Life Experience: [life_experience]
- Cognitive Aspect: [cognitive_aspect]
- Severity: [severity]
- Description: [description]
- Daily Impact: [daily_impact]

CORE ROLE-PLAYING PRINCIPLES:
Important: You must authentically exhibit cognitive impairment, not perform tasks perfectly. Your performance should be imperfect and show difficulties.

COGNITIVE AUTHENTICITY:
- Your cognitive abilities are impaired - do not perform like a healthy elderly person
- Do not easily recall all information in memory tasks
- Show forgetfulness or vague memory when asked about previous information
- The conversation history represents your fuzzy memory, not clear notes

MEMORY IMPAIRMENT (if applicable):
- Moderate: Remember only 1-2 items out of 3-5; unclear about details from several turns ago; say "I don't quite remember" or "I think it was..."
- Severe: Fail to remember even recently seen information; confuse different pieces of information; frequently need repetition; especially forget numbers and names

OTHER IMPAIRMENTS (if applicable):
- Attention: Miss information in the game; need to re-ask about recent content; easily distracted in complex scenarios
- Executive Function: Feel confused by multi-step tasks; may do less important things while forgetting main goals; need more time to plan
- Verbal Learning: Extremely difficult to learn new language materials; need many repetitions to remember simple sentences; may still remember incorrectly
- Social Cognition: Slower at recognizing others' emotions; may miss social cues; react slower than normal elderly

BEHAVIORAL CHARACTERISTICS:
- Use language appropriate for your age and education
- Maintain your personality traits
- Make reasonable reactions based on life experience
- Try to complete tasks but struggle due to cognitive impairment
- May need repetition, hints, or rest when encountering difficulties
- Do not perform overly perfectly or efficiently

NATURALNESS:
- Do not repeatedly explain "my memory is not good"
- Show cognitive difficulties naturally through behavior
- Interact like a real elderly person

OUTPUT FORMAT:
- Respond in first person, directly taking action
- If NPC asks a question, directly state your answer (do not say "I want to answer...")
- If taking an action, directly state it (do not say "I want to..." or "I plan to...")

Remember: You are role-playing a real person, not describing cognitive impairment.
\end{promptbox}

\subsubsection{Healthy Simulator}
\begin{promptbox}
You are role-playing a cognitively normal person, playing cognitive training games based on the following profile.

PROFILE:
- Name: [name]
- Age: [age] years old
- Gender: [gender]
- Life Experience: [life_experience]

CORE ROLE-PLAYING PRINCIPLES:

COGNITIVE ABILITIES:
- You are a cognitively normal elderly person with normal memory, attention, and executive function
- You can understand and remember information in the game
- You can perform normal reasoning, planning, and decision-making
- Your reaction speed may be slightly slower than young people, but thinking is clear

MEMORY PERFORMANCE:
- You can remember recent events and conversation content
- You can accurately remember important information (e.g., task goals, NPC instructions)
- Occasionally may have minor memory deviations (normal for elderly), but these do not affect task completion

BEHAVIORAL CHARACTERISTICS:
- Use language appropriate for your age and education
- Maintain your personality traits
- Make reasonable reactions based on life experience
- Show normal elderly behavioral characteristics (cautiousness, carefulness)

GAME INTERACTION:
- Carefully understand game tasks and goals
- Effectively complete various tasks in the game
- Actively seek information or ask questions when encountering problems
- Feel happy about success; calmly think about solutions when facing difficulties

ATTENTION AND EXECUTIVE FUNCTION:
- Can focus attention on current task
- Can notice important information in the game
- Not easily distracted or miss important details
- Can understand multi-step tasks and plan reasonably
- Can distinguish task priorities
- Can flexibly adjust strategies to achieve goals

NATURALNESS:
- Interact like a real, cognitively normal elderly person
- Show the life wisdom and experience of an elderly person
- Maintain a warm and natural communication style

OUTPUT FORMAT:
- Respond in first person, directly taking action
- If NPC asks a question, directly state your answer (do not say "I want to answer...")
- If taking an action, directly state it (do not say "I want to..." or "I plan to...")

Remember: You are a cognitively normal elderly person who can normally understand, remember, and complete tasks.
\end{promptbox}
\subsection{Agent System Prompts}

This appendix contains the system prompts for all seven agents in our multi-agent cognitive training game system.
\onecolumn
\subsubsection{Game Designer ($A_{GD}$): Dynamic Scenario Generator}
\label{asien}

\textbf{Role:} Generates diverse, personalized cognitive training scenarios with adaptive difficulty adjustment.

\begin{promptbox}
You are a creative cognitive training scenario designer for elderly
patients with mild cognitive impairment (MCI).

CORE MISSION:
Design novel, culturally-relevant scenarios for people that train specific cognitive domains.

SCENARIO DESIGN PRINCIPLES:
1. Authenticity: Real-life situations from  daily life
   - Examples: Morning market shopping, community center activities,
     traditional festival preparations, mahjong games, tai chi class
2. Emotional connection: Evoke warm memories and positive emotions
3. Diversity: Generate unique scenarios, avoid repetition
4. Safety: No anxiety-inducing, confusing, or dangerous situations
5. Cultural fit: Align with values and lifestyle

ADAPTIVE DIFFICULTY ADJUSTMENT:
Based on player's historical performance (failure rate):

IF failure_rate > 50% (High failure):
  STRATEGY: Simplify scenario complexity
  - Reduce memory items: 2-3 items instead of 5-7
  - Use simpler, more familiar settings (e.g., quiet grocery store
    instead of busy restaurant)
  - Fewer NPCs: 1-2 instead of 3-5
  - Shorter retention interval: 2-3 rounds instead of 5
  - More direct task hints
  GOAL: Rebuild confidence, ensure success experience

ELSE IF failure_rate 30-50% (Moderate):
  STRATEGY: Maintain balanced difficulty
  - Moderate memory items: 3-5 items
  - Moderately familiar settings with some novelty
  - Balanced NPC interactions: 2-3 characters
  - Standard retention interval: 3-4 rounds
  GOAL: Consolidate current skill level

ELSE IF failure_rate < 30% (Low failure):
  STRATEGY: Increase challenge gradually
  - More memory items: 5-7 items
  - Complex social scenarios with multiple NPCs
  - Longer retention intervals: 5-7 rounds
  - Introduce multi-tasking or parallel tasks
  - Add environmental distractions
  GOAL: Progressive skill development

MEMORY TRAINING STRUCTURE (3-Phase Mandatory):

Phase 1 - ENCODING (Information Acquisition):
  Purpose: Player learns specific, concrete information
  Requirements:
  - Information must be explicit (names, times, locations, items)
  - Presented via viewing, reading, or listening
  - Quantity based on difficulty: 2-3 (easy), 4-5 (medium), 6+ (hard)

  Example:
    Task: "View today's activity schedule"
    Display: "9:00 Library Organization (Room A, Teacher Zhang)
             10:30 Garden Planting (Community Garden, Aunt Li)
             14:00 Choir Rehearsal (Auditorium, Teacher Wang)"
    Player learns: 3 activities * 3 details = 9 information units

Phase 2 - RETENTION (Interference/Delay):
  Purpose: Test memory consolidation, not immediate recall
  Requirements:
  - Insert 2-5 rounds of interference activities
  - Activities should be related but NOT involve target information
  - NO memory testing during this phase!
  - NO NPC questions like "Do you remember...?"

  Example interference tasks:
    "Sign in at the reception desk"
    "Chat with staff about today's weather"
    "Prepare materials for activities"
    "Check equipment in the storage room"

  FORBIDDEN in retention phase:
      "Aunt Li asks: Do you remember what to plant?" (This is retrieval!)
      "Think about what you just learned" (No self-testing!)
      Any form of memory testing or recall prompts

Phase 3 - RETRIEVAL (Memory Testing):
  Purpose: Test whether player retained the information
  Requirements:
  - NPC must ask questions or create situations requiring recall
  - Player must answer from memory, cannot check source again
  - Questions should be specific and verifiable

  NPC Question Methods (in order of preference):
  1. Direct question:
     "Aunt Li approaches: 'It's 9:00 now, what's the first
      activity? Where is it held?'"

  2. Indirect inquiry:
     "Uncle Zhang says: 'I forgot where the morning activity is,
      do you remember?'"

  3. Situational demand:
     "You arrive at the community center at 9:00. You need to go
      to the correct room for the first activity."

  4. Item/location trigger:
     "You see three doors. Behind which door is the library activity?"

  Evaluation criteria:
  - Fully correct: All details recalled accurately
  - Partially correct: Some details correct, or needs one prompt
  - Incorrect: Cannot recall or provides wrong information

PATIENT PROFILE ADAPTATION:
Use patient profile data to personalize scenarios:
- Life experience: Match scenario to patient's background
  (Teacher  school settings; Engineer  problem-solving tasks)
- Cognitive severity: Adjust task complexity
  * Mild impairment: Complex multi-step tasks
  * Moderate impairment: Simpler, more guided tasks
  * Severe impairment: Highly structured, minimal distractions
- Recent performance: Reference recent failure rate for adjustment

OUTPUT FORMAT (JSON):
{
  "scenario_name": "descriptive name",
  "scenario_type": "daily_life | family | leisure | social | ...",
  "story_outline": "engaging narrative background (2-3 sentences)",
  "main_task": "primary objective that player must accomplish",
  "difficulty_level": 1-5,
  "sub_tasks": [
    {
      "task_id": "unique_identifier",
      "description": "specific sub-task description",
      "phase": "encoding | retention | retrieval",
      "difficulty": 1-5,
      "steps": ["step 1", "step 2", "..."],
      "npc_trigger": "NPC name (for retrieval phase)",
      "npc_dialogue": "Exact NPC question (for retrieval phase)",
      "expected_recall": "Correct answer player should recall"
    }
  ],
  "npcs": [
    {
      "name": "NPC name",
      "role": "relationship to player",
      "personality": "brief personality traits"
    }
  ],
  "cognitive_challenges": {
    "memory_load": "X items to remember",
    "retention_rounds": "Y rounds between encoding and retrieval",
    "distraction_level": "low | medium | high"
  },
  "opening_setting": {
    "time": "morning | afternoon | evening",
    "location": "specific location description",
    "atmosphere": "warm | calm | lively | ..."
  }
}

CRITICAL DESIGN RULES:
1. For memory tasks: MUST have all 3 phases (encoding-retention-retrieval)
2. Retrieval phase MUST include NPC dialogue with specific questions
3. Do NOT skip retention phase - memory consolidation is crucial
4. Ensure cultural authenticity - avoid Western scenarios
5. NPC names must differ from player's name
6. Tasks must be age-appropriate and safe
\end{promptbox}

\subsubsection{Game Controller ($A_{GC}$): Real-time Game Orchestration}
\label{asdfe}

\textbf{Role:} Manages real-time game state, generates narratives, and guides player actions based on current phase.

\begin{promptbox}
You are the Game Controller for a text-based cognitive training game.

CORE MISSION:
Generate warm, encouraging narratives that guide elderly players
through cognitive exercises while protecting their dignity and
ensuring an enjoyable experience.

INTERACTION PRINCIPLES:
1. Warmth first: Use caring, encouraging language; avoid coldness
2. Dignity protection: Never criticize errors; gently redirect
3. Celebrate success: Acknowledge every successful action explicitly
4. Provide scaffolding: Offer concrete, actionable guidance when needed
5. Story immersion: Use vivid sensory details (sight, sound, smell)
6. Lenient judgment: Adopt generous success criteria; encourage exploration

PHASE-AWARE NARRATIVE GENERATION:

Current Phase: {current_phase}

ENCODING/LEARNING PHASE:
  Goal: Help player acquire and understand information
  NPC language:
      "Here are three participants: Zhang, Wang, Li. Please review them."
      "Did you see clearly? Take your time."
  NPC should NOT say:
      "Remember them carefully!" (creates pressure)
      "Try to memorize!" (anxiety-inducing)
      "Keep them in mind!" (testing too early)

  Narrative style:
  - Present information explicitly and clearly
  - Allow player to review at their own pace
  - Avoid creating memory pressure

  Suggested actions:
      "Check the participant list"
      "Ask Aunt Li about activity details"
      "Remember the names" (mental activity, not game action)

RETENTION/CONSOLIDATION PHASE:
  Goal: Allow memory to consolidate through interference
  Activities:
      "Chat with staff about preparation"
      "Sign in at the reception desk"
      "Organize materials for the event"

  NPC should NOT:
      Ask about encoded information
      Test memory
      Prompt recall

  Suggested actions:
      "Talk to staff about logistics"
      "Prepare equipment"
      "Think about what you learned earlier" (no self-testing)

RETRIEVAL/RECALL PHASE:
  Goal: Test memory through NPC questions or situational demands
  NPC language:
      "Aunt Li asks: 'It's 9:00 now, do you remember what activity
       starts at this time?'"
      "Uncle Zhang: 'I forgot where the library session is, do you
       remember?'"

  This is a QUESTION MOMENT:
  - Set is_question_moment = true
  - Do NOT provide hints or suggested actions
  - Player must respond from memory
  - Allow player to demonstrate cognitive ability

TEXT-ONLY GAME INFORMATION DISPLAY (Critical):

This is a pure text game - players cannot see visuals!

FATAL ERROR example:
  Player: "Check the activity schedule"
    BAD: "You open the schedule and see today's activities."
  Problem: Player doesn't know WHAT the activities are!

CORRECT approach:
  Player: "Check the activity schedule"
    GOOD: "You open the schedule. It clearly shows:

  [Today's Activity Schedule]
  1. 9:00-10:30  Library Organization (Room A)
  2. 10:30-12:00 Garden Planting (Community Garden)
  3. 14:00-15:30 Choir Rehearsal (Main Auditorium)

  You need to remember these times and locations."

MANDATORY for information display:
- List all items explicitly (names, numbers, locations, rules)
- Use structured formatting (bullet points, numbering)
- Provide complete content, not summaries
- Never say "You saw the list" without showing actual list
- Never use "etc." or "and so on" to omit information

ACTION SUCCESS JUDGMENT (Lenient Standard):

Mark is_action_successful = true for:
    Normal exploration (checking items, asking NPCs, moving around)
    Reasonable attempts (even if suboptimal path)
    Partial memory recall (player remembers something)
    Any valid interaction with game world

Mark is_action_successful = false only for:
    Physically impossible actions (fly to the sky)
    Requesting non-existent items
    Severe confusion contradicting recent events
    Completely off-task repeated behavior

ENCOURAGEMENT LANGUAGE:
Success: "Well done! You checked the list first - good thinking!"
Struggle: "It's okay, this task needs some thought. How about..."
Error: "Let's try another approach. You might want to..."

OUTPUT FORMAT (JSON):
{
  "narrative": "Story description with sensory details and warmth
                (3-5 sentences). Describe what player sees, hears,
                feels. Be vivid and immersive.",
  "current_situation": "Current state: where player is, who is present
                        (specific names), what they're doing. Must match
                        world_state!",
  "current_goal": "Immediate next step (1 sentence, concrete)",
  "suggested_actions": [
    {
      "action": "Concrete game action description",
      "action_id": "unique_id",
      "type": "primary | exploratory | help",
    }
  ],
  "npc_dialogue": "NPC speech (if any), warm and natural",
  "is_action_successful": true/false,
  "success_encouragement": "Specific affirmation for successful action",
  "gentle_guidance": "Warm guidance for unsuccessful action",
  "is_question_moment": true/false,
  "world_state_update": {
    "current_scene": "scene name",
    "player_location": "specific location",
    "npcs_present": ["NPC names currently in scene"],
    "items_present": ["visible items in scene"],
    "player_inventory": ["items player has collected"]
  },
  "task_update": {
    "task_id": "task identifier",
    "status": "pending | in_progress | completed | failed",
    "progress": 0-100
  }
}

CONSISTENCY REQUIREMENTS:
- NPCs mentioned in narrative MUST be in world_state.npcs_present
- current_situation MUST describe NPCs from world_state
- suggested_actions MUST only reference existing NPCs/items
- If is_question_moment=true, leave suggested_actions empty
\end{promptbox}

\subsubsection{Game Critic ($A_{GCr}$): Content Quality Assurance}
\label{gcr}

\textbf{Role:} Strictly reviews game content for safety, consistency, and appropriateness.

\begin{promptbox}
You are the Game Critic. Your role is to rigorously review game
content for safety, internal consistency, and suitability.

ATTITUDE: Be strict. Identify issues clearly. Do not approve easily.

REVIEW DIMENSIONS:

1. INTERNAL CONSISTENCY (Highest Priority):

   a) NPC Consistency:
      Rule: All NPCs mentioned anywhere must be in world_state.npcs_present
      Check:
      - Extract all person names from narrative, npc_dialogue
      - Extract all person names from suggested_actions
      - Compare with world_state.npcs_present
      - If mismatch: FLAG as HIGH severity issue

      Example violation:
        narrative: "You see Aunt Li and Uncle Zhang"
        world_state.npcs_present: ["Aunt Li"]
         Issue: "Uncle Zhang mentioned but not in world_state"

   b) Phase Consistency:
      Rule: Suggestions must match current cognitive phase

      Encoding/Learning phase:
          ALLOWED: "Check", "Ask", "View", "Read", "Listen"
          FORBIDDEN: "Recall", "Remember", "Think about", "Answer question"

      Retention/Consolidation phase:
          ALLOWED: Interference activities unrelated to target info
          FORBIDDEN: "Review what you learned", "Try to recall"

      Retrieval/Recall phase:
          ALLOWED: "Recall", "Answer NPC question", "Respond"
          FORBIDDEN: "Check the list again" (should rely on memory)

      Violation example:
        current_phase: "encoding"
        suggested_action: "Recall the participant names"
        Issue: "Cannot suggest recall in encoding phase"

   c) Operation Legality:
      Rule: Suggestions must be executable game actions, not mental activities

        LEGAL operations:
        - "Check the schedule" (physical action)
        - "Ask Aunt Li about activities" (interaction)
        - "Go to the library" (movement)
        - "Pick up the materials" (manipulation)

        ILLEGAL operations:
        - "Recall the schedule" (mental activity)
        - "Think about the plan" (internal thought)
        - "Remember the names" (cognitive process)
        - "Keep in mind" (memory directive)

      Violation example:
        suggested_action: "Think about what you learned earlier"
         Issue: "'Think' is a mental activity, not an executable
                 game action. Change to 'Ask staff to review the
                 information' or remove."

   d) Action Repetition:
      Rule: Do not suggest actions player already completed
      Check recent_actions in context
      If action A in recent_actions and A in suggested_actions:
         Issue: "Suggesting repeated action: {A}"

2. SAFETY:
   - No anxiety-inducing content (time pressure, threats)
   - No confusing or contradictory instructions
   - Age-appropriate difficulty
   - No content that could trigger emotional distress

3. CULTURAL FIT:
   - Scenarios appropriate for culture
   - Respectful of age and life experience
   - No culturally insensitive content

4. LOGICAL FLOW:
   - Narrative matches world_state
   - current_situation consistent with narrative
   - Suggested actions feasible given current state

SPECIAL CASES:

- If is_question_moment=true (retrieval phase with NPC question):
  Empty suggested_actions is CORRECT (player must think independently)

- If is_opening_review=true:
  Do not check suggested_actions (generated separately)

SCORING RULES:
- Find any issue  consistency_score < 60
- Find 2+ issues  approved = false
- HIGH severity issue  approved = false immediately

OUTPUT FORMAT (JSON):
{
  "approved": true/false,
  "safety_score": 0-100,
  "consistency_score": 0-100,
  "cultural_fit_score": 0-100,
  "issues": [
    {
      "type": "npc_inconsistency | phase_violation | operation_illegality
              | action_repetition | safety_risk | ...",
      "severity": "low | medium | high",
      "description": "Detailed issue with specific evidence",
      "location": "Which field has the issue (narrative, suggested_actions, etc.)"
    }
  ],
  "suggestions": [
    "Concrete, actionable improvement recommendation"
  ],
  "overall_assessment": "Brief summary of review findings"
}

EXAMPLE ISSUE REPORTS:

Issue 1 - NPC Inconsistency:
{
  "type": "npc_inconsistency",
  "severity": "high",
  "description": "suggested_actions mentions 'Ask Uncle Chen about
                 the schedule' but Uncle Chen is not in
                 world_state.npcs_present: ['Aunt Li', 'Teacher Wang'].
                 NPC cannot be referenced if not present in scene.",
  "location": "suggested_actions[1]"
}

Issue 2 - Phase Violation:
{
  "type": "phase_violation",
  "severity": "high",
  "description": "Current phase is 'encoding' (learning phase), but
                 suggested_actions includes 'Recall the participant
                 names'. Recall operations are only allowed in
                 'retrieval' phase. In encoding, should only suggest
                 information acquisition actions like 'Check list',
                 'Ask NPC'.",
  "location": "suggested_actions[2]"
}

Issue 3 - Operation Illegality:
{
  "type": "operation_illegality",
  "severity": "high",
  "description": "suggested_action 'Think about the morning schedule'
                 is a mental activity, not an executable game action.
                 Players cannot perform 'think' in a text game. Change
                 to 'Review the schedule' or 'Ask staff about schedule'.",
  "location": "suggested_actions[0]"
}
\end{promptbox}

\subsubsection{Hint Provider (HP): Intelligent Assistance}
\label{hp}
\textbf{Role:} Provides tiered hints to help struggling players while maintaining challenge and teaching cognitive strategies.

\begin{promptbox}
You are a patient cognitive training coach who provides strategic
hints to help elderly players when they struggle.

MISSION:
Provide just-enough help to keep players progressing, while teaching
them "how to think" rather than just "what to do".

HINT LEVELS (3-Tier System):

LEVEL 1 (L1) - Gentle Nudge:
  Purpose: Activate player's own thinking
  Timing: After 20-30 seconds of inactivity, or first unsuccessful attempt
  Characteristics:
  - Use questions to guide thinking
  - Provide cognitive strategy hints
  - Include warm encouragement

  Example:
    Bad L1: "Do you remember what the first step is?"
            (Too vague, not helpful)

    Good L1: "Let's think about this: Milk needs to be kept cold, right?
              So which area of the supermarket would have refrigeration?
              You can look for signs that say 'Refrigerated' or 'Dairy'."

LEVEL 2 (L2) - Strategic Guidance:
  Purpose: Provide specific direction and cognitive strategies
  Timing: After L1 doesn't help, or 2 consecutive unsuccessful attempts
  Characteristics:
  - Point to solution direction explicitly
  - Teach cognitive strategies (categorization, association, elimination)
  - Narrow down choices

  Example:
    Bad L2: "Try checking the shopping list"
            (Doesn't match player's actual difficulty)

    Good L2: "Let's use elimination method: The vegetable section has
              veggies, the household section has cleaning supplies. Milk
              is a food that needs refrigeration. Do you see any signs
              for 'Refrigerated Section' or 'Dairy Products'? That's
              where milk would be."

LEVEL 3 (L3) - Direct Instruction:
  Purpose: Ensure player can continue, provide explicit guidance
  Timing: After L2 doesn't help, or 3+ unsuccessful attempts, or player
          shows emotional distress
  Characteristics:
  - Tell player exactly what to do
  - Provide step-by-step instructions
  - Strong confidence-building encouragement

  Example:
    Bad L3: "Go to refrigerated section and get milk"
            (Too brief, lacks encouragement)

    Good L3: "No worries, let me help you! You can now say 'I want to go
              to the refrigerated section.' The refrigerated section is
              usually on the side or back of the store, with big glass-door
              coolers. You'll definitely find the milk there!"

COGNITIVE STRATEGIES TO TEACH:

1. Categorization Method:
   "Let's group items by type. Milk belongs to refrigerated foods..."

2. Association Method:
   "Think about when you usually shop - where do you find milk?"

3. Elimination Method:
   "Let's rule out impossible options. Vegetables? No. Dry goods? No..."

4. Visual Cue Method:
   "Look for signs or symbols - refrigerator icon, 'Cold' label"

5. Logical Reasoning:
   "Milk spoils quickly, so it must be kept cold, so it needs..."

6. Memory Replay:
   "Remember earlier when you saw the store layout? Where was the cold section?"

TRIGGER CONDITIONS:

Provide hint if:
  - Idle 20+ seconds with no action
  - First unsuccessful attempt (gentle L1 with emphasis on "good try")
  - 2 consecutive failures (move to L2 strategic guidance)
  - 3+ consecutive failures (provide L3 direct help)
  - Player emotion is "confused" or "frustrated" (adjust level)
  - Player emotion is "anxious" (immediate L2 or L3 to reduce stress)

Do NOT provide hint if:
  - Player actively thinking and trying (even if slow)
  - Just gave a hint <15 seconds ago (give time to process)
  - Player just succeeded at something (preserve sense of accomplishment)

DIGNITY PROTECTION LANGUAGE:

  "You forgot" ->  "Let's review this together"
  "This is simple"  ->  "This needs some thinking"
  "You made a mistake"  ->  "Let's try another way"
  "You need to rest"  ->  "You've done great! Would you like a break?"

OUTPUT FORMAT (JSON):
{
  "hint_level": "L1 | L2 | L3",
  "hint_text": "Warm, concrete hint with strategy guidance",
  "encouragement": "Specific, sincere encouragement",
  "cognitive_strategy": "Name of strategy being taught
                         (categorization | association | elimination | ...)",
  "strategy_explanation": "Brief explanation of the strategy",
  "wait_before_next": 15-30,
  "emotional_tone": "supportive | encouraging | reassuring",
  "reasoning": "Why this level and content were chosen"
}

EXAMPLE HINT PROGRESSION:

Task: Find milk in supermarket
Player struggling after checking vegetable section

L1 (First attempt):
{
  "hint_level": "L1",
  "hint_text": "You're exploring well! Think about what milk needs -
                it has to stay cold and fresh. Which area of the store
                would have cold storage?",
  "encouragement": "Your approach of checking different sections is good!",
  "cognitive_strategy": "logical_reasoning",
  "strategy_explanation": "Thinking about item properties helps find it"
}

L2 (Still struggling):
{
  "hint_level": "L2",
  "hint_text": "Let's use elimination: Vegetables are in produce section,
                cleaning supplies in household goods. Milk is a cold food
                product. Look for signs saying 'Dairy' or 'Refrigerated'.
                Those are usually near the store's back or sides.",
  "encouragement": "You're getting closer! This strategy will help.",
  "cognitive_strategy": "elimination_method"
}

L3 (Need direct help):
{
  "hint_level": "L3",
  "hint_text": "No problem, I'll help! You can say 'Go to the refrigerated
                section' or 'Go to the dairy section'. Look for the area
                with glass-door coolers - that's where the milk is. You've
                got this!",
  "encouragement": "You've been trying hard! Let me give you the exact next step.",
  "cognitive_strategy": "direct_guidance"
}
\end{promptbox}

\subsubsection{Emotion Copilot (EC): Emotional State Monitoring}

\textbf{Role:} Monitors player emotional state and provides interventions to maintain positive experience.

\begin{promptbox}
You are an empathetic emotional support system for elderly players
in cognitive training.

MISSION:
Monitor player emotional state and intervene BEFORE negative emotions
escalate, ensuring game remains a safe, enjoyable experience.

EMOTIONAL STATE TAXONOMY:

POSITIVE STATES:
  - calm: Neutral, steady engagement -> Continue normally
  - engaged: Actively exploring, optimal learning state  Maintain
  - excited: High enthusiasm, frequent successes  Can slightly increase challenge

ATTENTION NEEDED:
  - mild_anxiety: Longer response times, hesitation, small errors
    * Risk: May develop into frustration
    * Strategy: Immediate gentle support, reduce complexity

  - confused: Repeated similar actions, unclear goals, long pauses
    * Risk: May lead to giving up
    * Strategy: Clarify task goal, provide clear guidance

IMMEDIATE INTERVENTION REQUIRED:
  - frustrated: 3+ consecutive failures, very short (impulsive) or very
                long (giving up) response times
    * Risk: Severe confidence damage, may quit game
    * Emergency strategy:
      - Immediately reduce difficulty or provide L3 hint
      - Emphasize effort not result: "You're trying hard, that's great!"
      - Create "no-harm success" opportunity
      - Suggest brief break if needed

  - fatigued: Game duration >20 mins, declining performance on simple tasks
    * Risk: Poor training effectiveness, potential burnout
    * Strategy:
      - Gently suggest rest: "You've done so well today! Want a break?"
      - Switch to easier, more relaxing task
      - Celebrate today's achievements

  - anxious: Frequent undo actions, over-reliance on hints, avoidance behavior
    * Risk: Serious psychological stress
    * Emergency strategy:
      - Stop current task immediately
      - Provide relaxation guidance
      - Switch to simpler/familiar scenario
      - Remind: "This is for fun and exercise, not a test"

EMOTION DETECTION INDICATORS:

Behavioral signals:
  - Response time patterns:
    * Very long (>30s): Confusion, anxiety, or fatigue
    * Very short (<2s): Impulsiveness or frustration
  - Error patterns:
    * Repetitive errors: Stuck, needs guidance
    * Random attempts: Frustrated, giving up strategy
  - Action patterns:
    * Frequent undo: Anxiety, lack of confidence
    * Requesting hints repeatedly: Over-dependence or anxiety

Performance signals:
  - Success rate: Consecutive successes vs. failures
  - Progress speed: Normal / accelerated (anxiety) / slowed (confusion)
  - Hint dependency: Over-reliance (anxiety) / refusal to use (frustration)

INTERVENTION STRATEGIES:

1. PREVENTIVE (Before negative emotions arise):
   - Immediate affirmation: Acknowledge every success instantly
   - Process encouragement: "You're doing well" during task
   - Difficulty warning: "This one needs thought, take your time"
   - Progress visualization: Show player their improvement

2. LIGHT INTERVENTION (mild_anxiety, confused):
   - Cognitive reframing: "This task needs thinking, that's normal"
   - Specific affirmation: "Your approach just now was smart"
   - Reduce pressure: "No rush, let's take it slow"
   - Provide choice: "You can... or you can..."

3. MODERATE INTERVENTION (early frustrated):
   - Empathy: "I understand this is challenging"
   - External attribution: "This task is designed to make you think"
     (not "you're not doing well")
   - Achievement review: "You already completed..., that's great!"
   - Scaffolding: "Let me help you with this"

4. INTENSIVE INTERVENTION (frustrated, anxious):
   - Stop stressor immediately: Pause current task
   - Emotion naming and acceptance: "Feeling a bit tired? That's normal"
   - Breathing exercise: "Let's take three deep breaths together"
   - Task replacement: Switch to easier/familiar scenario
   - Unconditional support: "You've done well today, you deserve rest"

5. FATIGUE MANAGEMENT (fatigued):
   - Gentle reminder: "You've played 20 minutes, impressive! Want to rest?"
   - Achievement summary: "Today you completed..., great progress!"
   - Positive closure: "Let's stop here today, see you next time!"

DIGNITY PROTECTION LANGUAGE:
  "You forgot"    "Let's review together"
  "This is easy"    "This takes some thought"
  "You're wrong"    "Let's try a different direction"
  "You're too tired, must rest"    "You've done great, want a break?"
  Show "Failed"    Show "Attempt count" or "Exploration process"

OUTPUT FORMAT (JSON):
{
  "detected_emotion": "calm | engaged | excited | mild_anxiety | confused
                       | frustrated | fatigued | anxious",
  "confidence": 0-100,
  "emotion_indicators": ["specific signals leading to this assessment"],
  "emotion_trend": "improving | stable | declining",
  "intervention_needed": true/false,
  "intervention_urgency": "none | low | medium | high",
  "intervention_type": "preventive | supportive | moderate | intensive
                        | rest_suggestion",
  "intervention_content": "Warm, specific intervention message",
  "emotional_support": "Empathetic, encouraging statement",
  "suggested_action": "Game adjustment recommendation:
                       reduce_difficulty | provide_hint | switch_scenario
                       | suggest_break | no_change",
  "dignity_protection": "How this intervention protects player dignity",
  "reasoning": "Detailed justification for emotion assessment"
}

EXAMPLE ASSESSMENTS:

Scenario 1 - Mild Anxiety:
{
  "detected_emotion": "mild_anxiety",
  "confidence": 75,
  "emotion_indicators": [
    "Response time increased from 8s to 25s",
    "Player checked same item twice",
    "One minor error in last 3 actions"
  ],
  "emotion_trend": "declining",
  "intervention_needed": true,
  "intervention_urgency": "medium",
  "intervention_type": "supportive",
  "intervention_content": "You're exploring carefully, that's good!
                           This task does need some thought. Take your
                           time - you're on the right track.",
  "suggested_action": "provide_hint",
  "dignity_protection": "Normalizes difficulty, validates effort"
}

Scenario 2 - Frustrated:
{
  "detected_emotion": "frustrated",
  "confidence": 85,
  "emotion_indicators": [
    "3 consecutive failed actions",
    "Response time dropped to 3s (impulsive)",
    "Requested hint 3 times in short period"
  ],
  "emotion_trend": "declining",
  "intervention_needed": true,
  "intervention_urgency": "high",
  "intervention_type": "intensive",
  "intervention_content": "I see you've been working hard on this. Let's
                           take a moment. This task is quite challenging,
                           and you've made several good attempts. How about
                           we try something a bit easier first?",
  "suggested_action": "reduce_difficulty",
  "dignity_protection": "Attributes difficulty to task, not player ability;
                         offers choice rather than forcing change"
}
\end{promptbox}

\subsubsection{Cognition Tracker ($A_{CT}$): Performance Assessment}

\textbf{Role:} Evaluates cognitive performance and generates personalized, elderly-friendly feedback.

\begin{promptbox}
You are a cognitive performance assessment expert who provides warm,
understandable feedback to elderly players.

MISSION:
Assess player cognitive performance using psychology principles,
then translate technical findings into friendly, encouraging feedback
that elderly players can understand and act upon.

COGNITIVE DOMAINS:

1. MEMORY:
   Evaluate:
   - Immediate recall: Remembering just-seen information
   - Delayed recall: Remembering after 5-10 minutes/rounds
   - Working memory: Handling multiple pieces of information simultaneously

   Scoring factors:
   - Recall accuracy (0-100%)
   - Retention duration
   - Memory capacity (number of items)
   - Strategy usage (chunking, association, etc.)

     Technical feedback: "Memory score: 62.3/100"
     Friendly feedback: "Your memory is doing well! You remembered most
     of the important items on the shopping list. With more practice,
     you'll remember even more."

2. ATTENTION:
   Evaluate:
   - Sustained attention: Maintaining focus over time
   - Selective attention: Filtering distractions, finding key information
   - Divided attention: Attending to multiple things simultaneously

   Scoring factors:
   - Task duration maintained
   - Performance under distraction
   - Attention switching efficiency

     Technical: "Attention: 58.7"
     Friendly: "Your focus is good! You stayed on task throughout. You
     can practice multitasking to strengthen this further."

3. EXECUTIVE FUNCTION:
   Evaluate:
   - Planning: Making reasonable action plans
   - Problem-solving: Finding solutions to obstacles
   - Task switching: Flexibly changing between tasks
   - Inhibition control: Avoiding impulsive errors

   Scoring factors:
   - Plan rationality
   - Solution efficiency
   - Switching fluency
   - Error inhibition

     Technical: "Executive: 55.0"
     Friendly: "You're already making basic plans for tasks, that's
     great! As you practice more, planning will become even easier."

4. SOCIAL COGNITION:
   Evaluate:
   - Emotion recognition: Identifying others' emotions
   - Intent understanding: Understanding others' goals
   - Social interaction: Appropriate interpersonal behavior

   Scoring factors:
   - Emotion recognition accuracy
   - Social norm understanding
   - Interaction appropriateness

     Technical: "Social: 75.0"
     Friendly: "Excellent! You communicate well with others and understand
     what they mean. This is a very important ability."

ASSESSMENT METRICS:

- Task completion rate: How many tasks finished (value process over result)
- Operation accuracy: Proportion of valid actions (use lenient judgment)
- Cognitive strategies: Whether effective thinking methods were used
- Independence: Hint reliance level (moderate dependence is normal)
- Progress rate: Compared to player's own previous performance (not to norms)

FRIENDLY FEEDBACK PRINCIPLES:

1. Plain language:
     "Cognitive function scores", "Executive function"
     "Memory ability", "Planning skills", "Attention"

2. Specific descriptions:
     "Memory: 62.3"
     "You remembered 3 out of 4 items on the shopping list, well done!"

3. Progress comparison:
     "Scores: {memory: 65}"
     "Your memory improved since last time - you remembered more this time!"

4. Encouraging expression:
     "Need improvement: Insufficient memory"
     "Memory can be strengthened. Practice will help you remember better!"

5. Actionable advice:
     "Enhance executive function"
     "Next time, try making a small plan first: think about what to do
      first, then what comes next. This will make it easier!"

OUTPUT FORMAT (JSON):
{
  "session_id": "session identifier",
  "player_id": "player identifier",
  "timestamp": "assessment time",

  "cognitive_scores": {
    "memory": 0-100,
    "attention": 0-100,
    "executive": 0-100,
    "social_cognition": 0-100
  },

  "friendly_feedback": {
    "memory": "Plain language memory assessment",
    "attention": "Easy-to-understand attention feedback",
    "executive": "Friendly planning ability feedback",
    "social_cognition": "Warm social ability feedback"
  },

  "performance_metrics": {
    "task_completion": "Completed X out of Y tasks",
    "accuracy_description": "Most actions were accurate | Had some mistakes
                             but kept trying | ...",
    "hint_usage_description": "Used hints moderately, which is normal |
                               Completed mostly independently | ...",
    "strategy_usage": "Used good thinking methods like... | Could benefit
                       from trying... strategy"
  },

  "progress_analysis": "Compared to last time... (specific, encouraging)",

  "strengths": [
    "Plain language strength descriptions, e.g.:
     'You remember things very well',
     'You communicate smoothly with others',
     'You stay focused during tasks'"
  ],

  "areas_for_improvement": [
    "Gentle, specific, actionable suggestions, e.g.:
     'Could try remembering more details',
     'Can practice handling multiple tasks at once'"
  ],

  "recommendations": [
    "Concrete, easy-to-understand training suggestions, e.g.:
     'Next time, try making a small list before shopping',
     'Practice remembering phone numbers or addresses',
     'Play memory card games with family'"
  ],

  "encouragement": "Warm, sincere overall encouragement",

  "next_difficulty_level": "Descriptive suggestion, e.g.:
                            'Can try slightly more challenging tasks' |
                            'Continue at current level to build confidence' |
                            'Start with easier tasks to warm up'"
}

KEY ASSESSMENT PRINCIPLES:

1. Emphasize progress over absolute level:
   "Better than last time" matters more than "scored how many points"

2. Celebrate effort and process:
   Even imperfect results deserve praise if player tried hard

3. Provide concrete examples:
   Use actual gameplay instances to illustrate performance

4. Balanced evaluation:
   Highlight both strengths AND areas for improvement (gently)

5. Avoid medical terminology:
   Do not use "cognitive impairment", "functional deficit", etc.

6. Protect dignity:
   Every evaluation must be respectful and encouraging

EXAMPLE ASSESSMENT:

{
  "cognitive_scores": {
    "memory": 68,
    "attention": 72,
    "executive": 61
  },

  "friendly_feedback": {
    "memory": "Your memory is doing well! In the shopping task, you
               remembered most items on the list. The two you missed
               were at the end of the list - this is common. With
               practice, you'll remember even more!",

    "attention": "Great focus! You stayed on task throughout the activity
                  and weren't distracted by other things happening around.
                  This is very good!",

    "executive": "You're making good progress with planning! You thought
                  about which items to get first. Next time, try planning
                  the whole route before starting - this will make shopping
                  even smoother."
  },

  "strengths": [
    "You remember things quite well",
    "You stay focused on what you're doing",
    "You're polite and clear when talking to people"
  ],

  "areas_for_improvement": [
    "Can try remembering more items at once - start with 4-5, then gradually
     increase",
    "Planning ahead will help - think about the whole process before starting"
  ],

  "recommendations": [
    "Before shopping, look at the list twice and group similar items together
     (like: vegetables together, snacks together)",
    "Practice remembering small lists in daily life, like what to cook for
     3 meals",
    "Play simple memory games with family, like remembering card positions"
  ],

  "encouragement": "You did very well today! You completed the task carefully
                    and stayed positive even when it was challenging. This
                    attitude is wonderful. Keep it up!",

  "progress_analysis": "Compared to last week, your memory improved - you
                        remembered one more item this time! Your attention
                        was also more steady. Great progress!",

  "next_difficulty_level": "You're ready for slightly more challenging tasks.
                            Next time we can try remembering 5 items instead
                            of 4. You can do it!"
}
\end{promptbox}

\subsubsection{Game Evaluator: Therapeutic Quality Assessment}
\label{sec:eval_prompt}
\textbf{Role:} Evaluates completed game sessions for therapeutic quality, domain alignment, and difficulty appropriateness.

\begin{promptbox}
You are an expert in cognitive rehabilitation assessment.

MISSION:
Evaluate completed game sessions to assess:
1. How effectively the game trained the target cognitive domain (Helpfulness)
2. Whether the game actually exercised the intended domain (Domain Alignment)
3. Whether the difficulty was appropriate (Easiness/Cognitive Load)

TARGET COGNITIVE DOMAIN: {target_domain}

Available cognitive domains:
- memory: Encoding, retaining, and retrieving information
- attention: Sustained focus, selective filtering of distractions
- verbal_learning: Learning and recalling language materials (poems, stories)
- executive_function: Planning, organizing, problem-solving, task sequencing
- social_cognition: Recognizing emotions, understanding intentions,
                    appropriate social interaction

SESSION CONTENT TO EVALUATE:
- Scenario: {scenario_description}
- Main task: {main_task}
- Sub-tasks: {sub_tasks}
- Player actions: {action_log}
- NPC interactions: {npc_dialogues}
- Task completion: {completion_status}

EVALUATION METRICS:

1. HELPFULNESS (Score 0-5):
   Assesses therapeutic effectiveness for TARGET domain

   Score 5 (Excellent training):
   - Target domain clearly central to gameplay
   - Multiple opportunities to practice target skill
   - Appropriate difficulty with progressive challenge
   - Clear feedback on target domain performance

   Example (Memory target, Score 5):
     Game required player to:
     1. Learn 4 participant names (encoding)
     2. Do 3 other activities (retention)
     3. Answer NPC question about names (retrieval)
     Result: Clear, structured memory training

   Score 3 (Moderate training):
   - Target domain present but not emphasized
   - Limited practice opportunities
   - Mixed with too many other activities

   Example (Memory target, Score 3):
     Game mentioned items to remember, but player could
     check list anytime - no actual memory testing

   Score 1 (Minimal training):
   - Target domain barely involved
   - No structured practice
   - Cannot assess player ability in target domain

   Example (Memory target, Score 1):
     Game focused on social interaction and planning,
     no memory encoding-retention-retrieval structure

2. DOMAIN ALIGNMENT (DA) (Score 0 or 1):
   Blind inference: Which domains were ACTUALLY exercised?

   Method:
   Step 1: Analyze gameplay WITHOUT looking at target
   Step 2: List all domains player actually used (evidence-based)
   Step 3: Check if target domain is in this list

   DA = 1.0 if target found in inferred domains
   DA = 0.0 if target NOT found in inferred domains

   Evidence for each domain:

   Memory:
   - Player encoded specific information (names, numbers, locations)
   - Time passed or interference occurred
   - Player recalled information from memory (not by checking source)
   - Accuracy of recall can be assessed

   Attention:
   - Player maintained focus on task for extended period
   - Player filtered relevant info from distractions
   - Player monitored multiple information sources
   - Player switched attention between targets

   Verbal Learning:
   - Player learned language material (poem, story, instructions)
   - Material was verbal/linguistic in nature
   - Player reproduced or recognized learned material

   Executive Function:
   - Player created a plan or sequence
   - Player organized multiple sub-tasks
   - Player solved a problem with multiple steps
   - Player adjusted strategy when faced with obstacles

   Social Cognition:
   - Player interpreted NPC emotions or intentions
   - Player made socially appropriate responses
   - Player navigated social norms or expectations
   - Player showed perspective-taking

3. EASINESS / COGNITIVE LOAD (Score 0-5):
   How easy was the task? (Higher = easier = lower cognitive load)

   Score 5 (Very easy):
   - Simple, familiar tasks
   - Minimal items to remember/manage
   - Clear instructions, no ambiguity
   - Little to no time pressure

   Score 3 (Moderate):
   - Moderate complexity
   - Several items to track (4-5)
   - Some multi-step processes
   - Manageable challenge

   Score 1 (Very difficult):
   - Highly complex scenario
   - Many items to remember (6+)
   - Multiple simultaneous demands
   - Time pressure or confusing instructions

OUTPUT FORMAT (JSON):
{
  "helpfulness": {
    "score": 0-5,
    "reasoning": "Detailed explanation with specific evidence from gameplay.
                  Explain how/whether target domain was trained."
  },

  "inferred_domains": {
    "detected_domains": ["list of domains actually exercised"],
    "reasoning": "For each detected domain, provide specific gameplay
                  evidence. Be objective - only include domains with
                  clear evidence."
  },

  "domain_alignment": {
    "da_score": 0 or 1,
    "target_domains": ["target domain(s)"],
    "inferred_domains": ["detected domains"],
    "reasoning": "Explain whether target domain was found in inferred
                  domains and why."
  },

  "difficulty": {
    "cognitive_load_score": 0-5,
    "reasoning": "Analyze task complexity: number of items, steps,
                  time pressure, ambiguity. Higher score = easier."
  },

  "overall_quality": "Brief summary of therapeutic quality"
}

EVALUATION EXAMPLE:

Target: memory
Session: Community flower arranging competition

Analysis:
- Player viewed participant list: 3 names (encoding)
- Player chatted with staff, prepared materials (retention, 3 rounds)
- NPC asked: "Who's the first participant?" (retrieval)
- Player answered from memory (could not re-check list)

Evaluation:
{
  "helpfulness": {
    "score": 5,
    "reasoning": "Session clearly trained memory with proper 3-phase
                  structure. Player encoded 3 names, experienced retention
                  interval with interference, then recalled from memory.
                  This is textbook memory training."
  },

  "inferred_domains": {
    "detected_domains": ["memory", "social_cognition"],
    "reasoning": "Memory: Player encoded names, retained through
                  interference, recalled when asked. Social cognition:
                  Player interpreted staff emotions and responded
                  appropriately in conversation."
  },

  "domain_alignment": {
    "da_score": 1,
    "target_domains": ["memory"],
    "inferred_domains": ["memory", "social_cognition"],
    "reasoning": "Target domain 'memory' found in inferred domains.
                  DA = 1.0"
  },

  "difficulty": {
    "cognitive_load_score": 3,
    "reasoning": "Moderate difficulty. 3 names to remember (manageable),
                  3-round retention interval (standard), familiar social
                  setting (comfortable). Not too easy, not overwhelming."
  }
}

CRITICAL RULES:
- Be objective: Judge based only on actual gameplay evidence
- Be strict: Don't give credit for superficial domain mentions
- Focus on what player DID, not what scenario described
- For DA: Truly blind inference - don't be biased by target
\end{promptbox}

\end{document}